\documentclass[master, english]{hgbthesis}

\usepackage{tikz}
\usetikzlibrary{arrows}
\usepackage{mathrsfs}
\usepackage{caption}
\usepackage{makeidx}
\usepackage{etaremune}
\usepackage[textwidth=13.5cm]{geometry}
\usepackage{stmaryrd}
\usepackage[titletoc,title,page]{appendix}
\usepackage[utf8]{inputenc}
\usepackage{adjustbox}
\usepackage{maple2e}
\usepackage{etoolbox}
\usepackage{tabularx}
\usepackage{array}
\newcolumntype{L}[1]{>{\raggedright\let\newline\\\arraybackslash\hspace{0pt}}m{#1}}
\newcolumntype{C}[1]{>{\centering\let\newline\\\arraybackslash\hspace{0pt}}m{#1}}
\newcolumntype{R}[1]{>{\raggedleft\let\newline\\\arraybackslash\hspace{0pt}}m{#1}}
\usepackage{hyperref}
\usepackage{nomencl}
\makenomenclature

\usepackage{etoolbox}
\renewcommand\nomgroup[1]{%
	\item[\bfseries
	\ifstrequal{#1}{A}{Abbreviations}{%
	\ifstrequal{#1}{N}{Number Sets}{%
	\ifstrequal{#1}{R}{Rings/Fields/Algebra}{%
	\ifstrequal{#1}{O}{Operators}{%
	\ifstrequal{#1}{V}{Variables}{%
	\ifstrequal{#1}{S}{Other Symbols}{%
}}}}}}
]
\goodbreak
}

\makeatletter
\patchcmd{\@chap@pppage}{\thispagestyle{plain}}{\thispagestyle{empty}}{}{}
\makeatother
\usepackage{titlesec}    
\titleformat{\chapter}[display]
{\normalfont%
    \huge
    \bfseries}{\chaptertitlename\ \thechapter}{20pt}{%
    \huge 
    }

\makeatletter
\renewcommand\part{%
	\if@openright
	\cleardoublepage
	\else
	\clearpage
	\fi
	\thispagestyle{empty}
	\if@twocolumn
	\onecolumn
	\@tempswatrue
	\else
	\@tempswafalse
	\fi
	\null\vfil
	\secdef\@part\@spart}
\makeatother
\RequirePackage[utf8]{inputenc}		

\logofile{arr}

\DefineParaStyle{Maple Output}
\DefineCharStyle{2D Math}
\DefineCharStyle{2D Output}

\makeindex
\begin{document}

\title{Definite Sums of\\[1ex] Hypergeometric Terms\\[1ex]and\\[1ex] Limits of P-Recursive Sequences}
\author{Hui Huang}
\studiengang{Institute for Algebra\\[1ex]Johannes Kepler University Linz}
\studienort{Linz}
\abgabedatum{2017}{01}{01}	


\frontmatter
\newgeometry{left=20mm,total={200mm,257mm},top=10mm}
\thispagestyle{empty}
\begingroup
%
%
\newif\ifeng
\engfalse
%
%
%
\def\title{Definite Sums of\\Hypergeometric Terms\\and Limits of\\P-Recursive Sequences}
%
%
\def\type{0}
%
%
%
%
%
\def\degree{Doktorin der Naturwissenschaften}
%
%
%
%
%
\def\study{Naturwissenschaften}
%
%
%
%
%
\def\name{Hui Huang}
%
%
%
\def\institute{Institut f\"ur Algebra}
%
%
%
%
\def\supervisor{Univ.-Prof.\ Dr.\  \\Manuel Kauers}
\newif\ifsupvismale
\supvismaletrue
%
\def\secondexaminer{Prof.\ Dr.\ Ziming Li}
\newif\ifsecexmale
\secexmaletrue
%
%
%
\def\assist{Prof.\ Dr.\ Ziming Li}
%
%
%
\def\date{Januar 2017}
%
%
\def\ifundefined#1{\expandafter\ifx\csname#1\endcsname\relax}
\DeclareFontShape{OT1}{cmss}{m}{n}
  {<5><6><7><8><9><10><10.95><12><14.4><17.28><20.74><24.88><29.86><35.83>%
   <42.99><51.59><67><77.38>cmss10}{}
\DeclareFontShape{OT1}{cmss}{bx}{n}
  {<5><6><7><8><9><10><10.95><12><14.4><17.28><20.74><24.88><29.86><35.83>%
   <42.99><51.59><67><77.38>cmssbx10}{}
\makeatletter
\def\Huge{\@setfontsize\Huge{29.86pt}{36}}
\makeatother
\unitlength 1cm
\sffamily
\begin{picture}(16.7,0)
\ifeng
 \put(11.5,-2.5){\includegraphics[width=5.2cm]{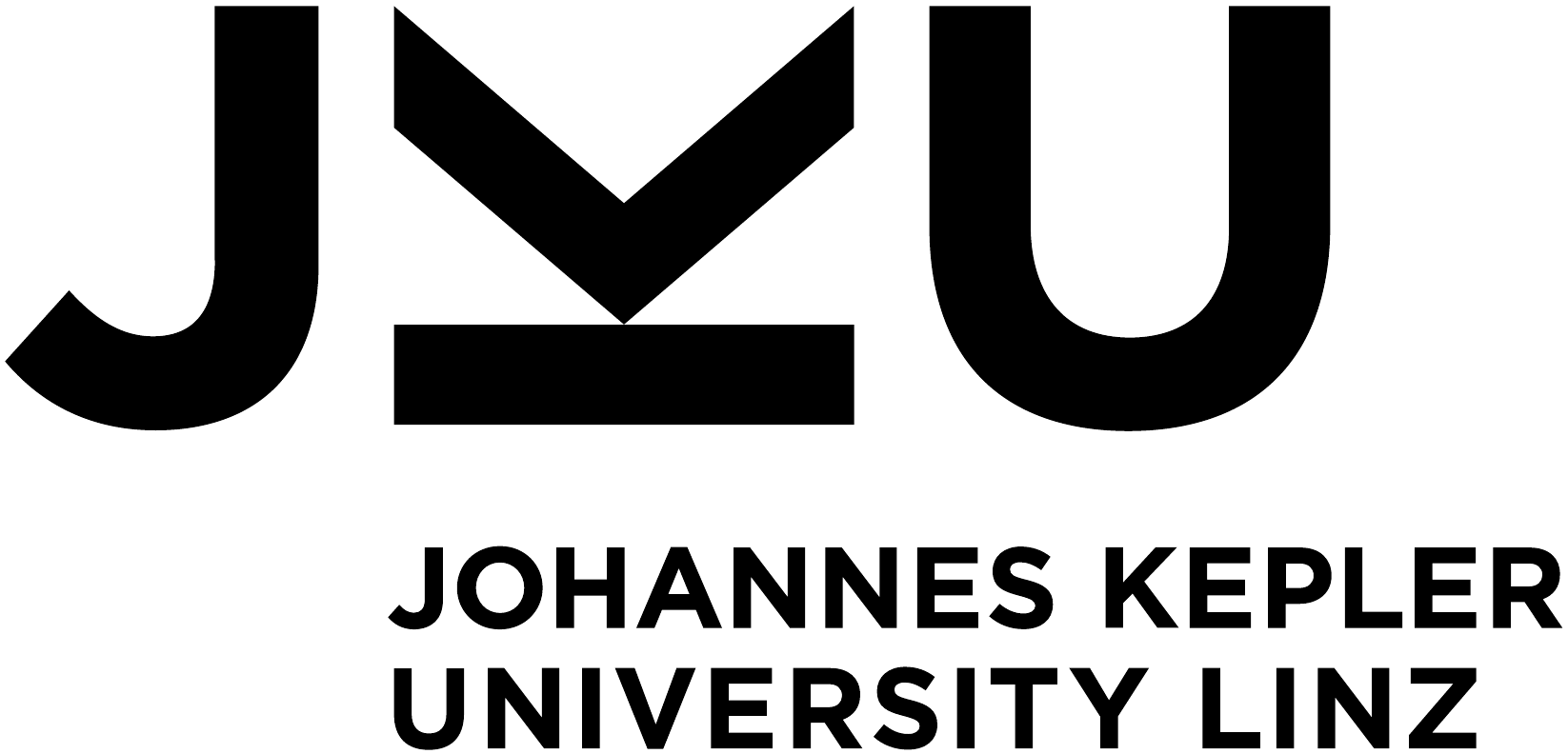}}
\else
 \put(11.5,-2.5){\includegraphics[width=5.2cm]{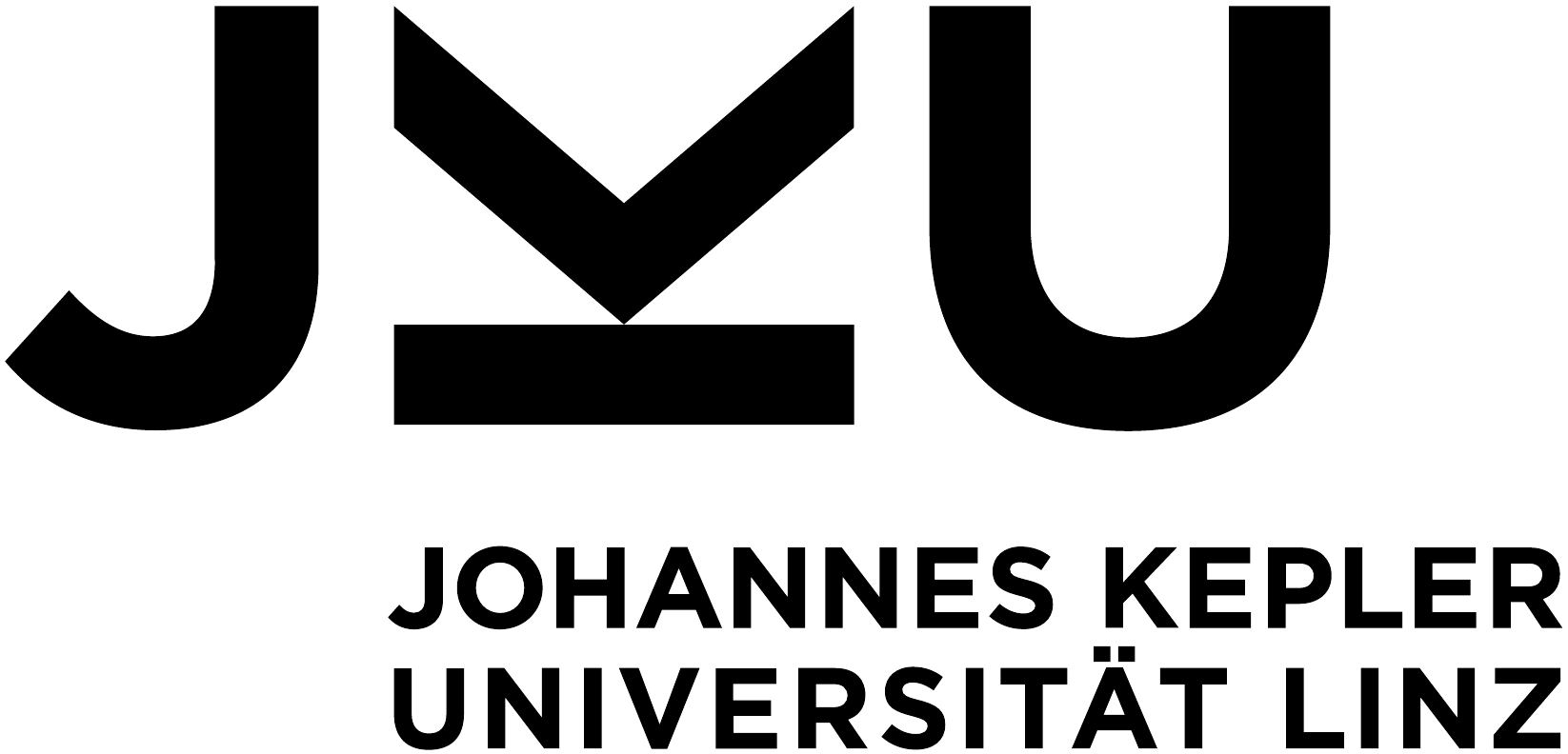}}
\fi
\put(12.9,-4.2){\begin{minipage}[t]{3.9cm}\footnotesize%
\ifeng
 Submitted by\\
\else
 Eingereicht von\\
\fi
{\bfseries\name}%
\vskip 4mm%
\ifeng
 Submitted at\\
\else
 Angefertigt am\\
\fi
{\bfseries\institute}%
\vskip 4mm%
\ifcase\type%
 \ifeng
  Supervisor and\\ First Examiner\\
 \else
  \ifsupvismale%
   Betreuer und\\ Erstbeurteiler\\
  \else
   Betreuerin und\\ Erstbeurteilerin\\
  \fi
 \fi
 {\bfseries\supervisor}%
 \vskip 4mm%
 \ifeng
  Second Examiner\\
 \else
  \ifsecexmale%
   Zweitbeurteiler\\
  \else
   Zweitbeurteilerin\\
  \fi
 \fi
 {\bfseries\secondexaminer}%
\else
 \ifeng
  Supervisor\\
 \else
  \ifsupvismale%
   Betreuer\\
  \else
   Betreuerin\\
  \fi
 \fi
 {\bfseries\supervisor}%
\fi
\vskip 4mm%
\ifundefined{assist}\else
 \ifeng
  Co-Supervisor\\
 \else
  Mitbetreuung\\
 \fi
 {\bfseries\assist}%
\vskip 4mm%
\fi
\date
\end{minipage}}
\put(12.9,-23.5){\begin{minipage}[t]{3.9cm}\footnotesize%
{\bfseries JOHANNES KEPLER\\
\ifeng
 UNIVERSITY
\else
 UNIVERSIT\"AT
\fi
LINZ}\\
Altenbergerstra{\ss}e 69\\
4040 Linz, \"Osterreich\\
www.jku.at\\
DVR 0093696
\end{minipage}}
\put(0,-12.2){\begin{minipage}[b]{12cm}\Huge\bfseries\title\end{minipage}}
\put(0,-17.2){\includegraphics[width=4.4cm]{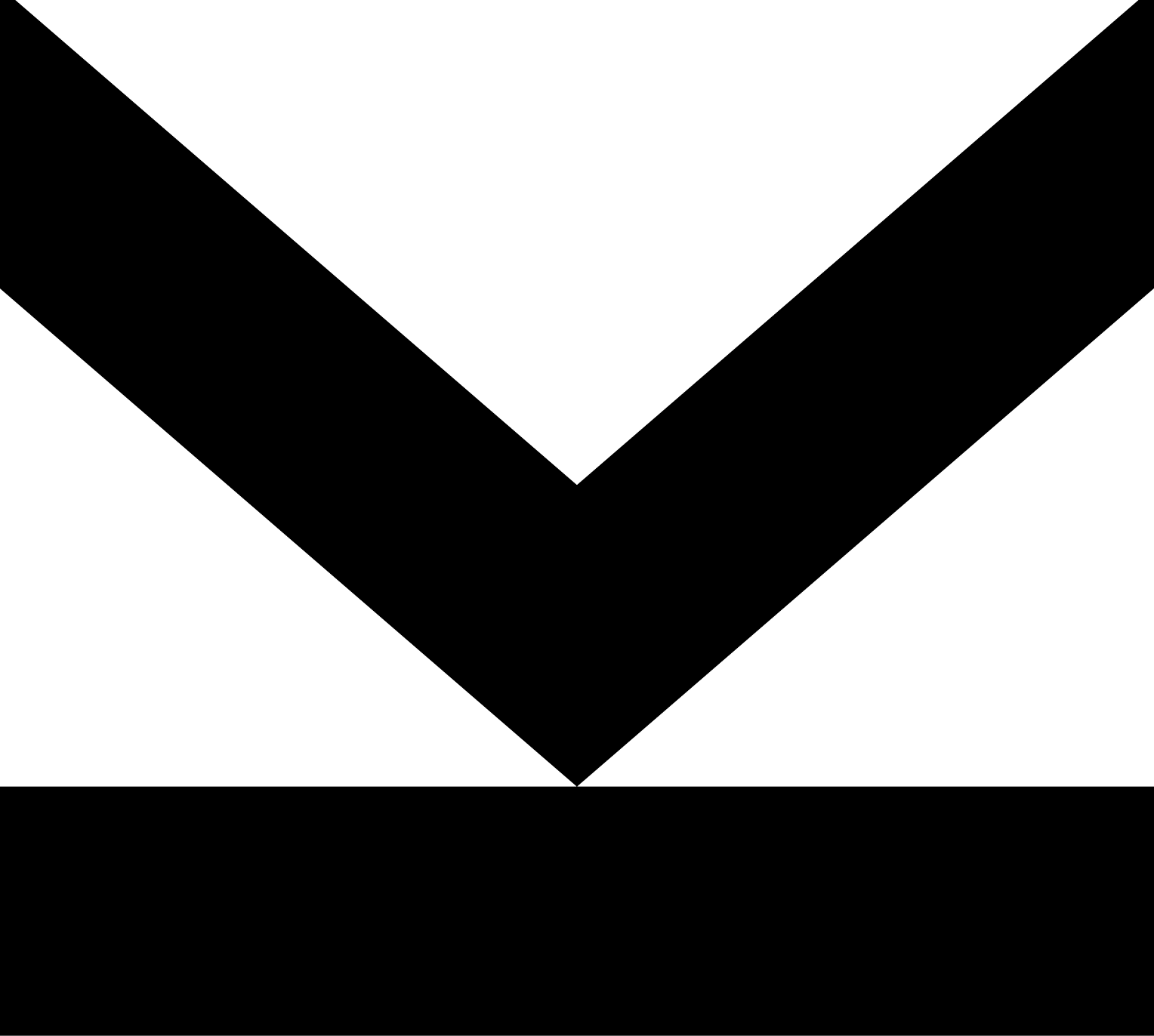}}
\put(0,-18.3){\begin{minipage}[t]{12cm}%
\ifeng
 {\large\ifcase\type Doctoral \or Diploma \or Master \fi Thesis}%
 \vskip 2mm%
 to obtain the academic degree of%
 \vskip 3mm%
 {\large\degree}
 \vskip 3mm%
 in the \ifcase\type Doctoral \or Diploma \or Master's \fi Program
\else
 {\large\ifcase\type Dissertation\or Diplomarbeit\or Masterarbeit\fi}%
 \vskip 2mm%
 zur Erlangung des akademischen Grades%
 \vskip 3mm%
 {\large\degree}
 \vskip 3mm%
 im \ifcase\type Doktoratsstudium \or Diplomstudium\or Masterstudium\fi
\fi
\vskip 3mm%
{\large\study}
\end{minipage}}
\end{picture}
\endgroup
\restoregeometry

\maketitle

\frontmatter
\newgeometry{top=20mm,textwidth=14cm}

\setcounter{page}{1}
\chapter{Abstract}\label{CH:abstract}

The ubiquity of the class of D-finite functions and P-recursive
sequences in symbolic computation is widely recognized. This class is
defined in terms of linear differential and difference equations with
polynomial coefficients. In this thesis, the presented work consists
of two parts related to this class.

In the first part, we generalize the reduction-based creative
telescoping algorithms to the hypergeometric setting.
This generalization allows to deal with definite sums of
hypergeometric terms more quickly.

The \ap reduction computes an additive decomposition of a given
hypergeometric term, which extends the functionality of Gosper's
algorithm for indefinite hypergeometric summation. We modify this
reduction so as to decompose a hypergeometric term as the sum of a
summable term and a non-summable one. Properties satisfied by the
output of the original reduction carry over to our modified
version. Moreover, the modified reduction does not solve any auxiliary
linear difference equation explicitly.
	
Based on the modified reduction, we design a new algorithm to compute
minimal telescopers for bivariate hypergeometric terms.  This new
algorithm can avoid the costly computation of certificates, and
outperforms the classical Zeilberger algorithm no matter whether
certificates are computed or not according to the computational
experiments.
	
We further employ a new argument for the termination of the above new
algorithm, which enables us to derive order bounds for minimal
telescopers.  Compared to the known bounds in the literature, our
bounds are sometimes better, and never worse than the known ones.

In the second part of the thesis, we study the class of
D-finite numbers, which is closely related to D-finite functions and
P-recursive sequences.  It consists of the limits of convergent
P-recursive sequences.  Typically, this class contains many well-known
mathematical constants in addition to the algebraic numbers.  Our
definition of the class of D-finite numbers depends on two subrings of
the field of complex numbers.  We investigate how different choices
of these two subrings affect the class.  Moreover, we show that
D-finite numbers over the Gaussian rational field are essentially the
same as the values of D-finite functions at non-singular algebraic number arguments
(so-called the regular holonomic constants).  
This result makes it easier to recognize certain numbers as belonging to this
class.

%

\chapter{Zusammenfassung}\label{CH:zusammenfassung}

Die Allgegenwart der Klasse der D-finiten Funktionen und der
P-rekursiven Folgen im Gebiet
des Symbolischen Rechnens ist allgemein bekannt. Diese Klasse ist
definiert durch lineare
Differential- und Differenzengleichungen mit polynomiellen
Koeffizienten. Die Ergebnisse
dieser Arbeit bestehen aus Teilen, die mit dieser Klasse zu tun haben.

Im ersten Teil verallgemeinern wir die reduktions-basierten Algorithmen
f\"ur
\emph{creative telescoping} auf den hypergeometrischen Fall. Diese
Verallgemeinerung
erlaubt eine effizientere Behandlung von definiten Summen
hypergeometrischer Terme.

Die Abramov-Petkov{\v s}ek-Reduktion berechnet eine additive Zerlegung
eines gegebenen
hypergeometrischen Terms, durch die die Funktionalit\"at des
Gosper-Algorithmus f\"ur
indefinite hypergeometrische Summen erweitert. Wir adaptieren diese
Reduktion so, dass
sie einen hypergeometrischen Term in einen summierbaren und einen
nichtsummierbaren
Term zerlegt. Eigenschaften des Outputs der urspr\"unglichen Zerlegung
bleiben f\"ur
unsere modifizierte Version erhalten. Dar\"uber hinaus braucht man bei
der modifizierten
Reduktion keine lineare Hilfsrekurrenz explizit zu l\"osen.

Ausgehend von der modifizierten Reduktion entwickeln wir einen neuen
Algorithmus zur
Berechnung minimaler Telescoper f\"ur bivariate hypergeometrische Terme.
Dieser neue
Algorithmus can die teure Berechnung von Zertifikaten vermeiden, und
gem\"a\ss\ unserer
Experimente l\"auft er schneller als der klassische
Zeilberger-Algorithmus, egal ob man
Zertifikate mitberechnet oder nicht.

Wir verwenden au\ss erdem ein neues Argument f\"ur die Terminierung der
genannten neuen
Algorithmen, das es uns erlaubt, Schranken f\"ur die Ordnung des
minimalen Telescopers
herzuleiten. Verglichen mit den bekannten Schranken in der Literatur
sind unsere Schranken
manchmal besser und nie schlechter als die bekannten.

Im zweiten Teil der Arbeit untersuchen wir die Klasse der D-finiten
Zahlen, die eng
verwandt mit D-finiten Funktionen und P-rekursiven Folgen ist. Sie
besteht aus den
Grenzwerten der konvergenten P-rekursiven Folgen. Typischerweise
enth\"alt diese Klasse
neben den algebraischen Zahlen viele weitere bekannte mathematische
Konstanten. Unsere
Definition der Klasse der D-finiten Zahlen h\"angt von zwei Unterringen
des K\"orpers
der komplexen Zahlen ab. Wir untersuchen, wie die Klasse von der Wahl
dieser zwei
Unterringe abh\"angt. Au\ss erdem zeigen wir, dass die D-finiten Zahlen
\"uber dem
K\"orper der Gau\ss schen rationalen Zahlen im wesentlichen dieselben
Zahlen sind,
die auch als Werte von D-finiten Funktionen an nicht-singul\"aren
algebraischen
Argumenten auftreten (die sogenannten regul\"aren holonomen Konstanten).
Dieses Resultat
erleichtert es, gewisse Zahlen als Elemente der Klasse zu erkennen.

%

\chapter*{Acknowledgments}\label{CH:acknowledgments}

I would like to express my deepest gratitude to my two co-supervisors:
Manuel Kauers and Ziming Li, for their academic guidance, constant
support and sincere advices. I thank Manuel, for giving me the
opportunity to benefit from his immense knowledge, excellent
programming skills, amazing scientific insights and high enthusiasm
for math. I thank Ziming, for sharing his rigorous scientific
attitude, assisting with mathematical and other matters, training my
speaking skills over and over again with great patience, and also
providing countless valuable suggestions.

My special thanks go to Shaoshi Chen, from whom I profited a lot. I
thank him very much for his useful suggestions and constructive
comments, which subsequently improved my work considerably. Besides, I
am impressed with his obsession with books and high enthusiasm for
math.

I was very lucky to be a student of the two lectures \lq\lq Computer
Algebra for Concrete Mathematics\rq\rq\ and \lq\lq Algorithmic
combinatorics\rq\rq\ given by Peter Paule.  I thank him for his enlightening
lessons and valuable encouragement.  I also learned much from
discussions with Hao Du, Ruyong Feng, Christoph Koutschan and Stephen Melczer.
I would especially like to thank Hao Du for improving my code.

I really appreciate many members of the Key Laboratory of Mathematics
Mechanization for their help, particularly, Wen-tsun Wu for creating
this beautiful subject and Xiaoshan-Gao for making our lab a
magnificent place to work in.  I also appreciate all of my colleagues
at the Institute for Algebra as well as my former colleagues at
RISC. Special thanks are due to all secretaries for their assistance,
and to my friends: Zijia, Miriam, Peng, Ronghua, Rika, Liangjie, for
making my life in Linz awesome.

Moreover, I wish to thank Mark Giesbrecht, George Labahn and {\'E}ric
Schost, for accepting me as a Postdoctoral Fellow in the Symbolic
Computation Group at the University of Waterloo.

Most importantly, I would like to dedicate this work to my beloved
parents, who make all these possible. Thank you for your unbounded
love and care. I thank all my family, for their selfless support and
for their understanding and appreciation of my work.

\bigskip\noindent This work was supported by the Austrian Science Fund
(FWF) grant W1214-N15 (project DK13), two NSFC grants (91118001,
60821002/F02) and a 973 project (2011CB302401).

 \tableofcontents 

\mainmatter         

\chapter{Introduction}
\label{CH:intro}

\section{Background and motivation}\label{SEC:background}

Using computer instead of human thought is one of the main themes in
the study of symbolic computation for the past century. In particular,
finding algorithmic solutions for problems about special functions is
one of the very popular topics nowadays.

As an especially attractive class of special functions, D-finite
functions have been recognized long
ago~\cite{Stan1980,Lips1989,Zeil1990b,SaZi1994,Mall1996,Stan1999}.
They are interesting on the one hand because each of them can be
easily described by a finite amount of data, and efficient algorithms
are available to do exact as well as approximate computations with
them.  On the other hand, the class is interesting because it covers a
lot of special functions which naturally appear in various different
context, both within mathematics as well as in applications.

The defining property of a {\em D-finite} function%
\index{D-finite! function}\index{function! D-finite} is that it satisfies a linear
differential equation with polynomial coefficients.  This differential
equation, together with an appropriate number of initial terms,
uniquely determines the function at hand.  Similarly, a sequence is
called {\em P-recursive} (or rarely, {\em D-finite})%
\index{P-recursive sequence}\index{sequence! P-recursive} if it
satisfies a linear recurrence equation with polynomial coefficients.
Also in this case, the equation together with an appropriate number of
initial terms uniquely determine the object.

The set of P-recursive sequences covers a lot of important
combinatorial sequences, including C-finite sequences, hypergeometric
terms and sequences whose generating functions are algebraic (called
algebraic sequences in this thesis).
Rather than talking about sequences themselves, our main interest
focus on their definite sums and limits. This thesis is divided
into two components.
\[\star\, \star\, \star\, \star\, \star\]
{\bf Part I. Hypergeometric terms.}  The set of hypergeometric terms
is a basic and powerful class of P-recursive sequences. It is defined
to be the nonzero solutions of first-order (partial) difference
equations with polynomial coefficients. Many familiar functions are
hypergeometric terms, for instance, nonzero rational functions,
exponential functions, factorial terms, binomial coefficients, etc. In
the study of symbolic summation, there are mainly two kinds of
problems related to hypergeometric terms.

\begin{problem}[Hypergeometric summation]\label{PROB:summation}
  \index{hypergeometric! summation} 
  Investigate whether or not the following sum is expressible 
  in simple \lq\lq closed form\rq\rq,
  \begin{equation}\label{EQ:summation}
    \sum_{k=a}^b f(n,k),\quad f(n,k) \text{ is a bivariate hypergeometric term in } n,k,
  \end{equation}
  where $a, b$ are fixed constants independent of all variables. By a
  closed form, we mean a linear combination of a fixed number of
  hypergeometric terms, where the fixed number must be a constant
  independent of all variables.
\end{problem}

\begin{problem}[Hypergeometric identities]\label{PROB:identity}
  \index{hypergeometric! identity} 
  Prove the following identity
  \begin{equation}\label{EQ:identity}
    \sum_{k= a}^{b} f(n,k) = h(n), \quad f(n,k) \text{ is a bivariate hypergeometric term in } n,k,
  \end{equation}
  where $a, b$ are fixed constants independent of all variables, and
  $h(n)$ is a known univariate function.
\end{problem}
Analogous to the first fundamental theorem of calculus,
Problem~\ref{PROB:summation} could be solved in terms of indefinite
summation provided that there exists a so-called \lq\lq
anti-difference\rq\rq. More precisely, we compute a hypergeometric
term $g(n,k)$ such that
\[f(n,k) = g(n,k+1) - g(n,k),\] 
and then Problem~\ref{PROB:summation} easily follows by the
telescoping sum technique. To our knowledge, the first complete
algorithm for indefinite summation was designed by
Gosper~\cite{Gosp1978} in 1978, namely the famous Gosper's
algorithm. To address the case when Gosper's algorithm is not
applicable, i.e., there exists no such $g$, Wilf and Zeilberger
developed a constructive theory in a series of 
articles~\cite{WiZe1990a,WiZe1990b,WiZe1992a,WiZe1992b,Zeil1990a,Zeil1990b,Zeil1991}
in early 1990s. This theory came to be known as Wilf-Zeilberger's
theory\index{Wilf-Zeilberger's theory}, whose main idea is to
construct a so-called telescoper for~$f$ to derive a difference
equation with polynomial coefficients satisfied by
\eqref{EQ:summation}, and then applying Petkov{\v s}ek's
algorithm~\cite{Petk1992}, which detects the existence of the
hypergeometric terms solutions, to this equation gives the final
answer for Problem~\ref{PROB:summation}.

On the other hand, Wilf-Zeilberger's theory also works for
Problem~\ref{PROB:identity}. To be precise, after deriving a
difference equation satisfied by the left-hand side of~\eqref{EQ:identity} 
from a telescoper as for Problem~\ref{PROB:summation}, 
we verify that $h$ satisfies the same equation and then 
\eqref{EQ:identity} easily follows by checking the initial values.

Wilf-Zeilberger's theory not only provides an algorithmic method to
solve the problems about hypergeometric summations or identities, but
also gives a constructive way to find new combinatorial identities. In
terms of algorithms, Wilf-Zeilberger's theory is a strong fundamental
tool for combinatorics and also the theory of special functions.

From the above discussion, one sees that the key step of
Wilf-Zeilberger's theory is to construct a telescoper. This process is
referred to as {\em creative telescoping}.\index{creative telescoping|(} %
To be more specific, for a bivariate hypergeometric
term $f(n,k)$, the task consists in finding some nonzero recurrence
operator $L$ and another hypergeometric term~$g$ such that
\begin{equation}\label{EQ:telescoper}
L \cdot f(n,k) = g(n,k+1)-g(n,k).
\end{equation}
It is required that the operator $L$ does not contain $k$ or the shift
operator $\sigma_k$, i.e., it must have the form $L =
e_0+e_1\sigma_n+\dots+e_\rho \sigma_n^\rho$ for some $e_0, \dots,
e_\rho$ that only depend on $n$. If $L$ and $g(n,k)$ are as above, we
say that $L$ is a \emph{telescoper}\index{telescoper} for~$f(n,k)$,
and $g(n,k)$ is a \emph{certificate}\index{certificate} for $L$.

As outlined in the introduction of \cite{CHKL2015}, we can distinguish
four generations of creative telescoping algorithms.

{\bf The first generation}~\cite{Fase1947, Zeil1990b, PWZ1996,
  ChSa1998} dates back to the 1940s, and the algorithms were based on
elimination techniques. {\bf The second generation}~\cite{Zeil1990a,
  AlZe1990, Zeil1991, PWZ1996} started with what is now known as
Zeilberger's (fast) algorithm\index{Zeilberger's algorithm}. The
algorithms of this generation use the idea of augmenting Gosper's
algorithm\index{Gosper's algorithm} for indefinite summation (or
integration) by additional parameters $e_0, \dots, e_\rho$ that are
carried along during the calculation and are finally instantiated, if
at all possible, such as to ensure the existence of a certificate $g$
in \eqref{EQ:telescoper}. These algorithms have been implemented in
many computer algebra programs, for example {\sc
  Maple}\index{Maple@{\sc Maple}}~\cite{ACGL2004} and {\sc
  Mathematica}\index{Mathematica@{\sc Mathematica}}~\cite{PaSc1995}. 
See~\cite{PWZ1996} for details about the first two generations.

{\bf The third generation}~\cite{MoZe2005,ApZe2006} was initiated by
Apagodu and Zeilberger. In a sense, they applied a second-generation
algorithm by hand to a generic input and worked out the resulting
linear system of equations for the parameters $e_0, \dots, e_\rho$ and
the coefficients inside the certificate $g$. Their algorithm then
merely consists in solving this system. This approach is interesting
not only because it is easier to implement and tends to run faster
than earlier algorithms, but also because it is easy to analyze. In
fact, the analysis of algorithms from this family gives rise to the
best output size estimates for creative telescoping known so
far~\cite{ChKa2012a,ChKa2012b,CKK2014}.
A disadvantage is that these algorithms may not always find the
smallest possible output.

{\bf The fourth generation} of the creative telescoping algorithms,
so-called reduction-based algorithms\index{reduction-based},
originates from~\cite{BCCL2010}. The basic idea behind these
algorithms is to bring each term $\sigma_n^if$ of the left-hand side
of \eqref{EQ:telescoper} into some kind of normal form modulo all
terms that are differences of other terms. Then to find
$e_0,\dots,e_\rho$ amounts to finding a linear dependence among these
normal forms. The key advantage of this approach is that it separates
the computation of the $e_i$ from the computation of~$g$. This is
interesting because a certificate is not always needed, and it is
typically much larger (and thus computationally more expensive) than
the telescoper, so we may not want to compute it if we don't have to.
With previous algorithms there was no way to obtain telescopers
without also computing the corresponding certificates, but with fourth
generation algorithms there is.  So far this approach has only been
worked out for several instances in the differential
case~\cite{BCCL2010,BLS2013,BCCLX2013}. The goal of the first part of
the present thesis is to give a fourth-generation algorithm for the
shift case, namely for the classical setting of hypergeometric
telescoping.  \index{creative telescoping|)}
\[\star\, \star\, \star\, \star\, \star\]
{\bf Part II. D-finite numbers.}%
\index{algebraic! number}\index{number! algebraic}%
\index{algebraic! function}\index{function! algebraic} In a sense, the theory of
D-finite functions generalizes the theory of algebraic functions. Many
concepts that have first been introduced for the latter have later
been formulated also for the former. In particular, every algebraic
function is D-finite (Abel's theorem)\index{Abel's theorem}, and many
properties the class of algebraic function enjoys carry over to the
class of D-finite functions.

The theory of algebraic functions in turn may be considered as a
generalization of the classical and well-understood class of algebraic
numbers. The class of algebraic numbers suffers from being relatively
small. There are many important numbers, most prominently the numbers
$\e$ and~$\pi$, which are not algebraic.

Many larger classes of numbers have been proposed, let us just mention
three examples.  The first is the class of periods\index{period} (in the
sense of Kontsevich and Zagier~\cite{KoZa2001}).  These numbers are
defined as the values of multivariate definite integrals of algebraic
functions over a semi-algebraic set.  In addition to all the algebraic
numbers, this class contains important numbers such as~$\pi$, all zeta
constants (the Riemann zeta function evaluated at an integer)%
\index{zeta constant}\index{Riemann zeta function}\index{function! Riemann zeta}
and multiple zeta values\index{multiple zeta values}, but it is so far
not known whether for example~$\e$, $1/\pi$ or Euler's
constant~$\gamma$ are periods (conjecturally they are not).  
The second example is the class of all numbers that appear as values of
so-called G-functions (in the sense of Siegel~\cite{Sieg2014}) at
algebraic number arguments~\cite{FiRi2014,FiRi2016a}.  The class of
G-functions\index{G-function} is a subclass of the class of D-finite
functions, and it inherits some useful properties of that class.
Among the values that G-functions can assume are $\pi$, $1/\pi$,
values of elliptic integrals and multiple zeta values, but it is so
far not known whether for example~$\e$, Euler's constant $\gamma$ or a
Liouville number are such a value (conjecturally not).

Another class of numbers is the class of holonomic constants, studied
by Flajolet and Vall{\' e}e~\cite[\S 4]{FlVa2000}.  (We thank Marc
Mezzarobba for pointing us to this reference.)  A number is {\em
  holonomic}\index{holonomic constant} if it is equal to the (finite)
value of a D-finite function at an algebraic point.  The number is
further called a {\em regular holonomic constant} if the evaluation
point is an ordinary point of the defining differential equation of
the given D-finite function; otherwise it is called a {\em singular
  holonomic constant}.  Typical examples of the regular holonomic
constants are $\pi$, $\log(2)$, $\e$ and the polylogarithmic value
$\Li_4(1/2)$; while several famous constants like Ap{\' e}ry's
constant~$\zeta(3)$, Catalan's constant $\G$ are of singular type.

It is tempting to believe that there is a strong relation between
holonomic constants and limits of convergent P-recursive sequences.
To make this relation precise, we introduce the class of D-finite
numbers in this thesis.  Let~$R$ be a subring of~$\set C$ and $\set F$ be a subfield
of~$\set C$.  A complex number $\xi$ is called
\emph{D-finite}\index{D-finite! number}\index{number! D-finite}
(w.r.t.~$R$ and~$\set F$) if it is the limit of a convergent sequence
in $R^{\set N}$ which is P-recursive over~$\bF$.  We denote
by~$\cD_{R,\bF}$ the set of all D-finite numbers with respect to $R$
and $\set F$.

It is clear that $\cD_{R,\set F}$ contains all the elements of~$R$,
but it typically contains many further elements.  For example, let $i$
be the imaginary unit\index{imaginary unit}, then $\cD_{\set Q(i)}$
contains many (if not all) the periods and, as we will see below, many
(if not all) the values of G-functions.  In addition, it is not hard
to see that $\e$ and $1/\pi$ are D-finite numbers.  According to
Fischler and Rivoal's work~\cite{FiRi2016a}, also Euler's
constant~$\gamma$ and any value of the Gamma function at a rational
number are D-finite.  (We thank Alin Bostan for pointing us to this
reference.)

The definition of D-finite numbers given above involves two subrings
of $\set C$ as parameters: the ring to which the sequence terms of the
convergent sequences are supposed to belong, and the field to which
the coefficients of the polynomials in the recurrence equations should
belong.  Obviously, these choices matter, because we have, for
example, $\cD_{\set R,\set R}=\set R\neq\set C=\cD_{\set C,\set C}$.  Also, since
$\cD_{\set Q,\set Q}$ is a countable set, we have $\cD_{\set Q,\set Q}\neq\cD_{\set
  R,\set R}$.  On the other hand, different choices of $R$ and $\set F$ may
lead to the same classes.  For example, we would not get more numbers
by allowing $\set F$ to be a subring of~$\set C$ rather than a field,
because we can always clear denominators in a defining recurrence.
One of our goals is to investigate how $R$ and $\set F$
can be modified without changing the resulting class of D-finite
numbers.

As a long-term goal, we hope to establish the notion of D-finite
numbers as a class that naturally relates to the class of D-finite
functions in the same way as the classical class of algebraic numbers
relates to the class of algebraic functions.

\section{Main results and outline}\label{SEC:results}
This section is intended to provide an outline of the thesis and the
main results.

In Chapter~\ref{CH:htprelim}, we recall basic notions and facts about
hypergeometric terms.

In Chapter~\ref{CH:apreduction}, our starting point is the \ap
reduction for hypergeometric terms introduced in
\cite{AbPe2001a,AbPe2002b}. Unfortunately the reduced forms obtained
by this reduction are not sufficiently \lq\lq normal\rq\rq\ for our
purpose. Therefore, we present a modified version of the reduction
process, which does not solve any auxiliary linear difference equation
explicitly like the original one and totally separates the summable
and non-summable parts of a given hypergeometric term. The outputs of
the \ap reduction and our modified version share the same required
properties. According to the experimental comparison, the modified
reduction is also more efficient than the original one.

Chapter~\ref{CH:rfproperties} is mainly used to connect univariate
hypergeometric terms with bivariate ones for later use. We explore
some important properties of discrete residual forms by means of
rational normal forms~\cite{AbPe2002b}. Furthermore, we show that the
residual forms are well-behaved with respect to taking linear
combinations.

We translate terminology concerning univariate hypergeometric terms to
bivariate ones in Chapter~\ref{CH:telescoping}. Based on the modified
version of \ap reduction in Chapter~\ref{CH:apreduction}, we present a
new algorithm to compute minimal telescopers for bivariate
hypergeometric terms. This new algorithm keeps the key feature of the
fourth generation, that is, it separates the computations of
telescopers and certificates. Experimental results illustrate that
the new algorithm is faster than the classical Zeilberger's algorithm
if it returns a normalized certificate; and the new algorithm is much
more efficient if it omits certificates.

In Chapter~\ref{CH:bounds}, we present a new argument for the
termination of the new algorithm in Chapter~\ref{CH:telescoping}. This
new argument provides an independent proof of the existence of
telescopers and even enables us to obtain upper and lower bounds for
the order of minimal telescopers for hypergeometric terms. Compared to
the known bounds in the literature, our bounds are sometimes better
and never worse than the known ones. Moreover, we present a variant of
the new algorithm by combining our bounds, which improves the new
algorithm in some special cases.

In Chapter~\ref{CH:dfprelim}, we review basic notions and useful
properties of the class of D-finite functions and P-recursive
sequences mainly from~\cite{FlSe2009,KaPa2011}.

In Chapter~\ref{CH:dfinitenos}, we study the class of D-finite
numbers, defined as the limits of convergent P-recursive sequences.
In general, this class is much larger than the class of algebraic
numbers.
The definition of the class depends on two subrings of the field of
complex numbers. We investigate the possible choices of these two
subrings that keep the class unchanged.  Moreover, we connect this
class with the class of holonomic constants~\cite{FlVa2000} and show
that D-finite numbers over the Gaussian rational field are essentially
the same as the regular holonomic constants. With this result,
certain numbers are easily recognized as belonging to this class,
including many periods as well as many values of G-functions.

\section{Remarks}\label{SEC:remark}
The main results in Chapters~\ref{CH:apreduction} -- \ref{CH:telescoping}
are joint work with S.\ Chen, M.\ Kauers and Z. Li, which have been
published in \cite{CHKL2015}. The main results in
Chapter~\ref{CH:bounds} were published in \cite{Huan2016}. The main
results in Chapter~\ref{CH:dfinitenos} are joint work with M.\ Kauers,
and are in preparation~\cite{HuKa}.


\part[Definite Sums of Hypergeometric Terms]{Definite Sums of \\Hypergeometric Terms}
\chapter{Hypergeometric Terms}
\label{CH:htprelim}

In this chapter, we recall basic notions and facts on difference rings
(fields) and hypergeometric terms. In addition, we review the context
of summability and multiplicative decomposition for hypergeometric
terms. These topics are well-known and more details can be found in
\cite{Ore1930,Cohn1965}.
\section{Basic concepts}\label{SEC:ht}
Let $\bF$ be a field of characteristic zero, and $\bF(k)$ be the field
of rational functions in~$k$ over~$\bF$. Let $\sigma_k$ be the
automorphism that maps $r(k)$ to $r(k+1)$ for every rational function
$r\in \bF(k)$. The pair $(\bF(k),\sigma_k)$ is called a
\emph{difference field}\index{difference! field}. A \emph{difference
  ring extension}\index{difference! ring} of $(\bF(k),\sigma_k)$ is a
ring $\bD$ containing $\bF(k)$ together with a distinguished
endomorphism $\sigma_k: \bD\rightarrow \bD$ whose restriction to
$\bF(k)$ agrees with the automorphism defined before. An element $c\in
\bD$ is called a constant if~$\sigma_k(c) = c$. It is readily seen
that all constants in $\bD$ form a subring of~$\bD$, denoted by
$C_{\sigma_k, \bD}$. In particular, $C_{\sigma_k,\bD}$ is a field
whenever $\bD$ is one. Moreover, we have $C_{\sigma_k, \bF(k)} = \bF$ according to
\cite[Theorem~2]{AbPe2002a}. In
other words, the set of all constants in $\bF(k)$ w.r.t.~$\sigma_k$ is
exactly the field $\bF$.

Throughout the thesis, for a polynomial $p \in \bF[k]$, its
degree\index{degree} and leading coefficient%
\index{leading coefficient} are denoted by $\deg_k(p)$ and $\lc_k(p)$,
respectively. For convenience, we define the degree of zero to be $-\infty$.

\begin{definition}\label{DEF:ht}
  Let~$\bD$ be a difference ring extension of~$\bF(k)$. A nonzero
  element~$T \in \bD$ is called a {\em hypergeometric
    term}\index{hypergeometric! term} over~$\bF(k)$ if it is invertible and
  $\sigma_k(T) = r T$ for some~$r \in \bF(k)$. We call~$r$ the 
  {\em shift-quotient}\index{shift-! quotient} of~$T$ w.r.t.~$k$.
\end{definition}
In the following two chapters, whenever we mention hypergeometric
terms, they always belong to some difference ring extension~$\bD$ of
$\bF(k)$, unless specified otherwise.

\begin{example}\label{EX:ht}
  All nonzero rational functions are hypergeometric. Moreover, the
  following two classes of combinatorial functions are also
  hypergeometric.
  \begin{enumerate}
  \item (Exponential functions).%
  \index{exponential function}\index{function! exponential} $T = c^k$ where $c\in
    \bF\setminus\{0\}$. The shift-quotient of $T$ is~$\sigma_k(T)/T = c$.

    \smallskip
  \item (Factorial terms).\index{factorial term} $T = (a k)!$ with
    $a\in \bN$ and $a>0$. The shift-quotient of $T$ is $\sigma_k(T)/T
    =(a k+a)(a k+a-1)\cdots(ak+1)$.
  \end{enumerate}
\end{example}

One can easily show that the product of hypergeometric terms and the
reciprocal of a hypergeometric term are again hypergeometric. However,
the sum of hypergeometric terms is not necessarily hypergeometric.
For example, $2^k + 1$ is not a hypergeometric term although~$2^k$ and
$1$ both are; otherwise we would have $(2^{k+1}+1)/(2^{k}+1)\in \bF(k)$, 
and then a straightforward calculation would yield that $2^k\in \bF(k)$, a contradiction.

Recall~\cite{Ore1930,PWZ1996} that two hypergeometric terms~$T_1, T_2$
over $\bF(k)$ are called {\em similar}\index{similar} if there exists
a rational function~$r\in \bF(k)$ such that~$T_1 = r T_2$. This is an
equivalence relation and all rational functions form one equivalence
class.  By Proposition~5.6.2 in~\cite{PWZ1996}, the sum of similar
hypergeometric terms is either hypergeometric or zero.

\section{Hypergeometric summability}\label{SEC:summability}
Analogous to indefinite integrals of elementary functions in calculus,
we consider indefinite sums of hypergeometric terms in shift
case. More precisely, given a hypergeometric term~$T(k)$, we compute
another hypergeometric term $G(k)$ such that
\[T(k) = G(k+1)-G(k).\] 
This motivates the notion of hypergeometric summability.%
\index{hypergeometric! summability|see {summable}}\index{summability|see {summable}}
\begin{definition}\label{DEF:summable}
  A univariate hypergeometric term~$T$ over $\bF(k)$ is called {\em
    hypergeometric summable},%
	\index{summable}\index{hypergeometric! summable|see {summable}} if there exists another hypergeometric
  term~$G$ such that
  \[T = \Delta_k(G),\quad \text{where}\ \Delta_k \ \text{denotes the
    difference of~$\sigma_k$ and the identity map}.\] We call $G$ an
  {\em indefinite summation}\index{indefinite summation} (or {\em
    anti-difference}\index{anti-difference|see {indefinite summation}}) of $T$. 
  If $T$ and $G$ are both rational functions,
  we also say $T$ is {\em rational summable}.%
  \index{rational! summable}\index{summable}
\end{definition}
We abbreviate \lq\lq hypergeometric summable\rq\rq\ as \lq\lq
summable\rq\rq\ in this thesis.
\begin{example}\label{EX:htsummable}
  All polynomials are summable. Moreover, we see that $k\cdot k!$ is
  summable since~$k\cdot k! = \Delta_k(k!)$, but $k!$ is not which
  will be shown in Example~\ref{EX:nonsummable}.
\end{example}

To solve the problem of indefinite summation, Gosper~\cite{Gosp1978}
developed a first complete algorithm which is known as Gosper's
algorithm\index{Gosper's algorithm}. This is a deterministic
procedure. It determines whether or not the input hypergeometric term
is summable, and then returns an indefinite summation if the answer is
yes. The basic idea is to reduce the summation problem to finding
polynomial solutions of a first-order difference equation with
polynomial coefficients.
\section{Multiplicative decomposition}\label{SEC:multidecom}
By~\cite{AbPe2001a,AbPe2002b}, every hypergeometric term admits a
multiplicative decomposition. This enables us to analyze a
hypergeometric term by rational functions.  To recall it, let us first
review the notion of shift-free polynomials and shift-reduced rational
functions~\cite[\S 1]{AbPe2001a}.
\begin{definition}\label{DEF:shiftfree}
  A nonzero polynomial $p\in \bF[k]$ is said to be {\em
    shift-free}\index{shift-! free} if for any nonzero integer $i$, we
  have $\gcd(p, \sigma_k^i(p))=1$.
\end{definition}
Consequently, no two distinct roots of a shift-free polynomial differ
by an integer. The following lemma indicates the relation between
shift-freeness and rational summability, whose proof can be found in
\cite[Proposition~1]{Abra1971}.
\begin{lemma}\label{LEM:rationalsummable}\index{rational! summable}
  Let $f = p/q$ be a rational function in $\bF(k)$, where $p,q\in
  \bF[k]$ are coprime and $\deg_k(p)<\deg_k(q)$. Further assume that
  $q$ is shift-free. If there exists a rational function $r\in \bF(k)$
  such that $f= \Delta_k(r)$, then $f=0$.
\end{lemma}

\begin{definition}\label{DEF:shiftreduced}
  A nonzero rational function $f = p/q\in \bF(k)$ with $p,q\in \bF[k]$
  coprime, is said to be {\em shift-reduced}\index{shift-! reduced} if
  for any integer $i$, we have $\gcd(p, \sigma_k^i(q))=1$.
\end{definition}
Some basic properties of shift-reduced rational functions are given
below.
\let\oldtheenumi=\theenumi
\let\oldlabelenumi=\labelenumi

\renewcommand{\theenumi}{\roman{enumi}}
\renewcommand{\labelenumi}{(\theenumi)}
\begin{lemma} \label{LEM:shiftreduced}
  Let~$f \in \bF(k)$ be shift-reduced.
  \begin{enumerate}
  \item\label{LEM:shiftreduced1} If there exists a nonzero rational function $r\in
    \bF(k)$ such that $f = \sigma_k(r)/r$, then~$r\in \bF$ and thus
    $f=1$.

   \smallskip
  \item\label{LEM:shiftreduced2} If $f\neq 1$ and there exists $r \in
    \bF[k]$ such that $f \sigma_k(r)-r = 0$, then $r=0$.
  \end{enumerate}
\end{lemma}
\begin{proof}
  \begin{enumerate}
  \item Suppose that $r = s/t \in \bF(k)\setminus\bF$, where $s, t$
  are coprime and at least one of them does not belong to
  $\bF$. W.l.o.g., we assume that $s \notin\bF$. Then there exists a
  nontrivial factor $p\in \bF[k]$ of $s$ such that $\deg_k(p) >
  0$. Let
  \[\ell = \min\{k \in \bZ: \sigma_k^{k}(p) \mid s\}\quad \text{and} \quad m = \max\{k \in \bZ: \sigma_k^k(p) \mid s\}. \]
  It follows that $m, \ell \geq 0$ and
  \begin{itemize}
  \item $\sigma_k^{-\ell}(p) \mid s $ but $\sigma_k^{-\ell}(p) \nmid \sigma_k(s)$;

    \smallskip
  \item $\sigma_k^{m+1}(p)\mid \sigma_k(s)$ but $\sigma_k^{m+1}(p)\nmid s$.
  \end{itemize}
  Since $s$ and $t$ are coprime, so are $\sigma_k(s)$ and
  $\sigma_k(t)$. Note that
  \[f = \frac{\sigma_k(r)}{r} = \frac{\sigma_k(s) t}{s \sigma_k(t)}.\]
  Hence $\sigma_k^{m+1}(p)$ is a nontrivial factor of the numerator of
  $f$ and $\sigma_k^{-\ell}(p)$ is a nontrivial factor of the
  denominator of $f$, a contradiction as $f$ is shift-reduced.
	
  \medskip
  \item Suppose that~$r \neq 0$. Then
  \[f =\frac{r}{\sigma_k(r)} = \frac{\sigma_k(1/r)}{1/r}.\] 
  Since~$f$ is unequal to one, $1/r$ does not belong to~$\bF$. 
  It follows from $(i)$ that~$f$ is not shift-reduced, a contradiction.
  \end{enumerate}
\end{proof}

According to~\cite{AbPe2001a, AbPe2002b}, every hypergeometric
term~$T$ admits a {\em multiplicative decomposition}%
\index{decomposition! multiplicative}\index{multiplicative decomposition}~$S H$, 
where~$S$ is in~$\bF(k)$ and~$H$ is another hypergeometric term whose
shift-quotient is shift-reduced. We call the shift-quotient~$K
:=\sigma_k (H) /H$ a {\em kernel}\index{kernel} of~$T$ w.r.t.~$k$
and~$S$ a corresponding {\em shell}\index{shell}. 
By Lemma~\ref{LEM:shiftreduced}~$(i)$, we know that $K=1$ if and only if~$T$ is a
rational function, which is then equal to~$cS$ for some constant $c\in
C_{\sigma_k, \bD}$. Here $\bD$ is a difference ring extension
of~$\bF(k)$.

Let~$T=SH$ be a multiplicative decomposition, where~$S$ is a rational
function and~$H$ a hypergeometric term with a kernel~$K$.  Assume
that~$T=\Delta_k(G)$ for some hypergeometric term~$G$. A
straightforward calculation shows that~$G$ is similar to~$T$. So there
exists~$r \in \bF(k)$ such that~$G = r H$. One can easily verify that
\begin{equation}\label{EQ:summable}
SH  = \Delta_k(rH)\, \Longleftrightarrow \, S = K \sigma_k(r)-r.
\end{equation}

\chapter[Additive Decomposition for Hypergeometric Terms]
{Additive Decomposition for\\ Hypergeometric Terms
\protect\footnotemark{}\protect
\footnotetext{The main results in this chapter are joint work 
with S.\ Chen, M.\ Kauers, Z.\ Li, published in~\cite{CHKL2015}.}}
\label{CH:apreduction}

Computing an indefinite summation of a given hypergeometric term is
one of the basic problems in the theory of difference equations.  In
terms of algorithms, Gosper's algorithm~\cite{Gosp1978} is the first
complete algorithm for solving this problem.  However, when there
exist no indefinite summations, Gosper's algorithm is not applicable
any more, but we still desire more information so as to handle
definite summations.  As far as we know, the first description of the
non-summable case was given by Abramov. In 1975,
Abramov~\cite{Abra1975} developed a reduction algorithm to compute an
additive decomposition of a given rational function, which was
improved later by Pirastu and Strehl~\cite{PiSt1995},
Paule~\cite{Paul1995}, and by Abramov himself~\cite{Abra1995a}, etc.  These
algorithms decompose a rational function into a summable part and a
proper fractional part whose denominator is shift-free and of minimal
degree. We refer to it as a {\em minimal additive decomposition}%
\index{decomposition! additive}\index{additive decomposition} of the 
given rational function.  According to
Lemma~\ref{LEM:rationalsummable}, the fractional part is in fact
non-summable. Hence a rational function is summable if and only if the
fractional part of a minimal decomposition is zero. In 2001, Abramov
and Petkov{\v s}ek~\cite{AbPe2001a,AbPe2002b} generalized these ideas
to the hypergeometric case. We call it the \ap reduction.
It preserves the minimality of additive decompositions. It loses,
however, the separation of summable and non-summable parts.  More
precisely, given a hypergeometric term~$T$, \ap
reduction\index{reduction! \ap}\index{\ap reduction} computes two
hypergeometric terms $T_1, T_2$ such that
\[T = \underbrace{\Delta_k(T_1)}_{\text{summable}} +
\underbrace{T_2}_{\text{possibly summable}},\]\index{summable}%
where $T_2$ is minimal in some sense. To determine the summability of
$T$, one needs to further solve an auxiliary difference
equation~\cite[\S 4]{AbPe2002b}. The discrepancy in the reductions for
the rational case and the hypergeometric case is unpleasant.

In this chapter, in order to obtain the consistency, we modify the \ap
reduction by a shift variant of the method developed by Bostan et
al.~\cite{BCCLX2013}.  The modified \ap reduction not only preserves
the minimality of the output additive decomposition, but also
decomposes a hypergeometric term as a sum of a summable part and a
non-summable part.  It laid a solid foundation for the new
reduction-based creative telescoping algorithm in
Chapter~\ref{CH:telescoping}.  Moreover, we implement the modified
reduction in {\sc Maple~18} and compare it
with the built-in Maple procedure \textsf{SumDecomposition}, which is
based on the \ap reduction. The experimental results illustrate that
the modified \ap reduction is more efficient than the original one.
%
\section{\ap reduction}\label{SEC:apred}
In the shift case, reduction algorithms for computing minimal additive
decompositions of rational functions have been well-developed. More
details can be found
in~\cite{Abra1971,Abra1975,Abra1995a,Paul1995,PiSt1995}. For this
reason, we will mainly focus on irrational hypergeometric terms.

The \ap reduction~\cite{AbPe2001a, AbPe2002b} is fundamental for the
first part of this thesis, which computes a minimal additive
decomposition of a given hypergeometric term. It can not only be used
to determine hypergeometric summability, but also provide some
description of the non-summable part when the given hypergeometric
term is not summable. In this sense, the \ap reduction is more useful
than Gosper's algorithm in some cases, as illustrated by the following
example.
\begin{example}\label{EX:apvsgosper}%
	\footnote{We thank Yijun Chen for providing this example.}
\index{Gosper's algorithm}\index{reduction! \ap}\index{\ap reduction}
  Consider a definite sum
  \[\sum_{k=0}^\infty T(k), \quad \text{where} \ T(k) = \frac1{(k^4 +
    k^2 +1) k!}.\] 
  Applying Gosper's algorithm shows that $T$ is not summable, and thus
  we cannot evaluate the sum in terms of indefinite
  summations. Applying the \ap reduction to $T$, however, yields
  \[T(k) =  \Delta_k\left(\frac{k^2}{2(k^2-k+1)k!}\right)+\frac1{2k!}.\] 
  Summing over $k$ from zero to infinity and using the telescoping sum
  technique leads to a \lq\lq closed form\rq\rq\ of the summation,
  \[\sum_{k=0}^\infty T(k) = \lim_{k\rightarrow \infty}\left(\frac{k^2}{2 (k^2-k+1) k!}\right) 
  - 0 + \sum_{k=0}^\infty \frac1{2 k!} = \frac12 e.\]
  Thus the given sum in fact admits a simple form.
\end{example}
To describe the Abramov-Petkov\v{s}ek reduction concisely, we need a
notational convention and a technical definition.
\begin{convention} \label{CON:notation} Let $T$ be a hypergeometric
  term over $\bF(k)$ with a kernel $K$ and a corresponding shell $S$. 
  Then $T=SH$, where $H$ is a hypergeometric term whose
  shift-quotient is $K$. Further write~$K =
  u/v$, where $u,v$ are nonzero polynomials in $\bF[k]$ with $\gcd(u,v)=1$.

  Moreover, we let~$\bU_T$ be the union of~$\{0\}$ and the set of
  summable\index{summable} hypergeometric terms that are similar
  to~$T$, and $\bV_K = \{ K \sigma_k(r)-r \mid r \in \bF(k) \}$.
\end{convention}
With the above convention, it is clear that~$\bU_T$ and $\bV_K$ are
both $\bF$-linear vector spaces and $\bU_T=\bU_H$ since $H$ is similar to
$T$.  Then \eqref{EQ:summable} translates into
\begin{equation} \label{EQ:congsum} 
  S H \equiv_k 0 \mod \bU_H \,
  \Longleftrightarrow \, S \equiv_k 0 \mod \bV_K.
\end{equation}
These congruences\index{congruence} enable us to shorten expressions.
\begin{definition}\label{DEF:prime}
  With Convention~\ref{CON:notation}, a nonzero polynomial $p$ in $\bF[k]$ is said to be 
  {\em strongly coprime}\index{strongly coprime} with~$K$ if 
  $\gcd(p,\sigma_k^{-i}(u) ) = \gcd( p, \sigma_k^{i}(v) ) = 1$ 
  for all $i \ge 0$.
\end{definition}	
The proof of Lemma~3 in~\cite{AbPe2001a} contains a reduction
algorithm whose inputs and outputs are given below.
%
\index{reduction! \ap}\index{\ap reduction}
 \begin{algo}[\ap Reduction]\label{ALG:apred}\leavevmode\null
 	
 	\noindent
 	{\bf Input}: Two rational functions $K, S\in \bF(k)$ as defined in Convention~\ref{CON:notation}.\\
 	{\bf Output}: A rational function $S_1\in \bF(k)$ and two polynomials $b, w\in \bF[k]$ such that 
 	$b$ is shift-free and strongly coprime with~$K$, 
 	and the following equation holds:
 	\begin{equation} \label{EQ:ap} 
 	S = K \sigma_k(S_1) - S_1 +
 	\frac{w}{ b\cdot \sigma_k^{-1}(u)\cdot v}.
 	\end{equation}
 \end{algo}
 The algorithm contained in the proof of Lemma~3 in~\cite{AbPe2001a}
 is described as pseudo code on page~4 of the same paper, in which the
 last ten lines make the denominator of the rational function
 $V$ in its output minimal in some technical sense. We shall not
 execute these lines. Then the algorithm will compute two rational
 functions $U_1$ and~$U_2$. They correspond to $S_1$ and 
 $w/(b\,\sigma_k^{-1}(u)\,v)$ in~\eqref{EQ:ap}, respectively.

 We slightly modify the output of the Abramov-Petkov\v{s}ek reduction
 so that we can analyze it more easily in the next section.  Note that
 $K$ is shift-reduced and $b$ is strongly coprime with~$K$.  Thus, $b$,
 $\sigma_k^{-1}\left( u\right)$ and $v$ are pairwise coprime.  By
 partial fraction decomposition,~\eqref{EQ:ap} can be rewritten as
\[S = K \sigma_k(S_1) - S_1  +  \left(\frac{a}{b} + \frac{p_1}{\sigma_k^{-1}(u)}+\frac{p_2}{v}\right),  \]
where $a, p_1, p_2 \in \bF[k]$.
Furthermore, set $r = p_1/\sigma_k^{-1}(u)$ and a direct calculation yields 
\[r  = K \sigma_k(-r) - (-r) + \frac{\sigma_k(p_1)}{v}.\]
Update $S_1$ to be $S_1 - r$ and set $p$ to be $\sigma_k(p_1) + p_2$. Then
\begin{equation} \label{EQ:sap}
S = K \sigma_k(S_1) - S_1 + \left( \frac{a}{b} + \frac{p}{v}\right).
\end{equation}
This modification leads to shell reduction specified below.

%
\index{reduction! shell}\index{shell reduction}
\begin{algo}[Shell Reduction]\label{ALG:shellred}\leavevmode\null
	
	\noindent
	{\bf Input}: Two rational functions $K, S\in \bF(k)$ as defined in Convention~\ref{CON:notation}.\\
	{\bf Output}: A rational function $S_1\in \bF(k)$ and three polynomials $a, b, p\in \bF[k]$ such that 
	$b$ is shift-free and strongly coprime with~$K$, 
	and that~\eqref{EQ:sap} holds.
\end{algo}
Shell reduction provides us with a necessary condition on summability.
\begin{prop} \label{PROP:shell} With Convention~\ref{CON:notation},
  let $a, b, p$ be polynomials in~$\bF[k]$ where~$b$ is shift-free and
  strongly coprime with $K$. Assume further that~\eqref{EQ:sap}
  holds. If~$T$ is summable\index{summable}, then $a/b$ belongs to
  $\bF[k]$.
\end{prop}
\begin{proof}
  Recall that $T=SH$ by Convention~\ref{CON:notation} and it has a
  kernel $K$ and a corresponding shell $S$.  It follows from
  \eqref{EQ:congsum} and~\eqref{EQ:sap} that
  $$T \equiv_k \left( \frac{a}{b}+\frac{p}{v}\right) H \mod \bU_H.$$
  Thus, $T$ is summable if and only if $\left( a/b+p/v\right) H$ is
  summable.
	
  Set $H^\prime=(1/v)H$, which has a kernel
  $K^\prime=u/\sigma_k(v)$. Note that since $b$ is strongly coprime with $K$,
  so is $K^\prime$. Applying \cite[Theorem~11]{AbPe2002b} to
  $\left(av/b+p\right)H^\prime$, which is equal to $\left(
    a/b+p/v\right) H$, yields that $(av/b+p)$ is a polynomial. Thus,
  $a/b$ is a polynomial because $b$ is coprime with $v$.
\end{proof}

The above proposition enables us to determine hypergeometric
summability directly in some instances.
\begin{example}\label{EX:nonsummable}
  Let $T = k^2 k!/(k+1)$. Then it has a kernel $K = k+1$ and the shell
  $S = k^2/(k+1)$. Shell reduction yields
  \[ S \equiv_k -\frac1{k+2}+ \frac{k}{v} \mod \bV_K,\]
  where $v  =  1$. By Proposition~\ref{PROP:shell}, 
  $T$ is not summable. By a similar argument as before, one sees that
  $k!$ is indeed not summable as mentioned in Example~\ref{EX:htsummable}.
\end{example}
\noindent 
Note that $a{/}b+p{/}v$ in \eqref{EQ:sap} can be nonzero for a summable $T$.
\begin{example}\label{EX:summable}
  Let $T = k \cdot k!$ whose kernel is $K = k+1$ and shell is $S = k$.
  Then
  $$S \equiv_k \frac{ k}{v} \mod \bV_K,$$
  where $v=1$. But $T$ is summable as it is equal to
  $\Delta_k\left(k!\right)$.
\end{example}
The above example illustrates that neither shell reduction nor the
Abramov-Petkov\v{s}ek reduction can decide summability directly when $a/b\in \bF[k]$
in~\eqref{EQ:sap}. One
way to proceed is, according to \cite{AbPe2002b}, to find a polynomial
solution of the auxiliary first-order linear difference equation $u
\sigma_k(z) - v z = a v/b+ p$, under the hypotheses of
Algorithm~\ref{ALG:shellred}. If there is a polynomial solution,
say~$f\in \bF[k]$, then $T = \Delta_k((S_1+f)H)$; otherwise $T$ is not
summable. This method reduces the summability problem to solving a
linear system over $\bF$.  We show in the next section how this can be
avoided so as to read out summability directly from a minimal
decomposition.

\section{Modified \ap reduction}\label{SEC:mapred}
\index{modified \ap reduction|(}\index{reduction! modified \ap|(}
After the shell reduction described in~\eqref{EQ:sap}, it remains to
check the summability of the hypergeometric term
$\left(a/b+p/v\right)H$. In the rational case, i.e., when the
kernel~$K$ is one, the rational function $a{/}b + p{/}v$
in~\eqref{EQ:sap} can be further reduced to~$a/b$ with $\deg_k(a) <
\deg_k(b)$, because all polynomials are rational summable. However, a
hypergeometric term with a polynomial shell is not necessarily
summable, for example, $k!$ has a polynomial shell but it is not
summable.

In this section, we define the notion of discrete residual forms for
rational functions, and present a discrete variant of the polynomial
reduction for hyperexponential functions%
\index{hyperexponential function} given in~\cite{BCCLX2013}. 
This variant not only leads to
a direct way to decide summability, but also reduces the number of
terms of~$p$ in~\eqref{EQ:sap}.

\subsection{Discrete residual forms}\label{SUBSEC:drf}
With Convention \ref{CON:notation}, we define an $\bF$-linear map
 \begin{align*}
 \begin{array}{cccc}
   \phi_K: & \bF[k] & \quad \rightarrow &\bF[k]\\[1ex]
    & p    & \quad\mapsto     & u \sigma_k(p) - v p,	
 \end{array}
 \end{align*}
 for all $p \in \bF[k]$. We call $\phi_K$ the {\em map for polynomial
   reduction w.r.t.~$K$}.  \index{map for polynomial reduction}
\begin{lemma} \label{LEM:direct}
  Let
  \[ \bW_K = \spanning_{\bF}\left\{k^{\ell} \mid \ell \in
    {\mathbb{N}}, \ell \neq \deg_k (p) \text{ for all nonzero } p \in
    \im\left(\phi_K \right) \right\}.\] 
  Then $\bF[k] = \im\left( \phi_K \right) \oplus \bW_K$.
\end{lemma}
\begin{proof}
  By the definition of $\bW_K$, $\im\left(\phi_K\right) \cap \bW_K =
  \{0\}$.  The same definition also implies that, for every
  nonnegative integer $m$, there exists a polynomial $f_m$
  in~$\im\left(\phi_K\right) \cup \bW_K$ such that the degree of $f_m$
  is equal to $m$.  The set $\{f_0, f_1, f_2, \ldots \}$ forms an
  $\bF$-basis of~$\bF[k]$. Thus $\bF[k] = \im\left( \phi_K \right)
  \oplus \bW_K$.
\end{proof}
In view of the above lemma, we call $\bW_K$ the {\em standard
  complement of $\im(\phi_K)$}.\index{standard complement} 
Note that if $K=1$, then $\phi_K= \Delta_k$ and $\bW_K= \{0\}$ 
since all polynomials are rational summable.
According to Lemma~\ref{LEM:direct}, every polynomial $p \in \bF$ can be
uniquely decomposed as $p=p_1+p_2$ where~$p_1 \in \im\left(\phi_K\right)$ and $p_2 \in \bW_K$.
\begin{lemma} \label{LEM:reducedform} With
  Convention~\ref{CON:notation}, let $p$ be a polynomial in $\bF[k]$.
  Then there exists a polynomial $q \in \bW_K$ such that $p/v
  \equiv_kq/v \mod \bV_K$.
\end{lemma}
\begin{proof}
  Let $q\in \bF[k]$ be the projection of $p$ on $\bW_K$. Then there
  exists $f$ in $\bF[k]$ such that $p = \phi_K(f) + q$, that is, $p =
  u \sigma_k(f) - vf + q$. So $p/v = K \sigma_k(f)-f+q/v$, which is
  equivalent to $p/v \equiv_kq/v \mod \bV_K$.
\end{proof}
\begin{remark} \label{REM:poly} 
  Replacing the polynomial $p$ in the above lemma by $vp$, 
  we see that, for every polynomial $p\in\bF[k]$,
  there exists $q \in \bW_K$ such that $p \equiv_kq/v \mod \bV_K$.
\end{remark}
By Lemma~\ref{LEM:reducedform} and Remark~\ref{REM:poly},
\eqref{EQ:sap} implies that
\begin{equation} \label{EQ:iap}
S \equiv_k\frac{a}{b} + \frac{q}{v} \mod \bV_K,
\end{equation}
where $a, b, q $ are polynomials in $\bF[k]$, $\deg_k(a)<\deg_k(b)$,
$b$ is shift-free and strongly coprime with $K$, and $q \in \bW_K$. The
congruence~\eqref{EQ:iap} motivates us to translate the notion of
(continuous) residual forms~\cite{BCCLX2013} into the discrete
setting.
\begin{definition} \label{DEF:residual} 
  With Convention~\ref{CON:notation}, we further let $f$ be a rational
  function in~$\bF(k)$. Another rational function $r$ in $\bF(k)$ is
  called a {\em (discrete) residual form}%
  \index{residual form}\index{discrete residual form|see {residual form}} of~$f$
  w.r.t.~$K$ if there exist $a,b,q$ in~$\bF[k]$ such that
  $$f \equiv_k r \mod \bV_K \quad \text{and} \quad r = \frac{a}{b}+\frac{q}{v},$$
  where $\deg_k(a)<\deg_k(b)$, $b$ is shift-free and strongly coprime
  with~$K$, and $q$ belongs to~$\bW_K$. For brevity, we just say that $r$ is
  {\em a residual form} w.r.t.~$K$ if $f$ is clear from the
  context. Moreover, we call $b$ the {\em significant
    denominator}\index{significant denominator} of $r$ if~$\gcd(a,b)
  =1$ and $b$ is monic, i.e., $\lc_k(b)=1$.
\end{definition}
Residual forms help us to decide summability, as shown below.
\begin{prop}\label{PROP:residual}\index{summable}
  With Convention~\ref{CON:notation}, we further assume that $r$ is a
  nonzero residual form w.r.t.~$K$. Then the hypergeometric term $rH$
  is not summable.
\end{prop}
\begin{proof}
%
  Suppose that $rH$ is summable. 
  Let $r=a/b+q/v$, where $a, b, q\in
  \bF[k]$, $\deg_k(a)<\deg_k(b)$, $b$ is shift-free and strongly coprime
  with~$K$, and~$q\in\bW_K$.  By Proposition~\ref{PROP:shell}, $a/b$
  is a polynomial.  Since $\deg_k(a)<\deg_k(b)$, we have $a=0$ and thus the
  term $(q/v)H$ is summable.  It follows from~\eqref{EQ:summable} that
  there exists a rational function $w\in\bF(k)$ such that $u\sigma_k(w)-vw=q$. Thus, $w
  \in \bF[k]$ by Theorem~5.2.1 in~\cite[page 76]{PWZ1996}, which
  implies that $q$ belongs to $\im\left(\phi_K\right)$. But $q$ also
  belongs to~$\bW_K$.  By Lemma~\ref{LEM:direct}, $q=0$ and then $r =
  0$, a contradiction.
\end{proof}
With Convention~\ref{CON:notation}, let $r$ be a residual form of the
shell $S$ w.r.t.~$K$.  Then 
\[SH \equiv_k rH \mod \bU_H\]
according to~\eqref{EQ:congsum} and~\eqref{EQ:iap}.  By
Proposition~\ref{PROP:residual}, $SH$ is summable if and only if
$r=0$.  Thus, determining the summability of a hypergeometric term~$T$
amounts to computing a residual form of a corresponding shell with
respect to a kernel of~$T$, which is studied below.

\subsection{Polynomial reduction}\label{SUBSEC:polyred}
\index{polynomial reduction|(}\index{reduction! polynomial|(} 
With Convention~\ref{CON:notation}, to compute a residual form of a rational function, 
we project a polynomial on~$\im(\phi_K)$ and also its standard complement $\bW_K$,
both defined in the previous subsection. If the given term $T$ is a
rational function, i.e., $K =1$, then this projection is trivial because
$\im(\phi)= \im(\Delta_k)=\bF[k]$ and $\bW_K =\{0\}$.

Now we assume $K\neq 1$ and let $\bB_K=\{ \phi_K(k^i) \mid i \in \bN \}$. 
Since $K\neq 1$, the $\bF$-linear map $\phi_K$ is injective by Lemma~\ref{LEM:shiftreduced}~$(ii)$.  
So~$\bB_K$ is an $\bF$-basis of~$\im\left( \phi_K \right)$, 
which allows us to construct an echelon basis of~$\im(\phi_K)$.  
By an echelon basis\index{echelon basis}, we mean an $\bF$-basis 
in which distinct elements have distinct degrees.
We can easily project a polynomial using an echelon basis and linear
elimination. %
\index{standard complement}\index{map for polynomial reduction}

To construct an echelon basis, we
rewrite $\im(\phi_K)$ as
\begin{equation*}
  \im(\phi_K) = \left\{u\Delta_k(p) - (v - u)p \mid p \in \bF[k]
  \right\}.
\end{equation*}
Set $\alpha_1 = \deg_k(u)$, $\alpha_2 = \deg_k(v)$, and $\beta =
\deg_k(v - u)$. Moreover, set
$$\tau_K = \frac{\lc_k(v - u)}{\lc_k(u)},$$
which is nonzero since $K\neq 1$ and let $p$ be a
nonzero polynomial in~$\bF[k]$.

\index{echelon basis|(}
We make the following case distinction.

\smallskip \noindent
{\em Case 1.} $\beta > \alpha_1$. Then $\beta = \alpha_2$, and
\begin{equation*}
  \phi_{K}(p) = - \lc_k(v - u)\lc_k(p)k^{\alpha_2 + \deg_k(p)} + \text{ lower terms}.
\end{equation*}
So $\bB_K$ is an echelon basis of $\im(\phi_{K})$, in which
$\deg_k(\phi_{K}(k^i))$ is equal to $\alpha_2 + i$ for all $i \in
\bN$.  Accordingly, $\bW_{K}$ has an echelon basis $\{1, k, \ldots,
k^{\alpha_2 - 1}\}$ and has dimension $\alpha_2$.

\smallskip \noindent
{\em Case 2.} $\beta = \alpha_1$. Then
\begin{equation*}
  \phi_{K}(p) = - \lc_k(v - u)\lc_k(p)k^{\alpha_1 + \deg_k(p)} + \text{ lower terms}.
\end{equation*}
So $\bB_K$ is an echelon basis of $\im(\phi_{K})$, in which
$\deg_k(\phi_{K}(k^i))$ is equal to $\alpha_1 + i$ for all $i \in
\bN$.  Accordingly, $\bW_{K}$ has an echelon basis $\{1, k, \ldots,
k^{\alpha_1 - 1}\}$ and has dimension $\alpha_1$.

\smallskip \noindent
{\em Case 3.} $\beta < \alpha_1 - 1$. If $\deg_k(p) = 0$, then $\phi_{K}(p) = (u - v)p$. Otherwise, we have
\begin{equation*}
  \phi_{K}(p) = \deg_k(p)\lc_k(u)\lc_k(p)k^{\alpha_1 + \deg_k(p) - 1} + \text{ lower terms}.
\end{equation*}
It follows that $\bB_K$ is an echelon basis of $\im(\phi_{K})$, in
which $\deg_k(\phi_{K}(1)) = \beta$ and
\begin{equation*}
  \deg_k(\phi_{K}(k^i)) = \alpha_1 + i - 1 \quad \text{ for all } i \geq 1.
\end{equation*}
Accordingly, $\bW_{K}$ has an echelon basis 
$\{1, \ldots, k^{\beta - 1}, k^{\beta + 1}, \ldots, k^{\alpha_1 - 1}\}$ 
and has dimension $\alpha_1-1$.

\smallskip \noindent
{\em Case 4.} $\beta = \alpha_1 - 1$ and $\tau_K$ is not a positive integer. Then
\begin{align}
  \phi_{K}(p) &= \left(\deg_k(p)\lc_k(u) - \lc_k(v -
    u)\right)\lc_k(p)k^{\alpha_1 + \deg_k(p) - 1} + \text{ lower
    terms}.\label{EQ:case4}
\end{align}
Accordingly, $\bB_K$ is an echelon basis of $\im(\phi_{K})$, in which
$\deg_k(\phi_K(k^i)) = \alpha_1 + i - 1$ for all $i \in \mathbb{N}$.
Accordingly, $\bW_{K}$ has an echelon basis $\{1, k, \ldots,
k^{\alpha_1 - 2}\}$ and has dimension $\alpha_1-1$.

\smallskip \noindent 
{\em Case 5.} $\beta = \alpha_1 - 1$ and $\tau_K$ is a positive integer. 
It follows from~\eqref{EQ:case4} that for~$i
\neq \tau_{K}$, we have $\deg_k(\phi_{K}(k^i)) = \alpha_1 + i -
1$. Moreover, for every polynomial~$p$ of degree $\tau_{K}$,
$\phi_{K}(p)$ is of degree less than $\alpha_1 + \tau_{K} - 1$. So any
echelon basis of~$\im(\phi_{K})$ does not contain a polynomial of
degree $\alpha_1 + \tau_{K} - 1$. Set
\begin{equation*}
  \bB_K^\prime = \left\{\phi_{K}(k^i) \mid i \in \bN, i \neq \tau_{K}\right\}.
\end{equation*}
Reducing $\phi_{K}(k^{\tau_{K}})$ by the polynomials in $\bB_K'$, we
obtain a polynomial $p'$ with degree less than $\alpha_1 - 1$. Since
$\bB_K$ is an $\bF$-basis and $\bB_K'\subset \bB_K$, $p' \neq 0$. 
Hence $\bB'_K \cup \{p'\}$ is an echelon basis of~$\im(\phi_{K})$. 
Consequently, $\bW_{K}$ has an echelon basis 
$\{1, k, \ldots, k^{\deg_k(p') -1}, k^{\deg_k(p') + 1}, 
\ldots, k^{\alpha_1- 2}, k^{\alpha_1 + \tau_{K} - 1}\}$.  
The dimension of $\bW_K$ is equal to $\alpha_1-1$.%
\index{echelon basis|)}

\begin{example}\label{EX:case5}
  Let $K = (k^4 + 1)/(k+1)^4$, which is shift-reduced. Then $\tau_{K}
  = 4$. According to Case~5, $ \im(\phi_{K})$ has an echelon basis
  \[ \left\{ \phi_K\left( p \right) \right\}\cup \left\{
    \phi_K\left(k^m \right) \mid m \in \bN, m \neq 4\right\},
  \]
  where $p=k^4 +k/3 + 1/2$, $\phi_K(p) = (5/3)k^2+2k+4/3$, and
  \[\phi_K\left(k^m\right) = (m-4)k^{m+3} + \text{lower terms}.\]
  Therefore, $\bW_{K}$ has a basis $\{1,k,k^7\}$.
\end{example}
From the above case distinction and example one observes that,
although the degree of a polynomial in the standard complement depends
on $\tau_K$, which may be arbitrarily high, the number of its terms
depends merely on the degrees of $u$ and $v$. We record this
observation in the next proposition.
\begin{prop}\label{PROP:termbound}
  With Convention \ref{CON:notation}, further let $\alpha_1 = \deg_k(u),
  \alpha_2 = \deg_k(v)$ and~$\beta =\max\{0, \deg_k(v - u)\}.$ Then there exists
  a set $\cP \subset \{k^i \mid i \in \mathbb{N}\}$ with
  $$|\cP| \leq \max \{\alpha_1, \alpha_2\}- \llbracket \beta \leq \alpha_1 -1\rrbracket$$
  such that every polynomial in $\bF[k]$ can be reduced modulo
  $\im(\phi_{K})$ to an $\bF$-linear combination of the elements in
  $\cP$. Note that here the expression $\llbracket \beta \leq \alpha_1
  -1 \rrbracket$ equals $1$ if $\beta \leq \alpha_1 -1$, otherwise it
  is $0$.
\end{prop}
\begin{proof}
  If $K=1$, then $\im(\phi_K)=\im(\Delta_k)=\bF[k]$ and $\alpha_1= \alpha_2=\beta=0$.
  Taking $\cP=\emptyset$ completes the proof. Otherwise $K\neq 1$.
  By the above case distinction, the dimension of $\bW_{K}$ over $\bF$
  is no more than $\max \{\alpha_1, \alpha_2\}-\llbracket \beta \leq
  \alpha_1 -1\rrbracket$. The lemma follows.
\end{proof}

When $K\neq 1$, the above case distinction enables one to find an infinite sequence
$p_0, p_1, \ldots$ in~$\bF[k]$ such that
\begin{equation*} 
  \bE_K = \left\{ \phi_K(p_i) | i \in \bN \right\}~\text{with $\deg_k \phi_K(p_i) < \deg_k \phi_K (p_{i+1}),$}
\end{equation*}
is an echelon basis of $\im\left(\phi_K\right)$.  This basis allows
one to project a polynomial on~$\im\left( \phi_K\right)$ and $\bW_K$,
respectively.  In the first four cases, the $p_i$'s can be chosen as
powers of $k$. But in the last case, one of the $p_i$'s is not
necessarily a monomial as shown in Example~\ref{EX:case5}. 

\goodbreak
Based on the above discussion, we have the following algorithm.  

%
\index{reduction! polynomial}\index{polynomial reduction}
\begin{algo}[Polynomial Reduction]\label{ALG:polyred}\leavevmode\null
	
	\noindent
	{\bf Input}: A polynomial $p\in \bF[k]$ and a shift-reduced rational function $K\in \bF(k)$.\\
	{\bf Output}: Two polynomials $f,q \in \bF[k]$ such that $q\in \bW_K$ and $p = \phi_K(f) + q$.
	
	\bigskip
	\step{1}{0} If $p=0$, then set $f=0$ and $q=0$; and return.
	
	\smallskip
	\step{2}{0} If $K =1$, then set $f = \Delta_k^{-1}(p)$ and $q=0$; and return.
	
	\smallskip
	\step 20 Set $d=\deg_k(p)$. 
	\step{}{0} Find the subset $\bP=\left\{p_{i_1}, \ldots, p_{i_s} \right\}$
		consisting of the preimages of all
	\step{}{0} polynomials in the echelon basis $\bE_K$ whose degrees are at most $d$.  
	
	\smallskip
	\step 30 For $j=s, s-1, \ldots, 1,$ perform linear elimination to 
	\step{}0 find $c_s,$ $c_{s-1},\ldots,c_1\in\bF$ such that 
		$p - \sum_{j=1}^s c_j \phi_K(p_{i_j})\in \bW_K.$
		
	\smallskip
	\step{4}{0} Set $f=\sum_{j=1}^s c_j p_{i_j}$ and $q=p- \phi_K(f)$; and return.
\end{algo}

\index{polynomial reduction|)}\index{reduction! polynomial|)}
Together with Algorithms~\ref{ALG:shellred} and~\ref{ALG:polyred}, we
are ready to present a modified version of the \ap reduction, which is
summarized as Algorithm~\ref{ALG:mapred}.  This modified reduction
determines summability without solving any auxiliary difference
equations explicitly.
\begin{algo}[Modified \ap Reduction]\label{ALG:mapred}\leavevmode\null
	
	\noindent
	{\bf Input}: A hypergeometric term $T$ over $\bF(k)$.\\
	{\bf Output}: A hypergeometric term $H$ with a kernel $K$
	and two rational functions $f, r \in \bF(k)$ such that $r$ is a residual form w.r.t.~$K$ and
	\begin{equation}\label{EQ:mapred} 
	T = \Delta_k(f H) + r H.
	\end{equation}
	
	\medskip
	\step{1}{0} Find a kernel $K$ and a corresponding shell $S$ of $T$.
	
	\smallskip
	\step 20 Apply Algorithm~\ref{ALG:shellred}, namely the shell reduction, 
	to $S$ w.r.t.~$K$ to
	\step{}0 find three polynomials $b, s, t \in \bF[k]$ and a rational function $g \in \bF(k)$
	\step{}0 such that $b$ is shift-free and strongly coprime with $K$, and
	\begin{equation} \label{EQ:prered}
	T = \Delta_k\left( g H\right) + \left( \frac{s}{b} + \frac{t}{v} \right) H,
	\end{equation}
	\step{}{0} where $\sigma_k(H)/H=K$ and $v$ is the denominator of $K$.
	
	\smallskip
	\step 30 Set $p$ and $a$ to be the quotient and remainder of $s$ and $b$, respectively.
	
	\smallskip
	\step{4}{0} Apply Algorithm~\ref{ALG:polyred}, namely
	the polynomial reduction, to $vp+t$ to 
	\step{}0 find $h \in
	\bF[k]$ and~$q \in \bW_K$ such that $vp+t = \phi_K(h) + q$.
	
	\smallskip
	\step{5}{0} Set $f=g+h$ and $r=a/b+q/v$; and return $H$, $f$ and $r$.
\end{algo}
\begin{theorem}\label{THM:mapred}
  With Convention~\ref{CON:notation}, Algorithm~\ref{ALG:mapred}
  computes a rational function~$f$
  in~$\bF(k)$ and a residual form~$r$ w.r.t.~$K$ such that
  \eqref{EQ:mapred} holds.  Moreover, $T$ is summable if and only if
  $r = 0$.
\end{theorem}
\begin{proof}
  Recall that $T=SH$, where $H$ has a kernel $K$ and $S$ is a rational
  function.  Applying shell reduction to $S$ w.r.t.~$K$
  yields~\eqref{EQ:prered}, which can be rewritten as
  \[ T = \Delta_k\left(g H\right) + \left( \frac{a}{b} +
    \frac{vp+t}{v} \right) H, \] 
  where $a$ and $p$ are given in step~$3$ of Algorithm~\ref{ALG:mapred}. 
  The polynomial reduction in step~$4$ yields that $vp+t=u\sigma_k(h)-vh + q$. 
  Substituting this into~\eqref{EQ:prered} gives
  \begin{align*}
    T & =  \Delta_k(gH) + \left(K \sigma_k(h)-h \right) H 
     + \left(\frac{a}{b} + \frac{q}{v}\right)H  \\[1ex]
    & = \Delta_k((g+h)H) + rH,
  \end{align*}
  where $r = a/b + q/v$. Thus,~\eqref{EQ:mapred} holds.  By
  Proposition~\ref{PROP:residual}, $T$ is summable if and only if $r$
  is equal to zero.
\end{proof}
\begin{example}\label{EX:mapnonsummable}
  Let $T$ be the same hypergeometric term as in
  Example~\ref{EX:nonsummable}. Then we know $K = k+1$ and $S =
  k^2/(k+1)$. Set $H = k!$. By the shell reduction in
  Example~\ref{EX:nonsummable},
  \begin{equation*}
    T = \Delta_k\left(\frac{-1}{k+1} H\right) + \left(\frac{-1}{k+2} 
        + \frac{k}{v}\right) H \quad \text{with} \ v = 1.
  \end{equation*}
  Applying the polynomial reduction to $(k/v)H$ yields 
  ${ (k/v)H = \Delta_k(1\cdot H)}$.  
  Combining the above steps, we decompose $T$ as
  $$
    T = \Delta_k\left(\frac{k}{k+1}H\right) - \frac1{k+2} H.
  $$
  So the input term $T$ is not summable, which is consistent with
  Example~\ref{EX:nonsummable}.
\end{example}
\begin{example}\label{ex:summable1}
  Let $T$ be the same hypergeometric term as in
  Example~\ref{EX:summable}. Then we know $K = k+1$ and $S = k$. Set
  $H = k!$. The shell reduction in Example~\ref{EX:summable} gives
  \[T = \Delta_k(0)+ \frac{k}{v} H\quad \text{with}\ v=1.\]
  By the polynomial reduction, $(k/v)H = \Delta_k\left(1\cdot H\right),$
  and hence $T = \Delta_k\left(k!\right)$, implying that $T$ is
  summable.
\end{example}

\begin{remark}\index{reduction! \ap}\index{\ap reduction}
  With the notation given in step~$5$ of Algorithm~\ref{ALG:mapred},
  we can rewrite~$rH$ as $\left(s_1/s_2 \right) G$, where $s_1=av+bq$,
  $s_2=b$, and $G=H/v$. It follows from the case distinction in this
  subsection that the degree of $s_1$ is bounded by $\lambda$ given
  in~\cite[Theorem~8]{AbPe2001a}. The polynomial $s_2$ is equal to $b$
  in~\eqref{EQ:ap} whose degree is minimal
  by~\cite[Theorem~3]{AbPe2001a}. Moreover, $\sigma_k(G)/G$ is
  shift-reduced because $\sigma_k(H)/H$ is. These are exactly the same
  required properties of the output of the \ap
  reduction~\cite{AbPe2001a}. In summary, the modified reduction
  preserves all required conditions for the outputs of the original
  reduction, namely, it also returns a minimal additive decomposition
  of a given hypergeometric term.
\end{remark}
It is remarkable that the modified \ap reduction also applies to
Example~\ref{EX:apvsgosper}. Moreover, compared to the original
reduction, the modified reduction not only further decomposes a
hypergeometric term into a summable part and a non-summable part, but
also provides a new method for proving identities in several examples.
\begin{example}\label{EX:apvsmap}
  Consider the following two famous combinatorial identities
  \[\sum_{k=0}^\infty {n\choose k} = 2^n \quad \text{and} \quad
  \sum_{k=0}^\infty {n \choose k}^2 = {2 n \choose n}.\] 
  Many methods can be used to prove the above identities. 
  In this example, we use the modified \ap reduction. 
  
  Referring to the first identity, we apply the modified reduction to the summand and get
  \[{n \choose k} = \Delta_k\left(-\frac12 {n \choose k}\right) +
  \frac{n+1}{2(k+1)} {n \choose k}.\] 
  Summing over~$k$ from zero to infinity and using the telescoping sum technique yields
  \begin{align*}
    \sum_{k=0}^\infty {n\choose k} 
    &= \lim_{k\rightarrow \infty} \left(-\frac12 {n\choose k} \right) 
    - \left(-\frac12\right) + \sum_{k=0}^\infty \frac{n+1}{2(k+1)} {n\choose k} \\[1ex]
    &=\frac12 + \frac12 \sum_{k=0}^\infty {n+1\choose k+1}= \frac12
    \sum_{k=0}^\infty {n+1\choose k}.
  \end{align*}
  Let~$F(n) = \sum_{k=0}^\infty {n\choose k}$. Then the above equation
  can be rewritten as a first-order difference equation about~$F(n)$,
  \[F(n+1) - 2F(n) = 0.\] 
  It is readily seen that~$2^n$ is a solution. 
  Since $2^{0}= 1 =F(0) $, we have $F(n) = 2^n$, which proves the first identity. 
  
  For the second identity, applying the modified reduction to the summand yields
  \[{n\choose k}^2 = \Delta_k\left(-\frac12\frac{n+2k+1}{2n+1}{n\choose k}^2\right)
  +\frac12\frac{(n+1)^3}{(2n+1)(k+1)^2}{n\choose k}^2.\]
  Along entirely similar lines as the first identity, we get a first-order difference equation
  \[(n+1)F(n+1)-2(2n+1)F(n)=0,\]
  where $F(n) = \sum_{k=0}^\infty {n\choose k}^2$. The second identity follows
  since ${2n\choose n}$ satisfies the same difference equation and has the same
  initial value at zero as $F(n)$.
  
  However, the \ap reduction applies to neither the first identity nor the second one.
\end{example}
\index{modified \ap reduction|(}\index{reduction! modified \ap|(}

\section{Implementation and timings}\label{SEC:timing}
We have implemented Algorithms~\ref{ALG:shellred} -- \ref{ALG:mapred} in
{\sc Maple~18}\index{Maple@{\sc Maple}}. The procedures are included in our Maple package
\textbf{ShiftReductionCT}.
\index{ShiftReductionCT@\textbf{ShiftReductionCT}} A detailed
description of this package is given in Appendix~\ref{APP:guide}.

In order to get an idea about the efficiency of our new procedures, 
we compared their runtime and memory requirements to the performance 
of known algorithms.
Since the comparisons of runtime\index{timing} and memory
requirements\index{memory requirement} almost have the same
indication, we only show that of runtime in this section. One can
refer to Appendix~\ref{APP:memory} for the memory requirements.  All
timings are measured in seconds on a Linux computer%
\index{Linux computer} with 388Gb RAM and twelve 2.80GHz Dual core processors.
The computations for this experiment did not use any parallelism.
For brevity, we denote
\begin{itemize}
\item \textsf{G}: the procedure \textsf{Gosper} in
  \textbf{SumTools[Hypergeometric]}, which is based on Gosper's
  algorithm;%
  \index{SumTools[Hypergeometric]@\textbf{SumTools[Hypergeometric]}! Gosper@\textsf{Gosper}}
  \index{Gosper@\textsf{Gosper}~{\em (SumTools[Hypergeometric])}}
  \index{Gosper's algorithm}

\smallskip
\item \textsf{AP}: the procedure \textsf{SumDecomposition} in
  \textbf{SumTools[Hypergeometric]}, 
  which is based on the \ap reduction;%
  \index{SumTools[Hypergeometric]@\textbf{SumTools[Hypergeometric]}! SumDecomposition@\textsf{SumDecomposition}}
  \index{SumDecomposition@\textsf{SumDecomposition}~{\em (SumTools[Hyper-\\geometric])}}
  \index{reduction! \ap}\index{\ap reduction}

\smallskip
\item \textsf{S}: the procedure~\textsf{IsSummable}
  in \textbf{ShiftReductionCT}, which determines hypergeometric
  summability in a similar way as Gosper's algorithm;%
  \index{ShiftReductionCT@\textbf{ShiftReductionCT}! IsSummable@\textsf{IsSummable}}
  \index{IsSummable@\textsf{IsSummable}~{\em (ShiftReductionCT)}}

\smallskip
\item \textsf{MAP}: the
  procedure~\textsf{ModifiedAbramovPetkovsekReduction} in \textbf{ShiftReductionCT}, 
  which is based on the modified reduction.%
  \index{ShiftReductionCT@\textbf{ShiftReductionCT}! ModifiedAbramovPetkovsekReduction@\textsf{ModifiedAbramovPetkovsekReduction}} %
  \index{ModifiedAbramovPetkovsekReduction@\textsf{ModifiedAbramovPetkovsekReduction}\\{\em (ShiftReductionCT)}}%
  \index{reduction! modified \ap}\index{modified \ap reduction}
\end{itemize}
We make the following two comparisons\index{comparison}. One is for
random hypergeometric terms, while the other is for summable
hypergeometric terms.
\begin{example}[Random hypergeometric terms]\label{EX:generaltest}
  Consider hypergeometric terms of the form
  \begin{equation}\label{EQ:generalht}
    T(k) = \frac{f(k)}{g_1(k) g_2(k)}\prod_{\ell=m_0}^k\frac{u(\ell)}{v(\ell)},
  \end{equation}
  where $f\in\bZ[k]$ of degree~20, $m_0\in \bF$ is fixed, $u, v$ are
  both the product of two polynomials in $\bZ[k]$ of degree one,
  $g_i=p_i\sigma_k^\lambda(p_i)\sigma_k^\mu(p_i)$ with $p_i\in\bZ[k]$
  of degree~10, $\lambda,\mu\in\bN$, and $\alpha,\beta\in\bZ$. For a
  selection of random terms of this type for different choices of
  $\lambda$ and~$\mu$, Table~\ref{TAB:generaltest} compares the
  timings of the four procedures described above.
  \begin{table}[!ht]
    \tabcolsep12pt
    \begin{center}
      \begin{tabular}{c|rr|rr}
        $(\lambda, \mu)$ & {\sf{G}} & {\sf{AP}} & {\sf{S}}  & {\sf{MAP}}  \\\hline
        $(0, 0)$ & 0.09 & 0.16 & 0.12 & 0.12 \\[.5ex]
        $(5, 5)$ & 0.36 & 3.99 & 0.37 & 0.45 \\[.5ex]
        $(10, 10)$ & 0.66 & 13.70 & 0.65 & 0.86 \\[.5ex]
        $(10, 20)$ & 4.05 & 40.82 & 1.41 & 2.53 \\[.5ex]
        $(10, 30)$ & 12.13 & 294.52 & 2.22 & 6.26 \\[.5ex]
        $(10, 40)$ & 19.09 & 564.71 & 3.31 & 14.11 \\[.5ex]
        $(10, 50)$ & 34.89 & 865.01 & 4.76 & 26.02 \\
        \hline
      \end{tabular}
    \end{center}
    \caption{Timing comparison of Gosper's algorithm, 
    the Abramov-Petkov\v{s}ek reduction and the modified version 
    for random hypergeometric terms (in seconds)}\label{TAB:generaltest}
  \end{table}
\end{example}
\begin{example}[Summable hypergeometric terms]\label{EX:summabletest}
  Consider the summable terms $\sigma_k(T)-T$, where~$T$ is of the
  form~\eqref{EQ:generalht}. Similarly, for the same choices of
  $\lambda$ and~$\mu$ as the previous example,
  Table~\ref{TAB:summabletest} compares the timings of the four
  procedures.
  \begin{table}[!h]
    \tabcolsep12pt
    \begin{center}
      \begin{tabular}{c|rr|rr}
        $(\lambda, \mu)$ & {\sf{G}} & {\sf{AP}} & {\sf{S}} & {\sf{MAP}}  \\\hline
        $(0, 0)$ & 1.13 & 2.34 & 1.27 & 1.26 \\[.5ex]
        $(5, 5)$ & 1.86 & 6.44 & 1.59 & 1.59 \\[.5ex]
        $(10, 10)$ & 2.22 & 13.78 & 1.63 & 1.63 \\[.5ex]
        $(10, 20)$ & 7.09 & 29.76 & 2.09 & 2.10 \\[.5ex]
        $(10, 30)$ & 19.61 & 57.63 & 2.34 & 2.33 \\[.5ex]
        $(10, 40)$ & 30.83 & 95.31 & 2.49 & 2.49 \\[.5ex]
        $(10, 50)$ & 64.69 & 168.72 & 2.69 & 2.69 \\
        \hline
      \end{tabular}
    \end{center}
    \caption{Timing comparison of Gosper's algorithm, 
    the \ap reduction and the modified version 
    for summable hypergeometric terms (in seconds)}\label{TAB:summabletest}
  \end{table}
  \vspace{-10pt}
\end{example}

\begin{figure}[h]
  \centering
    \includegraphics[scale=.43]{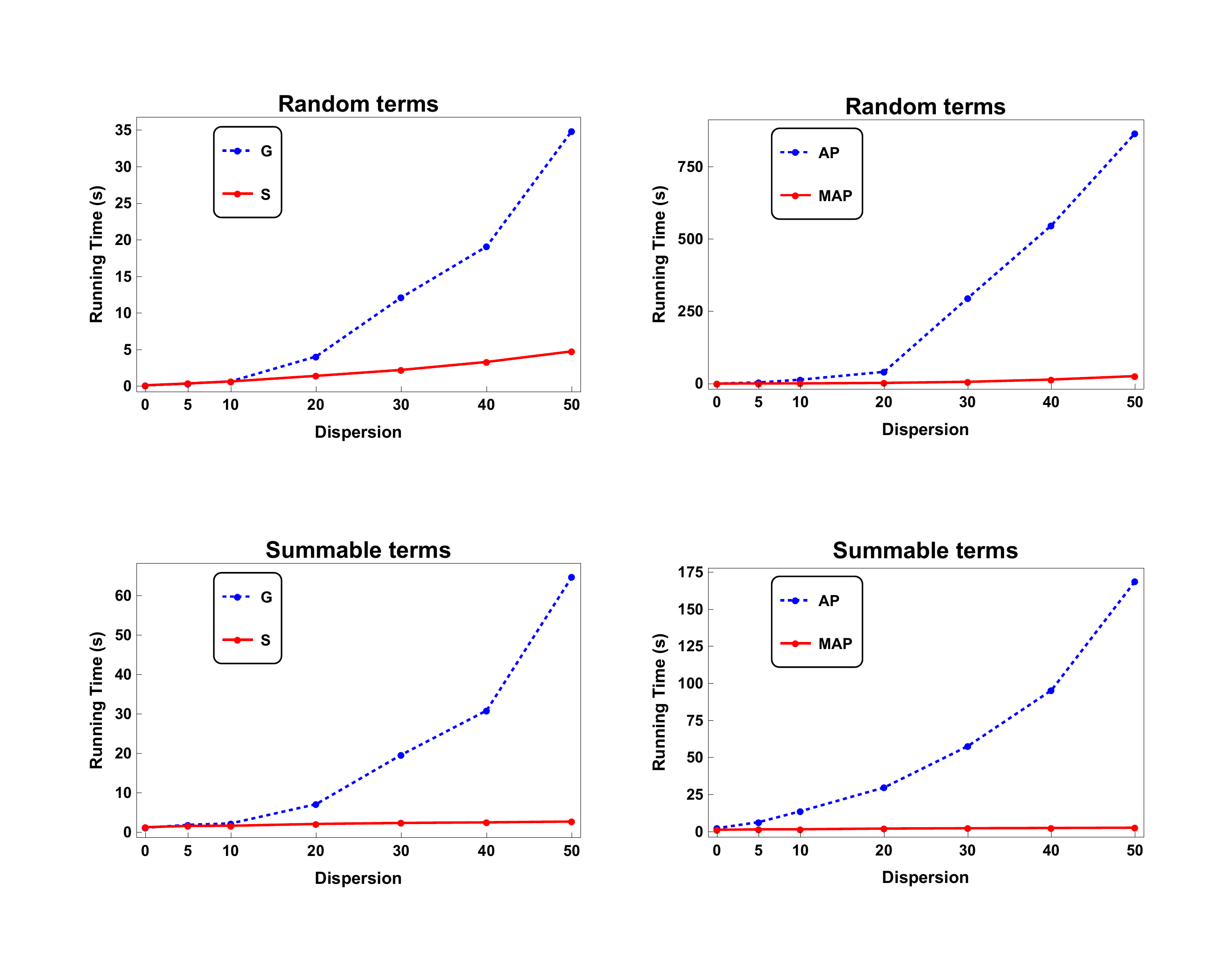}%
     \caption{
            Comparison of the effect of dispersion on Gosper's
            algorithm, the \ap reduction and the modified version for
            Examples~\ref{EX:generaltest} and~\ref{EX:summabletest}
          }\label{FIG:test}
\end{figure}

\newpage
Notice that~$\mu$ is the dispersion of $g_i$ and itself
in~\eqref{EQ:generalht} (see Definition~\ref{DEF:shiftcoprime}). From
Table~\ref{TAB:generaltest} and Table~\ref{TAB:summabletest}, we
observe that for different procedures, the effect of dispersion is
quite different. Figure~\ref{FIG:test} describes 
the effect of dispersion on the above four procedures in
Example~\ref{EX:generaltest} and Example~\ref{EX:summabletest}.

\chapter[Further Properties of Residual Forms]
{Further Properties of \\Residual Forms
\protect\footnotemark{}\protect
\footnotetext{The main results in this chapter are joint work 
with S.\ Chen, M.\ Kauers, Z.\ Li, published in~\cite{CHKL2015}.}}
\label{CH:rfproperties}

In Chapter~\ref{CH:apreduction}, we presented a modified version of
the \ap reduction, which decomposes a univariate hypergeometric term
into a summable part and a non-summable part. Moreover, the
non-summable part is described by a residual
form. In~\cite{BCCLX2013}, the authors used the Hermite
reduction\index{Hermite reduction} for univariate hyperexponential
functions to compute telescopers for bivariate hyperexponential
functions\index{hyperexponential function}. It allows one to separate
the computation of telescopers from that of certificates. We try to
translate their idea into the hypergeometric setting.

We call a bivariate nonzero term {\em  hypergeometric}
if its shift-quotients with respect to the two variables are both 
rational functions.  Given a hypergeometric term $T(n,k)$. 
Let $\sigma_n$ and $\sigma_k$ be the shift operators w.r.t.~$n$ and $k$, 
respectively. Applying the modified \ap reduction to $T$ as well as 
its shifts $\sigma_n(T), \dots, \sigma_n^i(T)$ w.r.t.~$k$, 
where $i$ is a nonnegative integer, we obtain
\[\sigma_n^j(T) \equiv_k r_j H\mod \bU_K\quad \text{for}\ j = 0, \dots, i,\]
where $H$ is another bivariate hypergeometric term whose
shift-quotient $K$ w.r.t.~$k$ is shift-reduced w.r.t.~$k$, and $r_j$
is a residual form w.r.t.~$K$. For univariate rational functions
$c_0(n), c_1(n), \dots, c_i(n)$, not all zero, we have
\[\sum_{j= 0}^i c_j \sigma_n^j(T) \equiv_k \sum_{j=0}^i c_j r_j H\mod \bU_K.\]
It is readily seen that $\sum_{j=0}^{i}c_j\sigma_n^j$ is a telescoper
for $T$ w.r.t.~$k$ if $\sum_{j=0}^i c_j r_j = 0$. Unfortunately, the
converse is false. This is because $\sum_{j=0}^i c_j r_j $ is not
necessarily a residual form, although all the $r_j$'s are. Thus
Theorem~\ref{THM:mapred} is not applicable.
This situation does not occur in the differential
case~\cite{BCCLX2013}.  To make Theorem~\ref{THM:mapred} applicable,
we need to find a way to make $\sum_{j=1}^ic_jr_j$ a residual form.

This chapter aims at connecting univariate hypergeometric
terms with bivariate ones for the next two chapters. In this chapter,
we present further properties of residual forms so as to estimate the
order bounds of telescopers in Chapter~\ref{CH:bounds}. To make the
modified reduction applicable to compute telescopers for
hypergeometric terms in Chapter~\ref{CH:telescoping}, we also show
that the linear combination of residual forms is well-behaved in terms
of congruences.
%
\section{Rational normal forms}\label{SEC:rnf}
\index{rational! normal form|(}
In this section, we recall the notion of rational normal forms from
\cite{AbPe2002b} and review the relation between distinct rational
normal forms of a rational function.
\begin{definition}\label{DEF:shift-equiv}
  Two polynomials $ p,q \in\bF[k]$ are called {\em
    shift-equivalent}\index{shift-! equivalent} w.r.t.~$k$ if there
  exists an integer $m$ such that $p = \sigma_k^m(q)$. We denote it by
  $p \sim_k q$.
\end{definition}
It is readily seen that $\sim_k$ is an equivalence relation. We call a
polynomial in~$\bF[k]$ {\em monic} if its leading coefficient
w.r.t.~$k$ is $1$.
\begin{definition}\label{DEF:rnf}
  Let $f$ be a rational function in~$ \bF(k)$. A rational function
  pair $(K,S)$ with~$K, S\in \bF(k)$ is called a {\em rational normal
    form}\index{rational! normal form} of $f$ if 
  \[f = K \cdot \frac{\sigma_k(S)}{S}\]
  and $K$ is shift-reduced.
\end{definition}
By Theorem~1 in \cite{AbPe2002b}, every rational function has a
rational normal form. It is not hard to see that there is a one-to-one
correspondence between multiplicative decompositions for a given
hypergeometric term and rational normal forms for the corresponding
shift-quotient. More precisely, for a hypergeometric term~$T$
over~$\bF(k)$, a rational function pair $(K,S)$ is a rational normal
form of $\sigma_k(T)/T$ if and only if $K$ is a kernel of~$T$ and $S$
a corresponding shell, if and only if $T$ has a multiplicative
decomposition $T=SH$ with~$H$ a hypergeometric term whose
shift-quotient is $K$.

In fact, a rational function can have more than one rational normal
form, as illustrated by the following example.
\begin{example}[Example~1 in~\cite{AbPe2002b}]\label{EX:rnfs}
  Consider a rational function
  \[f = \frac{k(k+2)}{(k-1)(k+1)^2(k+3)}.\] 
  It can be verified that the following rational function pairs
  \begin{align*}
    &\left(\frac{1}{(k+1)(k+3)},\, (k-1)(k+1)\right), \quad
    \left(\frac{1}{(k+1)^2},\, \frac{k-1}{k+2}\right), \\[1ex]
    &\left(\frac{1}{(k-1)(k-3)},\, \frac{k+1}{k}\right), \quad
    \left(\frac{1}{(k-1)(k+1)},\, \frac{1}{k(k+2)}\right).
  \end{align*}
  are all rational normal forms of~$f$.
\end{example}
The next theorem describes a relation between two distinct rational
normal forms of a rational function.\index{rational! normal form}
\begin{theorem}[Theorem~2 in \cite{AbPe2002b}]\label{THM:onetoone}
  Assume that~$(K, S), (K', S')\in \bF(k)^2$ are distinct 
  rational normal forms of a rational function in $\bF(k)$. 
  Write
  \[K=c\,\frac{u}{v}\quad \text{and} \quad K' = c'\,\frac{u'}{v'},\]
  where $c,c'\in \bF$, $u, u',v,v'\in \bF[k]$ are all monic, and
  $\gcd(u,v) = \gcd(u', v') = 1$. Then
  \begin{itemize}
  \item[$(i)$] $c = c'$;

    \smallskip
  \item[$(ii)$] $\deg_k(u) = \deg_k(u')$ and $\deg_k(v) = \deg_k(v')$;

    \smallskip
  \item[$(iii)$] there is a one-to-one correspondence $\phi$ between the
    multi-sets of nontrivial monic irreducible factors of $u$ and $u'$
    such that $p \sim_k \phi(p)$ for any nontrivial monic irreducible
    factor $p$ of $u$.

    \smallskip
  \item[$(iv)$] there is a one-to-one correspondence $\psi$ between the
    multi-sets of nontrivial monic irreducible factors of $v$ and $v'$
    such that $p \sim_k \phi(p)$ for any nontrivial monic irreducible
    factor $p$ of $v$.
  \end{itemize}
\end{theorem}
\index{rational! normal form|)}
\section{Uniqueness and relatedness of residual
  forms}\label{SEC:rfprop} \index{residual form} 
In this section, we will present two useful properties of residual
forms, which enables us to derive order bounds in
Chapter~\ref{CH:bounds}. For the notion of residual forms, one can
refer to Definition~\ref{DEF:residual}.

Unlike the differential case, a rational function may have more than
one residual form in the shift case. These residual forms, however,
are related to each other in some way. To describe it precisely, we
introduce the notion of shift-relatedness.
\begin{definition}\label{DEF:relatedness}
  Two shift-free polynomials $p, q \in \bF[k]$ are called {\em
    shift-related}\index{shift-! related}, denoted by $p \approx_k q$,
  if for any nontrivial monic irreducible factor $f$ of $p$, there
  exists a unique monic irreducible factor $g$ of $q$ with the same
  multiplicity as $f$ in~$p$ such that $f \sim_k g$, and vice versa.
\end{definition}
It is readily seen that~$\approx_k$ is an equivalence relation. The
following theorem describes the uniqueness of residual forms.
\begin{theorem}\label{THM:uniqueness}\index{significant denominator}\index{residual form}
  Let~$K\in \bF(k)$ be a shift-reduced rational function. Assume that
  $r_1, r_2$ are both residual forms of a same rational function
  in~$\bF(k)$ w.r.t.~$K$. Then the significant denominators of~$r_1$
  and~$r_2$ are shift-related to each other.
\end{theorem}
\begin{proof}
  Assume that~$r_1,r_2$ are of the forms
  \begin{equation*}\label{EQ:drfs}
    r_1 = \frac{a_1}{b_1} + \frac{q_1}{v} \quad \text{and}\quad r_2 = \frac{a_2}{b_2} + \frac{q_2}{v},
  \end{equation*}
  where for $i = 1,2$,~$a_i, b_i \in \bF[k]$, $\deg_k(a_i) <
  \deg_k(b_i)$,~$\gcd(a_i,b_i) = 1$,~$b_i$ is monic, shift-free and
  strongly coprime with~$K$, $q_i\in \bW_{K}$, and~$v$ is the
  denominator of~$K$. Since~$r_1, r_2$ are both residual forms of the
  same rational function, $r_1 \equiv_k r_2 \mod \bV_K$, which is
  equivalent to
  \[\frac{a_1}{b_1} \equiv_k\frac{a_2}{b_2}+\frac{q_2-q_1}{v} \mod
  \bV_K.\] 
  By \eqref{EQ:summable}, there exists $w\in \bF(k)$ so that
  \begin{equation}\label{EQ:related}
    \frac{a_1 v}{b_1}=u \sigma_k(w) - v w +\frac{a_2 v}{b_2}+(q_2-q_1).
  \end{equation}
  Let~$f\in \bF[k]$ be a nontrivial monic irreducible factor of~$b_1$
  with multiplicity~$\alpha>0$. If~$f^\alpha$ divides~$b_2$, then we
  are done. Otherwise, let~$\den(w)$ be the denominator of~$w$. Since
  $b_1$ is strongly coprime with $K$, we have $\gcd(f^\alpha,v)=1$. By
  \eqref{EQ:related} and partial fraction decomposition, $f^\alpha$
  either divides~$\den(w)$ or~$\sigma_k(\den(w))$. If $f^\alpha$
  divides~$\den(w)$, let
  \begin{equation*}
    m =  \max \{k \in \bZ \mid \sigma_k^{k}(f)^\alpha \text{ divides } \den(w)\},
  \end{equation*}
  and then $m \geq 0$. Since $b_1$ is strongly coprime with~$K$,
  $\gcd(\sigma_k^{m+1}(f)^\alpha, u) = 1$. Apparently,
  $\sigma_k^{m+1}(f)^\alpha$ divides $ \sigma_k(\den(w))$ but doesn't
  divide $\den(w)$ as $m$ is maximal. Note that $b_1$ is shift-free
  and $f\mid b_1$, thus $b_1$ is not divisible by
  $\sigma_k^{m+1}(f)^\alpha$. Hence~\eqref{EQ:related}
  implies~$\sigma_k^{m+1}(f)^\alpha$ is the required factor of~$b_2$.
  Similarly, we can show that~$\sigma_k^{\ell}(f)^\alpha$ with
  \[\ell = \min \{k \in \bZ \mid \sigma_k^{k}(f)^\alpha \text{ divides
  } \den(w)\} \leq -1,\]
  is the required factor of~$b_2$, if~$f^\alpha$ divides~$\sigma_k(\den(w))$.
	
  In summary, there always exists a monic irreducible factor of~$b_2$
  with multiplicity at least~$\alpha$ such that it is shift-equivalent
  to~$f$. Due to the shift-freeness of~$b_2$, this factor is
  unique. The same conclusion holds when we switch the roles of $b_1$
  and $b_2$. Therefore, $b_1\approx_k b_2$ by definition.
\end{proof}
For a given hypergeometric term, the above theorem reveals the
relation between two residual forms of the shell with respect to a
same kernel. To study the case with different kernels, we need the
following two lemmas.
\begin{lemma}\label{LEM:vrelatedness}
  Let~$(K,S)$ be a rational normal form of $f\in\bF(k)$ and~$r$ a
  residual form of~$S$ w.r.t.~$K$. Write $K = u/v \ \text{with}\ u,
  v\in \bF[k]\text{ and } \gcd(u,v)=1.$ Assume that $p$ is a
  nontrivial monic irreducible factor of~$v$ with
  multiplicity~$\alpha> 0$. Then the pair
  \[(K', S') = \left(\frac{u}{v'\sigma_k(p)^\alpha}, p^\alpha
    S\right)\] 
  is a rational normal form of~$f$, in which~$v'= v/p^\alpha$. 
  Moreover, there exists a residual form~$r'$ of~$S'$
  w.r.t.~$K'$ whose significant denominator equals that of~$r$.
\end{lemma}
\begin{proof}
  Since~$K$ is shift-reduced, so is~$K'$. The first assertion follows
  by noticing
  \[K\frac{\sigma_k(S)} {S} = \frac{u}{v'
    p^\alpha}\frac{\sigma_k(S)}{S} = \frac{u}{v'
    \sigma_k(p)^\alpha}\frac{\sigma_k(p^\alpha S) }{p^\alpha
    S}=K'\frac{\sigma_k(S')}{S'}.\] 
  Let $r$ be of the form $r = a/b+q/v$, 
  where $a, b, q \in \bF[k]$, $\deg_k(a) < \deg_k(b)$,
  $\gcd(a,b) = 1$, $b$ is monic, shift-free and strongly coprime
  with~$K$, and $q\in \bW_{K}$. Then there exists a rational function
  $g\in \bF(k)$ such that
  \[S = K \sigma_k(g) - g + \frac{a}{b} + \frac{q}{v'p^\alpha},\]
  which implies
  \begin{align*}
    S' &= p^\alpha S = p^\alpha K \sigma_k(g) - p^\alpha g + \frac{ap^\alpha }{b} + \frac{q}{v'}\\[1ex]
    &= \frac{u}{v'\sigma_k(p)^\alpha} \sigma_k(p^\alpha g) - p^\alpha g 
    + \frac{ap^\alpha}{b} + \frac{q\sigma_k(p)^\alpha}{v'\sigma_k(p)^\alpha}\\[1ex]
    &=K'\sigma_k(p^\alpha g) - p^\alpha g + \frac{ap^\alpha }{b} +
    \frac{q\sigma_k(p)^\alpha}{v'\sigma_k(p)^\alpha}
  \end{align*}
  Since $b$ is strongly coprime with $K$ and $\gcd(a,b)=1$, we have
  $\gcd(ap^\alpha, b)=1$. Using step~$3$ and step~$4$ in
  Algorithm~\ref{ALG:mapred} computes polynomials $a', q' \in \bF[k]$
  with $\deg_k(a') < \deg_k(b)$, $\gcd(a',b)=1$ and $q' \in \bW_{K'}$
  so that
  \[S' \equiv_k \frac{a'}{b}+ \frac{q'}{v' \sigma_k(p)^\alpha} \mod
  \bV_{K'}.\] Note that $b$ is strongly coprime with $K$, so $b$ is also
  strongly coprime with $K'$. Since $b$ is shift-free,
  $a'/b+ q'/(v' \sigma_k(p)^\alpha)$ is a residual form of $S'$
  w.r.t.~$K'$.
\end{proof}
\begin{lemma}\label{LEM:urelatedness}
  Let $(K,S)$ be a rational normal form of $f\in\bF(k)$ and $r$ a
  residual form of $S$ w.r.t.~$K$. Write $K = u/v \ \text{with}\ u,
  v\in \bF[k]\text{ and } \gcd(u,v)=1.$ Assume that $p$ is a
  nontrivial monic irreducible factor of $u$ with multiplicity
  $\alpha> 0$. Then the pair
  \[(K', S') = \left(\frac{u'\sigma_k^{-1}(p)^\alpha }{v},
    \sigma_k^{-1}(p)^\alpha S\right)\] 
  is a rational normal form of~$f$, in which $u'= u/p^\alpha$. 
  Moreover, there exists a residual form $r'$ of $S'$ w.r.t.~$K'$ 
  whose significant denominator equals that of~$r$.
\end{lemma}
\begin{proof}
  Similar to Lemma~\ref{LEM:vrelatedness}.
\end{proof}
\let\theenumi=\oldtheenumi
\let\labelenumi=\oldlabelenumi
\begin{prop} \label{PROP:sd} 
  Let $(K,S)$ be a rational normal form of $f\in \bF(k)$ 
  and $r$ a residual form of $S$ w.r.t.~$K$. 
  Then there exists a rational normal form $(\tilde K, \tilde S)$ 
  of~$f$ such that
  \begin{enumerate}
  \item $\tilde K$ has shift-free numerator and shift-free
    denominator;
		
    \smallskip
  \item there exists a residual form $\tilde r$ of $\tilde S$
    w.r.t.~$\tilde K$ whose significant denominator is equal to that
    of $r$.
  \end{enumerate}
\end{prop}
\begin{proof}
  Let $K=u/v$ with $u, v\in \bF[k]$ and $\gcd(u,v)=1$, and $b$ be the
  significant denominator of $r$.
	
  Assume that $v$ is not shift-free. Then there exist two nontrivial
  monic irreducible factors $p$ and $\sigma_k^m(p)$ $(m>0)$ of $v$
  with multiplicity $\alpha>0$ and $\beta>0$, respectively. W.l.o.g.,
  assume further that $\sigma_k^{\ell}(p)$ is not a factor of $v$ for
  all $\ell <0$ and $\ell >m$. By Lemma~\ref{LEM:vrelatedness}, $f$
  has a rational normal form $(K^\prime, S^\prime)$, in
  which~$K^\prime$ has a denominator of the form $\den(K')=v^\prime
  \sigma_k(p)^\alpha$, where $v^\prime=v/p^\alpha$, and the numerator
  remains to be $u$. Moreover, there exists a residual form of
  $S^\prime$ w.r.t.~$K^\prime$ whose significant denominator is $b$.
  If $m=1$, then $\sigma_k(p)$ is an irreducible factor of~$\den(K')$
  with multiplicity $\alpha+\beta$. Otherwise, it is an irreducible
  factor of~$\den(K')$ with multiplicity~$\alpha$.  More importantly,
  $\sigma_k^{\ell}(p)$ is not a factor of $\den(K')$ for all~$\ell
  <1$. Iteratively using the argument, we arrive at a rational normal
  form of $f$ such that $\sigma_k^m(p)$ divides the denominator of the
  new kernel with certain multiplicity but $\sigma_k^i(p)$ does not
  whenever $i\neq m$, and the numerator remains to be $u$.  Moreover,
  there exists a residual form of the new shell with respect to the
  new kernel whose significant denominator is equal to $b$. Applying
  the same argument to each irreducible factor, we can obtain a
  rational normal form of $f$ whose kernel has the numerator $u$ and a
  shift-free denominator, and whose shell has a residual form with
  significant denominator~$b$.
	
  With Lemma~\ref{LEM:urelatedness}, one can obtain a rational normal
  form of $f$ whose kernel has a shift-free numerator and whose shell
  has a residual form with significant denominator~$b$.
\end{proof}
A nonzero rational function is said to be {\em
  shift-free}\index{shift-! free} if it is shift-reduced and its
denominator and numerator are both shift-free. The relatedness of
residual forms with respect to different kernels is given below.
\begin{theorem} \label{THM:sf}\index{significant denominator}\index{residual form} 
  Let $(K,S), (K^\prime, S^\prime)$ be two rational normal forms 
  of $f\in \bF(k)$, and~$r, r^\prime$ residual forms of $S$ 
  (w.r.t.~$K$) and $S^\prime$ (w.r.t.~$K'$), respectively.  
  Then the significant denominators of
  $r$ and $r^\prime$ are shift-related.
\end{theorem}
\begin{proof}
  Let $b$ and $b^\prime$ be the significant denominators of $r$ and
  $r^\prime$, respectively.  By the above proposition, there exist two
  rational normal forms $(\tilde K, \tilde S)$ and $(\tilde{K}^\prime,
  \tilde{S}^\prime)$ of $f$ such that their kernels are shift-free and
  their shells have residual forms whose significant denominators are
  $b$ and $b^\prime$, respectively.
	
  According to Theorem~\ref{THM:onetoone}, the respective denominators
  $\tilde v$ and $\tilde{v}^\prime$ of $\tilde K$ and
  $\tilde{K}^\prime$ are shift-related. It follows that for a
  nontrivial monic irreducible factor $p$ of $\tilde v$ with
  multiplicity $\alpha>0$, there exists a unique factor
  $\sigma_k^\ell(p)$ with $\ell \in \bZ$ of $\tilde{v}^\prime$ with
  the same multiplicity. W.l.o.g., we may assume $\ell \le
  0$. Otherwise, we can switch the roles of two pairs $(\tilde K,
  \tilde S)$ and $(\tilde{K}^\prime, \tilde{S}^\prime)$.  If $\ell<0$,
  a repeated use of Lemma~\ref{LEM:vrelatedness} leads to a new
  rational normal form $(\tilde{K}^{\prime\prime},
  \tilde{S}^{\prime\prime})$ from $(\tilde{K}^\prime,
  \tilde{S}^\prime)$, such that $\tilde{K}^{\prime\prime}$ is
  shift-free with the same numerator as $\tilde{K'}$, and $p$ is a
  factor of the denominator of $\tilde{K}^{\prime\prime}$ with the
  same multiplicity~$\alpha$. Moreover, $\tilde{S''}$ has a residual
  form w.r.t.~$\tilde{K''}$ with significant denominator $b'$.
	
  Applying the above argument to each irreducible factor and using
  Lemma~\ref{LEM:urelatedness} for numerators in the same fashion, we
  can obtain two new rational normal forms whose kernels are equal and
  whose shells have respective residual forms with significant
  denominators $b$ and $b^\prime$. It follows from
  Theorem~\ref{THM:uniqueness} that $b$ and $b^\prime$ are
  shift-related.
\end{proof}
\section{Sum of two residual forms}\label{SEC:rfsum}
To compute a telescoper for a given bivariate hypergeometric terms by
the modified \ap reduction, we are confronted with the difficulty that
the sum of two residual forms is not necessarily a residual form, as
mentioned at the beginning of this chapter. This is because the least
common multiple of two shift-free polynomials is not necessarily
shift-free.

The goal of this section is to show that the sum of two residual forms
is congruent to a residual form modulo~$\bV_K$.
 \begin{example}\label{EX:rfsum}
   Let $K=1/k$,~$r=1/(2k+1)$ and~$s=1/(2k+3)$. Then both~$r$ and~$s$
   are residual forms w.r.t.~$K$, but their sum is not, because the
   denominator $(2k+1)(2k+3)$ is not shift-free.  However, we can
   still find an equivalent residual form. For example, we have
   \[
   r + s \equiv_k -\frac{1}{2(2k+1)} + \frac{1}{2k} \mod\bV_K.
   \]
   Note that the residual form is not unique. Another possible choice
   is
   \[
   r+s\equiv_k \frac1{3(2k+3)}+ \frac1{3k} \mod\bV_K.
   \]
 \end{example}
 
\begin{lemma}\label{LEM:rfaddclosed}
  With Convention~\ref{CON:notation}, let $r, s\in \bF(k)$ be two
  residual forms w.r.t.~$K$, i.e., $r$ and $s$ can be written as
  \[r = \frac{a}{f} + \frac{p}{v} \quad \text{and} \quad s =
  \frac{b}{g}+ \frac{q}{v},\] 
  where $a,f,b,g \in \bF[k]$, $\deg_k(a)<\deg_k(f)$, 
  $\deg_k(b)<\deg_k(g)$, $p,q\in \bW_K$, and $f,g$ are shift-free 
  and strongly coprime with~$K$. 
  Assume that $\gcd(a,f) = \gcd(b,g) = 1$. 
  Then for all $\lambda, \mu \in \bF$, 
  $\lambda r+\mu s$ is a residual form w.r.t.~$K$ if and only if the
  least common multiple of $f$ and $g$ is shift-free.
\end{lemma}
\begin{proof}
  Let $h$ be the least common multiple of $f$ and $g$. Then
  \begin{equation}\label{EQ:sum}
    \lambda r+\mu s = \frac{\lambda a(h/f)+\mu b(h/g)}{h} + \frac{\lambda p+\mu q}{v}.
  \end{equation}
	
  We first show the sufficiency. Assume that~$h$ is shift-free. It is
  clear that
  \[\deg_k(\lambda a(h/f) +\mu b (h/g)) < \deg_k(h).\]
  Since~$\bW_K$ is a $\bF$-vector space, we have $\lambda p +\mu q \in
  \bW_K$. Note that $f$ and $g$ are strongly coprime with~$K$, so is
  $h$. By definition, $\lambda r+\mu s$ is a residual form w.r.t.~$K$.
	
  To show the necessity, we suppose otherwise that~$h$ is not
  shift-free. Since $\lambda r+\mu s$ is a residual form w.r.t.~$K$,
  there exist~$b^*,h^*\in \bF[k]$ and $q^* \in \bW_K$
  with~$\deg_k(b^*) < \deg_k(h^*)$, and $h^*$ shift-free and strongly coprime with~$K$, 
  such that
  \[\lambda r+\mu s = \frac{b^*}{h^*}+\frac{q^*}{v}.\]
  It follows from~\eqref{EQ:sum} that
  \begin{equation}\label{EQ:mid}
    \frac{(\lambda a(h/f)+\mu b(h/g))v}{h}=\frac{b^*v}{h^*}+q^*-\lambda p-\mu q.
  \end{equation}
  Since~$h$ is not shift-free and $f,g$ are shift-free, there exist
  nontrivial monic irreducible factors~$p'$ and~$\sigma_k^\ell(p')$ of
  $h$ such that $p' \mid f$ and~$\sigma_k^\ell(p')\mid g$,
  where~$\ell$ is a nonzero integer. Because $\gcd(a,f) = \gcd(b,g) =
  1$ and~$h \mid fg$, so
  \begin{itemize}
  \item $p'\nmid (h/f)$ and $p'\nmid a$, but $p'\mid (h/g)$;
		
    \smallskip
  \item $\sigma_k^\ell(p') \nmid (h/g)$ and $\sigma_k^\ell(p) \nmid
    b$, but $\sigma_k^\ell(p') \mid (h/f)$.
  \end{itemize}
  Since $h$ is also strongly coprime with~$K$, $p'$ and
  $\sigma_k^\ell(p')$ are coprime with $v$. Thus they both divide the
  denominator of the left-hand side of~\eqref{EQ:mid}. By partial
  fraction decomposition, $p'$ and $ \sigma_k^\ell(p')$ both
  divide~$h^*$, a contradiction as $h^*$ is shift-free.
\end{proof}
To describe the shift-freeness of the least common multiple of two
polynomials, we introduce the following notions.
\begin{definition}\label{DEF:shiftcoprime}
  Let $f$ and $g$ be two nonzero polynomials in $\bF[k]$.  According
  to~\cite[\S 3]{AbPe2002b}, the {\em dispersion} \index{dispersion}
  of~$f$ and $g$ is defined to be the largest nonnegative integer~$\ell$ 
  such that $f$ and~$\sigma_k^\ell(g)$ have a
  nontrivial common divisor, or $-1$ if no such~$\ell$
  exists. Moreover, we say that $f$ and $g$ are {\em
    shift-coprime}\index{shift-! coprime} if $\gcd(f,\sigma_k^\ell(g))
  = 1$ for all nonzero integer $\ell$.
\end{definition}
It is clear that the least common multiple of two shift-free
polynomials is shift-free if and only if these two polynomials are
shift-coprime. Let $f$ and $g$ be two nonzero shift-free polynomials
in~$\bF[k]$. By polynomial factorization and dispersion computation
(see~\cite{AbPe2002b}), one can uniquely decompose
\begin{equation} \label{EQ:shiftcoprime} 
  g = \tilde{g} \sigma_k^{\ell_1}\left(p_1^{m_1} \right) \cdots
  \sigma_k^{\ell_\rho}\left(p_\rho^{m_\rho} \right),
\end{equation}
where $\tilde{g}$ is shift-coprime with $f$, $p_1, \ldots, p_\rho$ are
pairwise distinct and monic irreducible factors of $f$, $\ell_1,
\ldots, \ell_\rho$ are nonzero integers, $m_1, \ldots, m_\rho$ are
multiplicities of the factors $\sigma_k^{\ell_1}(p_1)$, $\ldots$,
$\sigma_k^{\ell_\rho}(p_\rho)$ in $g$, respectively. We refer
to~\eqref{EQ:shiftcoprime} as the \emph{shift-coprime decomposition}%
\index{decomposition! shift-coprime} \index{shift-! coprime decomposition} of $g$ w.r.t.~$f$.
\begin{remark} \label{REM:coprime} 
  The factors $\tilde g, \sigma_k^{\ell_1}\left(p_1^{m_1} \right)$, \ldots,
  $\sigma_k^{\ell_\rho}\left(p_\rho^{m_\rho}\right)$
  in~\eqref{EQ:shiftcoprime} are pairwise coprime, since $f$ and $g$
  are shift-free.
\end{remark}
To construct a residual form congruent to the sum of two given
residual ones, we need three technical lemmas.  The first one
corresponds to the kernel reduction%
\index{reduction! kernel}\index{kernel reduction} in~\cite{BCCLX2013}.
\begin{lemma} \label{LEM:kernelreduction} 
  With Convention~\ref{CON:notation}, assume that $p_1, p_2$ are in
  $\bF[k]$ and $m$ in~$\bN$. Then there exist $q_1, q_2$ in~$\bW_K$
  such that
 \[\frac{p_1}{\prod_{i=0}^m \sigma_k^i(v)}\equiv_k 
   \frac{q_1}{v}~{\rm mod}~\bV_K \quad \text{and} \quad
   \frac{p_2}{\prod_{j=1}^m \sigma_k^{-j}(u)}\equiv_k 
   \frac{q_2}{v}~{\rm mod}~\bV_K. \]
\end{lemma}
\begin{proof}
  To prove the first congruence, let $w_m= \prod_{i=0}^m \sigma_k^i(v)$.
 	
  We proceed by induction on $m$. If $m=0$, then the conclusion holds
  by Lemma~\ref{LEM:reducedform}.  Assume that the lemma holds for
  $m-1$ with $m>0$.  Consider the equality
  \[ \frac{p_1}{w_m} = K \sigma_k\left( \frac{s}{w_{m-1}}\right) -
  \frac{s}{w_{m-1}} + \frac{t}{w_{m-1}}, \] where $s, t \in \bF[k]$
  are to be determined. This equality holds if and only
  if \[\sigma_k(s) u+ (t-s) \sigma_k^{m}(v) = p_1.\] Since $u$ and
  $\sigma_k^{m}(v)$ are coprime, such $s$ and $t$ can be computed by
  the extended Euclidean algorithm. Thus, $p_1/w_m \equiv_kt/w_{m-1}
  \mod \bV_K. $ Consequently, $p_1/w_m$ has a required residual form
  by the induction hypothesis.
 	
  To prove the second congruence, we use the identity
  \[\frac{p_2}{\sigma_k^{-1}(u)} =
  K \sigma_k\left( - \frac{p_2}{\sigma_k^{-1}(u)} \right) - \left(-
    \frac{p_2}{\sigma_k^{-1}(u)}\right) + \frac{\sigma_k\left( p_2
    \right)}{v}, \] 
  which implies that $p_2/\sigma_k^{-1}(u)
  \equiv_k\sigma_k\left( p_2 \right)/v~{\rm mod}~\bV_K.$ 
  By Lemma~\ref{LEM:reducedform}, there exists a polynomial 
  $q_2 \in \bW_K$ such that $q_2/v$ is a residual form of
  $p_2/\sigma_k^{-1}(u)$ w.r.t. $K$.  
  Thus the conclusion holds for $m =0$.  
  Assume that the congruence holds for $m-1$ with $m>0$.  The
  induction can be completed as in the proof for $p_1/w_m$.
\end{proof}
The next lemma provides us with flexibility to rewrite a rational
function modulo~$\bV_K$.
\begin{lemma} \label{LEM:cong} 
  Let~$K \in \bF(k)$ be nonzero and shift-reduced. 
  Then for every rational function $f \in \bF(k)$ and
  every positive integer~$\ell$,
 $$ f \equiv_k \sigma_k^\ell(f) \prod_{i=0}^{\ell-1} \sigma_k^i(K) \equiv_k
 \sigma_k^{-\ell}(f) \prod_{i=1}^\ell
 \sigma_k^{-i}\left(\frac{1}{K}\right) \bmod \bV_K.$$
\end{lemma}
\begin{proof}
  Let's show the first congruence by induction
  on~$\ell$. For~$\ell=1$, the identity
  \[f = K \sigma_k(-f) - (-f) + \sigma_k(f) K\] 
  implies that~$f$ is congruent to~$\sigma_k(f) K$ modulo~$\bV_K$. 
  Assume that it holds for~$\ell-1$ with~$\ell >1$. 
  Set~$w_\ell= \prod_{i=0}^{\ell-1}\sigma_k^i(K)$. 
  Then by the induction hypothesis,
  \[f\equiv_k\sigma_k^{\ell-1}(f) w_{\ell-1}\mod \bV_K.\] 
  Moreover, $\sigma_k^{\ell-1}(f) w_{\ell-1}\equiv_k\sigma_k^{\ell}(f) w_{\ell}
  \mod \bV_K$ by the induction base, in which~$f$ is replaced
  with~$\sigma_k^{\ell-1}(f) w_{\ell-1}$. Hence, $f$ is congruent to
  $\sigma_k^{\ell}(f) w_{\ell}$ modulo~$\bV_K$.
 	
  The second congruence can be shown similarly. For the base case
  $\ell=1$, let~$r = \sigma_k^{-1}(f) \sigma_k^{-1}(1/K)$. Then the
  identity $f = K \sigma_k(r) -r + r$ implies that $f$ is congruent
  to~$r$ modulo~$\bV_K$.  We can then proceed as in the proof of the
  first congruence.
\end{proof}
\begin{lemma} \label{LEM:rfshift} 
  With Convention~\ref{CON:notation}, let~$a, b \in \bF[k]$ with $b\neq0$. 
  Assume that~$b$ is shift-free and strongly coprime with~$K$.  
  Assume further that~$\sigma_k^\ell(b)$ is strongly coprime with~$K$ 
  for some integer~$\ell$, then~$a/b$ has a residual 
  form~$c/\sigma_k^\ell(b) + q/v$ w.r.t.~$K$, where~$c \in
  \bF[k]$ with~$\deg_k(c) < \deg_k(b)$ and~$q \in \bW_K$.
\end{lemma}
\begin{proof}
  First, consider the case in which~$\ell \ge 0$.  If~$\ell=0$, then
  there exist two polynomials $c, p \in \bF[k]$
  with~$\deg_k(c)<\deg_k(b)$ such that $a/b=c/b+p$.  The lemma follows
  from Remark~\ref{REM:poly}.  Assume that~$\ell > 0$. By the first
  congruence of Lemma~\ref{LEM:cong},
  \[ \frac{a}{b} \equiv_k \sigma_k^\ell\left(\frac{a}{b}\right) \left(
    \prod_{i=0}^{\ell-1} \sigma_k^i(K) \right) =
  \frac{\sigma_k^\ell(a)}{\sigma_k^\ell(b)} \frac{\prod_{i=0}^{\ell-1}
    \sigma_k^i(u)}{\prod_{i=0}^{\ell-1} \sigma_k^i(v)} \mod \bV_K. \]
  Note that~$\sigma_k^\ell(b)$ is strongly coprime with~$v$ by
  assumption. Then it is coprime with the product~$v \sigma_k(v)
  \cdots \sigma_k^{\ell-1}(v)$. By partial fraction decomposition, we
  get
  \[ \frac{a}{b} \equiv_k\frac{\tilde{a}}{\sigma_k^\ell(b)} +
  \frac{\tilde{q}}{\prod_{i=0}^{\ell-1} \sigma_k^i(v) } \mod \bV_K,\]
  where $\tilde{a}, \tilde{q}\in \bF[k]$ and $\deg_k \tilde{a}< \deg_k
  (b)$. By the first congruence of Lemma~\ref{LEM:kernelreduction},
  the second summand in the right-hand side of the above congruence
  can be replaced by a residual form whose denominator is equal
  to~$v$. The first assertion holds.
 	
  The case $\ell<0$ can be handled in the same way, in which the
  second congruences of Lemmas~\ref{LEM:cong}
  and~\ref{LEM:kernelreduction} will be used instead of the first
  ones.
\end{proof}
 
\begin{remark}\label{REM:shiftsignificantden}
  With the assumptions of the above lemma, let $p$ be a nontrivial
  factor of $b$ with $\gcd(b', p) =1$ where $b' = b/p$. Assume that
  $\sigma_k^\ell(p)$ is also strongly coprime with~$K$. Then by partial
  fraction decomposition and Lemma~\ref{LEM:rfshift}, there
  exist~$c,q\in \bF[k]$ with $\deg_k(c) < \deg_k(b)$ and~$q\in \bW_K$
  such that $c/(b'\sigma_k^\ell(p)) + q/v $ is a residual form
  of~$a/b$ w.r.t.~$K$.
\end{remark}
We will refer to Lemma~\ref{LEM:rfshift} and
Remark~\ref{REM:shiftsignificantden} as {\em the shifting property of
  significant denominators}.\index{significant denominator} %
Now we are ready to present the main result of this section.
\begin{theorem} \label{THM:rfsum}\index{residual form} 
  With Convection~\ref{CON:notation}, let~$r$ and~$s$ be two residual 
  forms w.r.t.~$K$.  
  Then there exists a residual form~$t$ congruent to $s$ modulo $\bV_K$
  so that for all constants~$\lambda, \mu \in \bF$, the sum
  $\lambda r+ \mu t$ is a residual form w.r.t.~$K$ congruent
  to~$\lambda r + \mu s$ modulo~$\bV_K$.
\end{theorem}
\begin{proof}
  Since $r$ and $s$ are two residual forms w.r.t.~$K$, they can be
  written as
  \begin{equation}\label{EQ:rfform}
    r=\frac{a}{f}+\frac{p}{v} \quad \text{and} \quad
    s=\frac{b}{g}+\frac{q}{v},
  \end{equation}
  where~$a,f,b,g \in \bF[k]$,
  $\deg_k(a)<\deg_k(f)$, $\deg_k(b)<\deg_k(g)$, $p, q \in \bW_K$,
  and~$f, g$ are shift-free and strongly coprime with~$K$.
 	
  Assume that~\eqref{EQ:shiftcoprime} is the shift-coprime
  decomposition of~$g$ w.r.t.~$f$.  Define $P_i=\sigma_k^{\ell_i}(p_i)$
  for~$i=1$, \ldots,~$\rho$.  By Remark~\ref{REM:coprime} and partial
  fraction decomposition,
  \begin{equation} \label{EQ:adddecomp1} 
   \frac{b}{g} = \frac{b_0}{\tilde g} + \sum_{i=1}^\rho \frac{b_i}{P_i^{m_i}},
  \end{equation}
  where~$b_0, b_1, \ldots, b_\rho\in \bF[k]$,
  $\deg_k(b_0)<\deg_k(\tilde{g})$ and $\deg_k(b_i) < m_i
  \deg_k(p_i)$. Note that~$p_i=\sigma_k^{-\ell_i}(P_i)$, which is a
  factor of~$f$. Thus it is strongly coprime with~$K$. So we can apply
  Lemma~\ref{LEM:rfshift} to each fraction~$b_i/P_i^{m_i}$
  in~\eqref{EQ:adddecomp1} to get
  \begin{equation} \label{EQ:adddecomp2} 
    \frac{b}{g} \equiv_k
    \frac{b_0}{\tilde g} + \sum_{i=1}^\rho
    \frac{b_i^\prime}{p_i^{m_i}} + \frac{q^\prime}{v} \mod \bV_K,
  \end{equation}
  where~$b_1^\prime, \ldots, b_\rho^\prime \in \bF[k]$, $\deg_k(b_i')<
  m_i\deg_k(p_i)$ and~$q^\prime \in \bW_K$. 
 	
  Let~$h=\tilde g \prod_{i=1}^\rho p_i^{m_i}$. Then~$h$ is shift-free
  and strongly coprime with~$K$ as both~$f$ and~$g$ are.  Since~$f$ is
  shift-free, all its factors are shift-coprime with~$f$, so are
  the~$p_i$'s, and so is~$h$. Let~$t$ be the sum of~$q/v$ and the
  rational function in the right-hand side
  of~\eqref{EQ:adddecomp2}. Then there exist $b^*\in \bF[k]$ with
  $\deg_k(b^*)<\deg_k(h)$ and $q^*\in \bW_K$ such that
  \[t = \frac{b^*}{h}+\frac{q^*}{v}.\]
  Since~$f$ and~$h$ are shift-coprime, 
  their least common multiple is shift-free.
  Therefore,~$\lambda r + \mu t$ is a residual form w.r.t.~$K$ by
  Lemma~\ref{LEM:rfaddclosed}, and $\lambda r + \mu t$ is congruent
  to~$\lambda r + \mu s$ modulo $ \bV_K$.
\end{proof}
The above proof contains an algorithm, which can translate a residual form 
properly according to a given one, so that the resulting sum
is again a residual form. We outline this algorithm as follows.
\begin{algo}[Translation of Discrete Residual Forms]\label{ALG:transferdrf}\leavevmode\null
	
	\noindent
	{\bf Input}: A shift-reduced rational function $K\in \bF(k)$, 
		a polynomial $f\in \bF[k]$ which is shift-free and strongly coprime with~$K$, 
		and a residual form $s$ w.r.t.~$K$ of the form~\eqref{EQ:rfform}.\\
	{\bf Output}: A rational function $w\in \bF(k)$ and a residual form $t$ w.r.t.~$K$ such that
        \[s = K\sigma_k(w)-w+t,\]
        and the least common multiple of the given polynomial $f$ and the significant denominator of $t$ is shift-free.
        	
	\bigskip
	\step{1}{0} Compute the shift-coprime decomposition, say~\eqref{EQ:shiftcoprime}, of $g$ w.r.t.~$f$.
	
	\smallskip
	\step 20 Set $P_i = \sigma_k^{\ell_i}(p_i)$ for $i = 1, \dots, \rho$.

	\smallskip
	\step 30 Compute the partial fraction decomposition~\eqref{EQ:adddecomp1} of $b/g$.
		
	\smallskip
	\step{4}{0} Apply Lemma~\ref{LEM:rfshift} to each $b_i/P_i^{m_i}$ to find $w_i\in \bF(k)$ and
        $b_i^\prime, q_i^\prime \in \bF[k]$
        \step{}0 with $\deg_k(b_i')<m_i\deg_k(p_i)$ and $q_i^\prime \in \bW_K$ such that
        \[\frac{b_i}{P_i^{m_i}} = K\sigma_k(w_i)-w_i+ \frac{b_i^\prime}{p_i^{m_i}} + \frac{q_i^\prime}{v}.\]
        
        \smallskip
        \step 50 Set $w = \sum_{i=1}^\rho w_i$ and
        \[t = \frac{b_0}{\tilde g} + \sum_{i=1}^\rho
        \frac{b_i^\prime}{p_i^{m_i}} + \frac{\sum_{i=1}^\rho q_i^\prime+q}{v};\]
        \step{}0 and return.
\end{algo}
\chapter[Creative Telescoping for Hypergeometric Terms]
{Creative Telescoping for\\ Hypergeometric Terms
\protect\footnotemark{}\protect
\footnotetext{The main results in this chapter are joint work 
with S.\ Chen, M.\ Kauers, Z.\ Li, published in~\cite{CHKL2015}.}}
\label{CH:telescoping}

\index{creative telescoping|(} 
In the study of combinatorics, we often
encounter problems about evaluating definite sums or proving
identities of hypergeometric terms.  These terms are exactly nonzero
solutions of first-order (partial) difference equations with
polynomial coefficients.  Traditionally~\cite{Rior1979}, such problems
were solved case by case using methods that do not give rise to
general algorithms.  Based on a series of
work~\cite{WiZe1990a,WiZe1990b,WiZe1992a,WiZe1992b,Zeil1990a,Zeil1990b,Zeil1991}
in early 1990s, Wilf and Zeilberger developed a constructive theory,
which is now known as Wilf-Zeilberger's theory.%
\index{Wilf-Zeilberger's theory} This theory provides a systematic solution to a large class
of problems concerning hypergeometric summations and identities, and
has wide application in the areas of combinatorics and physics.  The
key step of Wilf-Zeilberger's theory is to compute a telescoper for a
given hypergeometric term. The efficiency of the computation
determines the utility of this theory.  During the past 26 years,
numerous algorithms have been developed for computing telescopers.  In
early 1990s, Zeilberger~\cite{Zeil1990b} first came up with an
algorithm based on elimination techniques.  This algorithm was
improved later by Takayama~\cite{Taka1992} and Chyzak,
Salvy~\cite{ChSa1998}, respectively.  In 1990,
Zeilberger~\cite{Zeil1990a} developed another algorithm, known as
Zeilberger's (fast) algorithm, based on a parametrization of Gosper's
algorithm.
15 years later, Apagodu and Zeilberger designed a new algorithm which
reduced the problem to solving a linear system.  The common feature of
the above algorithms is that there was no way to obtain a telescoper
without also computing a certificate.  In many
applications, however, certificates are not needed, and they typically
require more storage space than telescopers do.  It would be more
efficient to avoid computing certificates if we don't need them.  To
achieve this goal, Bostan et al.~\cite{BCCL2010} presented a new
algorithm for bivariate rational functions in the differential case,
based on the Hermite reduction\index{Hermite reduction}.  This
algorithm separates the computation of telescopers and the
corresponding certificates. So far, this approach has been generalized
to several instances including rational functions in three
variables~\cite{CKS2012}, multivariate rational
functions~\cite{BLS2013}, bivariate hyperexponential
functions~\cite{BCCLX2013} and bivariate algebraic
functions~\cite{CKK2016}. These algorithms turn out to be more
efficient than the classical algorithms in practice.  However, all
these algorithms only work for the differential case. %
\index{creative telescoping|)}

In this chapter, we discuss how to translate their ideas into the
hypergeometric setting.  Using the modified \ap reduction, we develop
a new creative telescoping algorithm.  This new algorithm separates
the computation of telescopers from that of certificates. We have
implemented the new algorithm in {\sc Maple~18} and compare it to the built-in Maple procedure
\textsf{Zeilberger} in the package \textbf{SumTools[Hypergeometric]},
which is based on Zeilberger's algorithm. The experimental results
indicate that the new algorithm is faster than the Maple procedure if
it returns a normalized certificate, and the new algorithm is much
more efficient if it omits the computation of certificates.

%

\section{Bivariate hypergeometric terms}\label{SEC:bht}
In this section, we translate terminology concerning univariate
hypergeometric terms to bivariate ones and introduce the notions of
telescopers as well as certificates for bivariate hypergeometric
terms. Moreover, we recall~\cite{WiZe1992a,Abra2003} an existence
criterion for telescopers.

Let~$\set K$ be a field of characteristic zero, and~$\set K(n,k)$ be the
field of rational functions in~$n$ and~$k$ over~$\set K$.  Let~$\sigma_n$
and $ \sigma_k$ be the shift operators%
\index{operator! shift}\index{shift operator} w.r.t.~$n$ and~$k$, respectively,
defined by
\[\sigma_n(f(n,k)) = f(n+1, k) \quad \text{and} \quad \sigma_k(f(n,k)) = f(n,k+1), \]
for any rational function~$f\in \set K(n,k)$. Clearly, $\sigma_n$ and
$\sigma_k$ are both automorphisms of~$\set K$.  The pair~$(\set K(n,k),
\{\sigma_n, \sigma_k\})$ forms a {\em partial difference
  field}.\index{difference! field} A \emph{partial difference ring
  extension}\index{difference! ring} of
$(\set K(n,k),\{\sigma_n,\sigma_k\})$ is a ring $\bD$ containing
$\set K(n,k)$ together with two distinguished endomorphism $\sigma_n$ and
$\sigma_k$ from $ \bD$ to itself, whose restrictions to~$\set K(n,k)$
agree with the two automorphisms defined before, respectively.

Analogous to the univariate case in Chapter~\ref{CH:htprelim}, an
element $c\in \bD$ is called a {\em constant} if it is invariant under
the applications of $\sigma_n$ and $\sigma_k$. It is readily seen that
all constants in $\bD$ form a subring of~$\bD$. Moreover, Theorem~2 in
\cite{AbPe2002a} yields that the set of all constants in $\set K(n,k)$
w.r.t.~$\sigma_n$ and $\sigma_k$ is exactly the field $\set K$.

\begin{definition}\label{DEF:bihyper}
  Let~$\bD$ be a partial difference ring extension of~$\set K(n,k)$.
  A nonzero element~$T\in \bD$ is called a \emph{hypergeometric term}%
  \index{hypergeometric! term} over~$\set K(n,k)$ if it is invertible and 
  there exist~$f, g\in \set K(n,k)$ such that $\sigma_n(T) = fT$ 
  and~$\sigma_k(T) = g T$. We call~$f$ and~$ g$ the
  \emph{shift-quotients}\index{shift-! quotient} of~$T$ w.r.t.~$n$
  and~$k$, respectively.
\end{definition}
In the rest of this chapter and also the next chapter, whenever we
mention hypergeometric terms, they always belong to some difference
ring extension~$\bD$ of~$\set K(n,k)$, unless specified otherwise.

Let~$\bF$ be the field~$\set K(n)$, and~$\bF\langle S_n \rangle $ be
the ring of linear recurrence operators%
\index{operator! recurrence}\index{recurrence operator} in~$n$, in which the
commutation rule is that~$S_n r = \sigma_n(r)S_n$ for all~$r\in \bF$.
The application of an operator~$L = \sum_{i=0}^\rho \ell_iS_n^i\in
\lsn$ to a hypergeometric term~$T$ is defined as
\[L(T) =\sum_{i=0}^\rho \ell_i\sigma_n^i(T).\]

\begin{definition}\label{DEF:tele}
  Let~$T$ be a hypergeometric term over~$\bF(k)$. A nonzero recurrence
  operator~$L \in \lsn$ is called a
  \emph{telescoper}\index{telescoper} for~$T$ w.r.t.~$k$ if there
  exists a hypergeometric term~$G$ such that
  \[L(T) = \Delta_k(G).\] 
  We call~$G$ a \emph{certificate}\index{certificate} of~$L$.
\end{definition}
In contrast to the differential case, telescopers for hypergeometric
terms do not always exist. To describe the existence of telescopers
concisely, we recall~\cite{Abra2003} the definitions of integer-linear
polynomials and proper terms.
\begin{definition}\label{DEF:integerlinear}
  An irreducible polynomial~$p\in \set K[n, k]$ is
  called~\emph{integer-linear}\index{integer-linear} over~$\set K$ if
  there exists~$P\in \set K[z]$ and $\lambda, \mu \in \bZ$ such that
  $p = P(\lambda n + \mu k)$. A polynomial in~$\set K[n, k]$ is
  called~\emph{integer-linear} over~$\set K$ if all of its irreducible
  factors are integer-linear. A rational function in~$\set K(n,k)$ is
  called~\emph{integer-linear} over~$\set K$ if its denominator and
  numerator are both integer-linear.
\end{definition}
\begin{definition}\label{DEF:proper}
  A hypergeometric term~$T$ over~$\set K(n,k)$ is called a
  \emph{factorial term}\index{factorial term} if the
  shift-quotients~$\sigma_n(T)/T$ and~$\sigma_k(T)/T$ are
  integer-linear over~$\set K$. A \emph{proper term}%
  \index{proper term} over~$\set K(n,k)$ is the product of a factorial term and a
  polynomial in~$\set K[n,k]$.
\end{definition}
We have the following existence criterion for telescopers according
to~\cite{WiZe1992a,Abra2003}.
\begin{theorem}[Existence criterion]\label{THM:existence}%
\index{existence criterion for telescopers}\index{telescoper}%
  Let $T$ be a hypergeometric term over~$\bF(k)$ and let $K = u/v$
  with $u,v\in \bF[k]$, $\gcd(u,v) =1$ be a kernel of $T$ w.r.t.~$k$
  and $S$ a corresponding shell of $T$. Assume that applying
  Algorithm~\ref{ALG:mapred}, i.e., the modified \ap reduction
  w.r.t.~$k$ to $T$ yields
  \begin{equation}\label{EQ:adddecom}
    T = \Delta_k(gH) + \left(\frac{a}{b}+\frac{q}{v}\right)H,
  \end{equation}
  where $g\in \bF(k)$, $H = T/S$, and $a/b+q/v$ is a residual form
  of~$S$ w.r.t.~$K$, that is,~$a,b\in \bF[k]$
  with~$\deg_k(a)<\deg_k(b)$, $b$ is shift-free and strongly coprime
  with~$K$ w.r.t.~$k$, and $q\in \bW_K$. Then $T$ has a telescoper
  w.r.t.~$k$ if and only if $b$ is integer-linear over~$\set K$.%
  \index{integer-linear}
\end{theorem}
\begin{proof}
  Since the kernel $K=\sigma_k(H)/H$ is shift-reduced w.r.t.~$k$, it
  follows from \cite[Theorem~8]{AbPe2001b} that $H$ is a factorial
  term over~$\bF(k)$. Thus $K$ is integer-linear over~$\set K$, and
  then so are the numerator~$u$ and the denominator~$v$.
	
  We first show the sufficiency. Since $b$ is integer-linear
  over~$\set K$, the term $H/(bv)$ is again a factorial term. Hence
  \[\left(\frac{a}{b} + \frac{q}{v}\right) H  = (av + bq) \frac{H}{bv}\]
  is a proper term, whose telescopers exist according to the
  fundamental lemma in~\cite{WiZe1992a}. By \eqref{EQ:adddecom}, $T$
  has a telescoper w.r.t.~$k$.
	
  To show the necessity, assume that $T$ has a telescoper
  w.r.t.~$k$. Then the term~$(a/b+q/v)H$ is proper by
  \cite[Theorem~10]{Abra2003}.  Thus $H/(bv)$ is a factorial term.
  Note that
  \[\frac{\sigma_k(H/(bv))}{H/(bv)} =
  \frac{u}{\sigma_k(v)}\frac{b}{\sigma_k(b)}.\] 
  Hence, $b/\sigma_k(b)$ is integer-linear over~$\set K$ as $u, v$ are
  integer-linear. Because $b$ is shift-free w.r.t.~$k$, so
  $\gcd(b,\sigma_k(b))\in \bF$. The assertion follows by noticing that
  all elements in $\bF$ are integer-linear.
\end{proof}

\section{Telescoping via reductions}\label{SEC:telescoping}
\index{creative telescoping|(} 
Let $T$ be a hypergeometric term over $\bF(k)$. 
If there exists a telescoper for $T$ w.r.t.~$k$ by
Theorem~\ref{THM:existence}, then all telescopers for $T$ w.r.t.~$k$
together with the zero operator form a left ideal of the principal
ideal ring $\lsn$. We refer to a generator of this ideal as a {\em
  minimal telescoper}%
\index{telescoper}\index{minimal telescoper|seealso {telescoper}} for $T$ w.r.t.~$k$. 
Roughly speaking, a minimal telescoper is a telescoper of the minimal order.

Since 1990, various
algorithms~\cite{Zeil1990a,Zeil1990b,Zeil1991,Le2003b,AbLe2005} have
been designed to compute a minimal telescoper for a given
hypergeometric term, typically the classical Zeilberger's
algorithm~\cite{Zeil1990a}.  When telescopers exist, Zeilberger's
algorithm\index{Zeilberger's algorithm} constructs a telescoper for a
given hypergeometric term~$T$ by iteratively using Gosper's algorithm
to detect the summability of~$L(T)$ for an ansatz
\[L=\sum_{i=0}^\rho \ell_i S_n^i\in \lsn,\]
where $\ell_i$ are indeterminates.
In order to get a telescoper, one needs to solve a linear system with
unknowns $\ell_i$ and also unknowns from the certificate.  
Any nontrivial solution gives rise to a telescoper and 
a corresponding certificate simultaneously.
There is no obvious way to avoid the computation 
of certificates in Zeilberger's algorithm.

In order to separate the computations of telescopers and certificates,
we follow the ideas in the continuous
case~\cite{BCCL2010,Chen2011,BLS2013,BCCLX2013}, and use the modified
\ap reduction to develop a creative telescoping algorithm.  The
algorithm is outlined below.

\index{creative telescoping}
\begin{algo}[Reduction-based creative telescoping]\index{reduction-based}\label{ALG:reductionct}\leavevmode\null
	
	\noindent
	{\bf Input}: A hypergeometric term $T$ over~$\bF(k)$.\\
	{\bf Output}: A minimal telescoper for~$T$ w.r.t.~$k$ and a corresponding certificate
	if telescopers exist; \lq\lq No telescoper exists!\rq\rq, otherwise. 
	
	\bigskip
	\step{1}0 Find a kernel~$K$ and shell~$S$ of~$T$ w.r.t.~$k$ 
	such that $T=SH$ 
	\step{}0 with~$K=\sigma_k(H)/H$.

	\smallskip
	\step{2}0 Apply the modified \ap reduction to~$T$ to get
	\begin{equation}\label{EQ:initred}
		T = \Delta_k(g_0H) + r_0H.
	\end{equation}
	\step{}0 If~$r_0=0$, then return~$(1, g_0H)$.

	\smallskip
	\step{3}0 If the denominator of~$r_0$ is not integer-linear, return \lq\lq No telescoper exists!\rq\rq.

	\smallskip
	\step{4}0 Set $N = \sigma_n(H)/H$ and $R = \ell_0 r_0$, where~$\ell_0$ is an indeterminate.

	\step{}0 For $i= 1,2,\dots$ do
	\substep{4.1}{2} View $\sigma_n(r_{i-1}) NH$ as
	a hypergeometric term with kernel~$K$ and
	\substep{}{2.1} shell~$\sigma_n(r_{i-1})N$. Using Algorithm~\ref{ALG:shellred}
	and Algorithm~\ref{ALG:polyred} w.r.t.~$K$, 
	\substep{}{2.1} find $g_i'\in \bF$ and a residual form $\tilde{r}_i$ w.r.t.~$K$ such
	that
	\[\sigma_n(r_{i-1})NH = \Delta_k(g_i'H ) + \tilde{r}_iH.\]
	\substep{4.2}{2} Set $\tilde{g}_i =
	\sigma_n(g_{i-1})N + g_i'$, so that
	\begin{equation}\label{EQ:midred}
		\sigma_n^i(T) = \Delta_k(\tilde{g}_iH)+\tilde{r}_i H.
	\end{equation}
	\substep{4.3}{2} Apply Algorithm~\ref{ALG:transferdrf} to $\tilde{r}_i$ w.r.t.~$K$ and $R$, 
        to find $g_i,r_i\in \bF(k)$
	\substep{}{2.1} such that $r_i \equiv_k \tilde{r}_i \mod \bV_K$,
	\begin{equation}\label{EQ:ithred}
		\sigma_n^i(T) = \Delta_k(g_i H) + r_i H,
	\end{equation}
	\substep{}{2.1} and $R + \ell_i r_i$ is a residual form
	w.r.t.~$K$, where $\ell_i$ is an indeterminate.
	
	\smallskip
	\substep{4.4}2 Update $R$ to $R+\ell_i r_i$.
	\substep{}{2.1} Find $\ell_j\in \bF$ such that
	$R =0$ by solving a linear system in $\ell_0, \dots, \ell_i$ 
	\substep{}{2.1} over~$\bF$. If there is a nontrivial solution, return
	\[\left(\sum_{j=0}^i \ell_j S_n^j,\ \sum_{j=0}^i\ell_j g_j H\right).\] 
	\\
\end{algo}

\begin{theorem}\label{THM:redct}\index{creative telescoping}\index{telescoper}
  Let~$T$ be a hypergeometric term over~$\bF(k)$. If~$T$ has a
  telescoper, then Algorithm~\ref{ALG:reductionct} terminates and
  returns a telescoper of minimal order for~$T$ w.r.t.~$k$.
\end{theorem}
\begin{proof}
  By Theorem~\ref{THM:mapred}, $r_0=0$ in step~$2$ implies that $T$ is
  summable, and thus~$1$ is a minimal telescoper for~$T$ w.r.t.~$k$.
  Now let~$r_0$ obtained from step~$2$ be of the form $r_0= a_0/b_0
  +q_0/v$,
  where $a_0, b_0, v \in \bF[k]$, $\deg_k(a_0) < \deg_k(b_0)$, $b_0$
  is strongly coprime with~$K$, $q_0 \in \bW_K$, and~$v$ is the
  denominator of~$K$. According to~\cite[Theorem~8]{AbPe2001b}, $K$ is
  integer-linear and so is~$v$. It follows that~$b_0$ is
  integer-linear if and only if~$b_0v$ is.  By
  Theorem~\ref{THM:existence},~$T$ has a telescoper if and only if the
  denominator of~$r_0$ is integer-linear.  Thus, steps~$2$ and~$3$ are
  correct.
	
  It follows from~\eqref{EQ:initred} and
  $\sigma_n(r_0H)=\sigma_n(r_0)NH$ that~\eqref{EQ:midred} holds for~$i=1$.  
  By Algorithm~\ref{ALG:transferdrf}, there exists a rational function~$u_1\in \bF(k)$
  and a residual form~$r_1$ w.r.t.~$K$ such that 
  \[\tilde{r}_1 = K\sigma_k(u_1) -u_1 + r_1,\quad \text{i.e.,}
  \quad r_1 \equiv_k {\tilde r}_1 \mod \bV_K,\]
  and $R+\ell_1 r_1$ is again a residual form 
  for all~$\ell_0, \ell_1  \in \bF$. Setting~$g_1=\tilde{g}_1+u_1$, we see
  that~\eqref{EQ:ithred} holds for~$i=1$.  By a direct induction
  on~$i$,~\eqref{EQ:ithred} holds in the loop of step~$4$.

  Assume that~$L = \sum_{i=0}^\rho c_i S_n^i$ is a minimal telescoper
  for~$T$ with $\rho \in \bN$, $c_i \in \bF$ and~$c_\rho \neq 0$.
  Then~$L(T)$ is summable w.r.t.~$k$. By
  Theorem~\ref{THM:mapred},~$\sum_{i=0}^\rho c_i r_i$ is equal to
  zero.  Thus, the linear homogeneous system (over~$\bF$) obtained by
  equating $\sum_{i=0}^\rho \ell_i r_i$ to zero has a nontrivial
  solution, which yields a minimal telescoper.
\end{proof}
\begin{remark}\label{REM:redCT}\index{telescoper}\index{certificate}
  Algorithm~\ref{ALG:reductionct} indeed separates the computation of minimal
  telescopers from that of certificates. In applications where
  certificates are irrelevant, we can drop step~$4.2$, and in
  step~$4.3$ we compute $g_i$ and $r_i$ with
  \[r_i\equiv_k\tilde r_i\mod\bV_K,\quad
  \sigma_n^i(r_{i-1})NH=\Delta_k(g_iH)+r_iH\] 
  and $R+\ell_i r_i$ is a residual form w.r.t.~$K$, 
  where $\ell_i$ is an indeterminate.
  Moreover, the rational function $g_i$ can be discarded, and we do
  not need to calculate $\sum_{j=0}^i\ell_j g_jH$ in the end.
\end{remark}
\begin{remark}\label{REM:simple}
  Instead of applying the modified reduction to $\sigma_n(r_{i-1})NH$
  in step~$4.1$, it is also possible to apply the reduction
  to~$\sigma_n^i(T)$ directly, but our experiments suggest that this
  variant takes considerably more time. This observation agrees with
  the advices given in~\cite[Example~6]{AbLe2005}.
\end{remark}
\index{creative telescoping} %
Since Algorithm~\ref{ALG:reductionct} performs the same function as
Zeilberger's algorithm, it is also applicable to the examples and
applications indicated in~\cite{PWZ1996}.  In other words, it can be
used to evaluate definite sums and prove identities of hypergeometric
terms efficiently.

\begin{example}\label{EX:reductionct}
  Consider the hypergeometric term $T = {\displaystyle
    \binom{n}{k}^3}.$
  Then the respective shift-quotients of~$T$ with respect to $n$ and
  $k$ are
  \[f = \frac{\sigma_n(T)}{T}= \frac{(n+1)^3}{(n+1-k)^3}\quad
  \text{and}\quad g = \frac{\sigma_k(T)}{T} =
  \frac{(n-k)^3}{(k+1)^3}.\] Since~$g$ is shift-reduced w.r.t.~$k$,
  its kernel is equal to~$g$ itself, and the corresponding shell
  is~$1$, implying that $H = T$ in step~$1$ of
  Algorithm~\ref{ALG:reductionct}.  In step~$4$,
  applying the modified \ap reduction to~$T, \sigma_n(T),
  \sigma_n^2(T)$, incrementally, yields
  \[ \sigma_k^{i}(T) = \Delta_k(g_iH) + \frac{q_i}{v} H, \]
  where~$i=0,1,2$, $v=(k+1)^3$,
  \begin{align*}
    q_0 & = \frac12(n+1)(n^2-n+3k(k-n+1)+1), \quad
    q_1 = (n+1)^3,   \\[1ex]
    q_2 &= \frac{(n+1)^3}{(n+2)^2} \left( 11n^2-12nk+17n+20+12k+12k^2
    \right),
  \end{align*}
  and $g_0,g_1,g_2 \in \bF(k)$ which are too complicated to be
  reproduced here.  By finding an~$\bF$-linear dependency among~$q_0,
  q_1, q_2$, we see that
  \[L = (n+2)^2 S_n^2 -(7n^2+21n+16)S_n -8(n+1)^2\] is a minimal
  telescoper for~$T$ w.r.t.~$k$.  For a corresponding certificate~$G$, one can choose to leave it as an unnormalized term
  \[G = (n+2)^2 g_2-(7n^2+21n+16)g_1-8(n+1)^2 g_0,\]
  or normalize it as one rational function according to the specific
  requirements.
\end{example}
\index{creative telescoping|)}
\section{Implementation and timings}\label{SEC:cttiming}
We have implemented Algorithm~\ref{ALG:reductionct} in 
{\sc Maple~18}\index{Maple@{\sc Maple}}. 
The procedure is named as \textsf{\rct} in the Maple package 
\textbf{ShiftReductionCT}.\index{ShiftReductionCT@\textbf{ShiftReductionCT}} 
See Appendix~\ref{APP:guide} for more details.

In this section, we compare the runtime\index{runtime|see {timing}}\index{timing} of
the new procedure to the performance of Zeilberger's algorithm. All
timings are measured in seconds on a Linux computer%
\index{Linux computer} with 388Gb RAM and twelve 2.80GHz Dual core processors.
No parallelism was used in this experiment. In addition, we also
compare the memory requirements\index{memory requirement} of all
procedures, which is shown in Appendix~\ref{APP:memory}.  For brevity,
we denote
\begin{itemize}
\item \textsf{Z}: the procedure
  \textbf{SumTools[Hypergeometric]}[\textsf{Zeilberger}], which is
  based on Zeilberger's algorithm;%
  \index{SumTools[Hypergeometric]@\textbf{SumTools[Hypergeometric]}! Zeilberger@\textsf{Zeilberger}}\index{Zeilberger's algorithm}
  \index{Zeilberger@\textsf{Zeilberger}~{\em (SumTools[Hypergeometric])}}

\smallskip
\item \textsf{RCT$_{tc}$}: the procedure~\textsf{ReductionCT} in
  \textbf{ShiftReductionCT}, which computes a minimal telescoper and
  a corresponding normalized certificate;

\smallskip
\item \textsf{RCT$_{t}$}: the procedure~\textsf{ReductionCT} in
  \textbf{ShiftReductionCT}, which computes a minimal telescoper
  without constructing a certificate.%
  \index{ShiftReductionCT@\textbf{ShiftReductionCT}! ReductionCT@\textsf{ReductionCT}}%
  \index{ReductionCT@\textsf{ReductionCT}~{\em (ShiftReductionCT)}}

\smallskip
\item \textsf{order}: the order of the resulting minimal telescoper.
  \index{order}
\end{itemize}

\begin{example}\label{EX:test}
  Consider bivariate hypergeometric terms of the form
  \[
  T=\frac{f(n,k)}{g_1(n+k)g_2(2n+k)}\frac{\Gamma(2\alpha
    n+k)}{\Gamma(n+\alpha k)}
  \]
  where $f\in\bZ[n,k]$ of degree~$d_2$, and for $i=1,2$,
  $g_i=p_i\sigma_z^\lambda(p_i)\sigma_z^\mu(p_i)$ with $p_i\in\bZ[z]$
  of degree~$d_1$ and $\alpha,\lambda,\mu\in\bN$.  For different
  choices of $d_1,d_2,\alpha,\mu,\lambda$, Table~\ref{TAB:cttest}
  compares the timings of the four procedures.
  \begin{table}[!ht]
    \tabcolsep12pt
    \begin{center}
      \begin{tabular}{l|r|rr|c}
        $(d_1, d_2, \alpha, \lambda, \mu)$ & {\sf{Z}} & {\sf{RCT$_{tc}$}} & {\sf{RCT$_{t}$}} & {\sf{order}}\\
        \hline
        $(1, 0, 1, 5, 5)$            &     17.12 &     5.00 &     1.80 & 4     \\[.5ex]
        $(1, 0, 2, 5, 5)$            &     74.91 &    26.18 &     5.87 & 6     \\[.5ex]
        $(1, 0, 3, 5, 5)$            &    445.41 &    92.74 &    17.34 &  7     \\[.5ex]
        $(1, 8, 3, 5, 5)$            &    649.57 &   120.88 &    23.59 &  7     \\[.5ex]
        $(2, 0, 1, 5, 10)$           &    354.46 &    58.01 &     4.93 & 4     \\[.5ex]
        $(2, 0, 2, 5, 10)$           &    576.31 &   363.25 &    53.15 & 6     \\[.5ex]
        $(2, 0, 3, 5, 10)$           &   2989.18 &  1076.50 &   197.75 &  7     \\[.5ex]
        $(2, 3, 3, 5, 10)$           &   3074.08 &  1119.26 &   223.41 &  7     \\[.5ex]
        $(2, 0, 1, 10, 15)$          &   2148.10 &   245.07 &    11.22 & 4     \\[.5ex]
        $(2, 0, 2, 10, 15)$          &   2036.96 &  1153.38 &   153.21 & 6     \\[.5ex]
        $(2, 0, 3, 10, 15)$          &  11240.90 &  3932.26 &   881.12 & 7     \\[.5ex]
        $(2, 5, 3, 10, 15)$          &  10163.30 &  3954.47 &   990.60 & 7     \\[.5ex]
        $(3, 0, 1, 5, 10)$           &  18946.80 &   407.06 &    43.01 & 6     \\[.5ex]
        $(3, 0, 2, 5, 10)$           &  46681.30 &  2040.21 &   465.88 & 8     \\[.5ex]
        $(3, 0, 3, 5, 10)$           & 172939.00 &  5970.10 &  1949.71 & 9     \\ \hline
      \end{tabular}
    \end{center}
    \vspace{-\smallskipamount}
    \caption{Timing comparison\index{comparison} of Zeilberger's algorithm 
    to reduction-based\index{reduction-based} creative telescoping with 
    and without construction of a certificate (in seconds)}\label{TAB:cttest}
\end{table}
\end{example}	
\begin{remark}\label{REM:unnormalized}\index{certificate}
  The difference between \textsf{RCT$_{tc}$} and \textsf{RCT$_{t}$}
  mainly comes from the time needed to bring the rational function $g$
  in the certificate $gH$ on a common denominator. When it is
  acceptable to keep the certificate as an unnormalized linear
  combination of rational functions, their timings are virtually the
  same.
\end{remark}
\chapter[Order Bounds for Minimal Telescopers]
{Order Bounds for \\Minimal Telescopers
\protect\footnotemark{}\protect
\footnotetext{The main results in this chapter 
are published in~\cite{Huan2016}.}}
\label{CH:bounds}

In the previous chapter, we have presented a reduction-based creative
telescoping\index{creative telescoping} algorithm 
for bivariate hypergeometric terms, namely
Algorithm~\ref{ALG:reductionct}. Roughly speaking, its basic idea is
as follows.  Using the modified \ap reduction from
Chapter~\ref{CH:apreduction}, we first reduce a given hypergeometric
term and its shifts to some required \lq\lq standard forms\rq\rq\
(called {\em remainders}\index{remainder} in the sequel), such that
the difference between the original function and its remainder is
summable. Then computing a telescoper amounts to finding a linear
dependence among these remainders.  In order to show that this
algorithm terminates, we show that for every summable term, its
remainder is zero. This ensures that the algorithm terminates by the
existence criterion given in Theorem~\ref{THM:existence}, and in fact
it will find the smallest possible telescoper, but it does not provide
a bound on its order. Another possible approach is to show that the
vector space spanned by the remainders has a finite dimension. Then,
as soon as the number of remainders exceeds this dimension, we can be
sure that a telescoper will be found.  This approach was taken
in~\cite{BCCLX2013,BLS2013,CKK2016}.  As a nice side result, this
approach provides an independent proof of the existence of
telescopers, and even a bound on the order of minimal telescopers.

In this chapter, we show that the approach for the differential case
also works for the shift case, i.e., the remainders in the shift case
also form a finite-dimensional vector space, so as to eliminate the
discrepancy. As a result, we obtain a new argument for the termination
of Algorithm~\ref{ALG:reductionct}, and also get new bounds for the
order of minimal telescopers for hypergeometric terms.  We do not find
exactly the same bounds that are already in the
literature~\cite{MoZe2005,AbLe2005}.  Comparing our bounds to the
known bounds in the literature, it appears that for \lq\lq
generic\rq\rq\ input (see Subsection~\ref{SUBSEC:azub} for a
definition), the values often agree (of course, because the known
bounds are already generically sharp).  However, there are some
special examples in which our bounds are better than the known
bounds. On the other hand, our bounds are never worse than the known
ones. In addition, we give a variant of Algorithm~\ref{ALG:reductionct} based on the new bounds. 
An experimental comparison is presented in the final section.

%
\section{Shift-homogeneous decompositions}\label{SEC:shifthom}
In this section, we generalize the notion of shift-equivalence in
Chapter~\ref{CH:rfproperties} to the bivariate case, and then derive a
useful decomposition for an integer-linear polynomial.

Using the same notations as the previous chapter, $\set K$ is a field
of characteristic zero, and $\set K(n,k)$ is the field of rational
functions in~$n$ and $k$ over~$\set K$. Let~$\sigma_n$ and~$\sigma_k$
be the shift operators w.r.t.~$n$ and $k$, respectively.
\begin{definition}\label{DEF:shiftequiv}
  Two polynomials $ p,q \in\set K[n,k]$ are called {\em
    shift-equivalent}\index{shift-! equivalent} w.r.t.~$n$ and~$k$ if
  there exist integers $\ell, m$ such that
  $q=\sigma_n^\ell\sigma_k^m(p)$. We denote it by $p \sim_{n,k} q$.
\end{definition}
Clearly $\sim_{n,k}$ is an equivalence relation. 
In particular, when $\ell = 0$ or $m=0$, the above definition
degenerates to Definition~\ref{DEF:shift-equiv}.  Thus $\sim_n$ or
$\sim_k$ implies $\sim_{n,k}$.  
Choosing the pure lexicographic order~$n\prec k $, we say a
polynomial is {\em monic} if its highest term has coefficient~$1$. A
rational function is said to be {\em shift-homogeneous}%
\index{shift-! homogeneous} if all non-constant monic irreducible factors of its
denominator and numerator belong to the same shift-equivalence class.

By grouping together the factors in the same shift-equivalence class,
every rational function $r\in \set K(n,k)$ can be decomposed into the
form
\begin{equation}\label{EQ:shifthomdecomp}
  r = c\, r_1 \dots r_s,
\end{equation}
where $c \in \set K$, $s \in \bN$, each $r_i$ is a shift-homogeneous
rational function, and any two non-constant monic irreducible factors
of $r_i$ and $r_j$ are pairwise shift-inequivalent whenever~$i \neq
j$. We call each $r_i$ a 
{\em shift-homogeneous component}\index{shift-! homogeneous component} of $r$ 
and \eqref{EQ:shifthomdecomp} a {\em shift-homogeneous
  decomposition}\index{decomposition! shift-homogeneous}%
\index{shift-! homogeneous decomposition} of $r$.  Noticing that the field $\set
K(n,k)$ is a unique factorization domain, one can easily show that the
shift-homogeneous decomposition is unique up to the order of the
factors and multiplication by nonzero constants.

Let $p\in \set K[n,k]$ be an irreducible
integer-linear\index{integer-linear} polynomial. Then it is of the
form $p=P(\lambda n + \mu k)$ for some $P \in \set K[z]$ and $\lambda, \mu\in
\bZ$, not both zero. W.l.o.g., we further assume that~$\mu \geq 0$
and~$\gcd(\lambda, \mu) =1$. Under this assumption,
making ansatz and comparing coefficients 
yield the uniqueness of~$P$
since $\bZ$ is a unique factorization domain. 
In view of this, we call the pair $(P, \{\lambda, \mu\})$ the
{\em univariate representation}\index{univariate representation} of 
the integer-linear polynomial~$p$.
By B{\' e}zout's relation, there exist
unique integers $\alpha, \beta$ with $|\alpha| <|\mu|$ and~$|\beta|
<|\lambda|$ such that $\alpha \lambda + \beta \mu =
1$. Define $\delta^{(\lambda, \mu)}$ to be $\sigma_n^\alpha
\sigma_k^\beta$. For brevity, we just write $\delta$ if $(\lambda,
\mu)$ is clear from the context. Note that $\delta(P(z))=P(z+1)$
with~$z=\lambda n + \mu k$, which allows us to treat integer-linear
polynomials as univariate ones.  
For a Laurent polynomial~$\xi=\sum_{i=\ell}^\rho m_i \delta^i$
in~$\bZ[\delta,\delta^{-1}]$ with~$\ell, \rho, m_i\in \bZ$ and $\ell
\leq\rho$, define
\[p^\xi = \delta^{\ell}(p^{m_{\ell}}) \delta^{\ell+1}(p^{m_{\ell+1}}) 
\cdots \delta^\rho(p^{m_\rho}).\]

Let $p, q\in \set K[n,k]$ be two irreducible integer-linear polynomials
of the forms
\[p=P(\lambda_1 n+\mu_1 k)\quad \text{and} \quad q=Q(\lambda_2 n+
\mu_2 k),\] 
where $(P,\{\lambda_1,\mu_1\})$ and $(Q,\{\lambda_2, \mu_2\})$ are the
univariate representations of $p$ and $q$, respectively.  Namely, $P,
Q\in \bK[z]$, $\lambda_1, \mu_1, \lambda_2, \mu_2 \in \bZ$, $\mu_1,
\mu_2\geq 0$ and $\gcd(\lambda_1, \mu_1) = \gcd(\lambda_2, \mu_2) =1$.
It is readily seen that $p \sim_{n,k} q$ if and only if $\lambda_1 =
\lambda_2$, $\mu_1=\mu_2$ and $q = p^{\delta^\ell}$ for some
integer~$\ell$, in which~$\delta = \delta^{(\lambda_1, \mu_1)}
=\delta^{(\lambda_2, \mu_2)}$.
 
 Given a shift-homogeneous and integer-linear rational function $r \in  \bK(n,k)$, let~$h$
 be a monic, irreducible and integer-linear polynomial in $\bK[n, k]$ with the property that
 all monic irreducible factors of the numerator and denominator of~$r$ are equal to
 some shift of~$h$ w.r.t.~$n$ and~$k$. Assume that the univariate representation
 of~$h$ is the pair~$(P_h, \{\lambda_h, \mu_h\})$. Then~$r$ can be written as~$c \, h^{\xi_h}$ for 
 some~$c \in \bK$  and~$\xi _h\in \bZ[ \delta^{-1}, \delta]$ with $\delta= \delta^{(\lambda_h,\mu_h)}$. We call~$(P_h, \{\lambda_h, \mu_h\}, \xi_h)$
 a {\em univariate representation} of~$r$. 
 Assume that~$(P_g, \{\lambda_g, \mu_g\}, \xi_g)$ is another univariate representation of~$r$ with~$g \in \bK[n, k]$.
 By the conclusion made in the preceding paragraph, we find that~$g = h^{\delta^\ell}$ for some~$\ell \in \bZ$, or, equivalently,
 $P_g(z)=P_h(z+\ell)$. Moreover,~$(\lambda_g, \mu_g)=(\lambda_h,\mu_h)$. It follows that~$\xi_g =  {\delta^\ell} \xi_h$. 
 In particular,~$\deg_z(P_h)$ is equal to~$\deg_z(P_g)$ and the nonzero coefficients of~$\xi_h$ are exactly the same as those of~$\xi_g$.  When the choice of~$h$ and~$g$ is insignificant, we say that a tuple~$(P, \{\lambda, \mu\}, \xi)$ is a {\em univariate representation} of~$r$ if the polynomial $P \in \bK[z]$ is irreducible and~$r(n,k)=cP(\lambda n + \mu k)^\xi$ for some $c\in \bK$. Note that the coefficients of~$\xi$
 are all nonnegative if~$r$ is a polynomial.
 
 Let~$r \in \bK(n, k)$ be integer-linear with the shift-homogeneous decomposition
 \[    r = c \, r_1 \cdots r_s. \]
 For $i=1, \ldots, s$, assume that~$U_i=(P_i, (\lambda_i, \mu_i), \xi_i)$ is a univariate representation of~$r_i$. Then we call the tuple
 $$\left(c, \, (U_1, \ldots, U_s)\right)$$
 a {\em univariate representation}\index{univariate representation} of~$r$. 

To avoid unnecessary duplication, we make a notational convention.
\begin{convention}\label{CON:convention}
  Let~$T$ be a hypergeometric term over $\set K(n,k)$ with a
  multiplicative decomposition $SH$, where $S\in \set K(n,k)$ and $H$ is
  a hypergeometric term whose shift-quotient~$K$ w.r.t.~$k$ is
  shift-reduced w.r.t.~$k$. 
  By~\cite[Theorem~8]{AbPe2001b}, we know~$K$ is integer-linear
  over~$\set K$. Write~$K = u/v$ where~$u, v\in \set K(n)[k]$
  and~$\gcd(u,v)=1$.
\end{convention}

\section{Shift-relation of residual forms}\label{SEC:shiftrelation}
In this section, we describe a relation among residual forms
\index{residual form} of a given hypergeometric term and its
shifts. This relation enables us to derive a shift-free common multiple of
significant denominators\index{significant denominator} of those
residual forms, provided that telescopers exist. The existence of this
common multiple implies that the residual forms span a
finite-dimensional vector space over~$\bK(n)$, and then lead to order
bounds for the minimal telescopers presented in the next section.

\begin{lemma}\label{LEM:xrfs}
  With Convention~\ref{CON:convention}, let~$r$ be a residual form
  of~$S$ w.r.t.~$K$. Then~$\sigma_n(K)$ and $\sigma_n(S)$ are a kernel
  and a corresponding shell of~$\sigma_n(T)$ w.r.t.~$k$,
  respectively. Moreover, $\sigma_n(r)$ is a residual form
  of~$\sigma_n(S)$ w.r.t.~$\sigma_n(K)$.
\end{lemma}
\begin{proof}
  By Convention~\ref{CON:convention}, $\sigma_n(T) =
  \sigma_n(S)\sigma_n(H)$ and $\sigma_n(K)$ is the shift-quotient
  of~$\sigma_n(H)$ w.r.t.~$k$. To prove the first assertion, one needs
  to show that~$\sigma_n(K)$ is shift-reduced w.r.t.~$k$. This can be
  proven by observing that, for any two polynomials $p_1, p_2 \in
  \set K(n)[k]$, we have $\gcd(\sigma_n(p_1),\sigma_n(p_2)) = 1$ if and only if
  $\gcd(p_1,p_2) = 1$.
	
  For the second assertion, since $r$ is a residual form w.r.t.~$K$,
  we write
  \[r = \frac{a}{b}+\frac{q}{v},\]
  where~$a,b,q\in\set K(n)[k]$, $\deg_k(a)<\deg_k(b)$, 
  $\gcd(a,b)=1$, $b$ is shift-free and strongly coprime with~$K$, 
  and~$q\in \bW_K$.  It is clear that
  $\deg_k(\sigma_n(a))<\deg_k(\sigma_n(b))$ and
  $\gcd(\sigma_n(a),\sigma_n(b)) = 1$. By the above observation,
  $\sigma_n(b)$ is shift-free and strongly coprime with~$\sigma_n(K)$.
	
  Note that~$\sigma_n\circ \deg_k = \deg_k \circ \,\sigma_n$ and
  $\sigma_n\circ \lc_k = \lc_k\circ \,\sigma_n$, where~$\lc_k(p)$ is
  the leading coefficient of $p\in \set K(n)[k]$ w.r.t.~$k$. So the
  standard complements~$\bW_K$ and~$\bW_{\sigma_n(K)}$ for polynomial
  reduction have the same echelon basis according to the case study in
  Subsection~\ref{SUBSEC:drf}. It follows from~$q\in \bW_K$ that
  $\sigma_n(q) \in \bW_{\sigma_n(K)}$. Accordingly, $\sigma_n(r)$ is a
  residual form of~$\sigma_n(S)$ w.r.t.~$\sigma_n(K)$.
\end{proof}
\begin{theorem}\label{THM:brelation}
  With Convention \ref{CON:convention}, for every nonnegative integer
  $i$ assume
  \begin{equation}\label{EQ:ithmapred}
    \sigma_n^i(T) = \Delta_k(g_i H) +  \left(\frac{a_i}{b_i} + \frac{q_i}{v}\right)H,
  \end{equation}
  where $g_i\in \set K(n,k)$, $a_i, b_i \in \set K(n)[k]$
  with~$\deg_k(a_i)<\deg_k(b_i)$, $\gcd(a_i,b_i)=1$, $b_i$ is
  shift-free w.r.t.~$k$ and strongly coprime with~$K$, and~$q_i$ belongs
  to $\bW_K$. Then $b_i$ is shift-related to $\sigma_n^i(b_0)$, i.e.,
  $b_i\approx_k \sigma_n^i(b_0)$.
\end{theorem}
\begin{proof}
  We proceed by induction on $i$. For $i = 0$, the reflexivity of the
  relation $\approx_{k}$ implies that $b_0\approx_k b_0$.  

  Assume that $b_{i-1}\approx_k \sigma_n^{i-1}(b_0)$ for $i\geq
  1$. Note that $K$ is also a kernel of $\sigma_n^{i-1}(T)$ and
  $\sigma_n^i(T)$ w.r.t.~$k$. Let $S_{i-1}$ and $S_{i}$ be the
  corresponding shells, respectively. Consider the equality
  \[\sigma_n^{i-1}(T) = \Delta_k(g_{i-1} H) +
  \left(\frac{a_{i-1}}{b_{i-1}} + \frac{q_{i-1}}{v}\right)H,\] 
  where $g_{i-1}\in \set K(n,k)$ and $a_{i-1}/b_{i-1}+q_{i-1}/v$ is a residual
  form of $S_{i-1}$ w.r.t.~$K$. Applying~$\sigma_n$ to both sides yields
  \begin{align*}
    \sigma_n^i(T) &= \sigma_n(\Delta_k(g_{i-1} H))+
    \sigma_n\left(\frac{a_{i-1}}{b_{i-1}} + \frac{q_{i-1}}{v}\right)\sigma_n(H)\\[1ex]
    &=\Delta_k(\sigma_n(g_{i-1}
    H))+\left(\frac{\sigma_n(a_{i-1})}{\sigma_n(b_{i-1})} +
      \frac{\sigma_n(q_{i-1})}{\sigma_n(v)} \right)\sigma_n(H)
  \end{align*}
  It follows from Lemma~\ref{LEM:xrfs} that $\sigma_n(K)$ and $
  \sigma_n(S_{i-1})$ are a kernel and the corresponding
  shell~$\sigma_n^i(T)$ w.r.t.~$k$, and
  $\sigma_n(a_{i-1})/\sigma_n(b_{i-1}) +
  \sigma_n(q_{i-1})/\sigma_n(v)$ is a residual form of~$S_i$
  w.r.t.~$\sigma_n(K)$.  By~\eqref{EQ:ithmapred} with~$i = 1$, we know that
  $a_i/b_i + q_i/v$ is a residual form of~$S_i $ w.r.t.~$K$. By
  Theorem~\ref{THM:sf}, $b_i\approx_k \sigma_n(b_{i-1})$. Thus
  $b_i\approx_k \sigma_n^i(b_0)$ by the induction hypothesis.
\end{proof}
Using the relation revealed in the above theorem, we can derive a
common multiple as promised at the beginning of this section. 
To this end, we need two simple lemmas.

The first lemma says that, with Convention~\ref{CON:convention}, for
any~$f \in \set K(n)[k]$, there always exists~$g \in \set K(n)[k]$ such that
$f\approx_k g$ and $g$ is strongly coprime with~$K$.
\begin{lemma}\label{LEM:stronglyprime}
  With Convention~\ref{CON:convention}, assume that~$p$ is an
  irreducible polynomial in~$\set K(n)[k]$. Then there exists an
  integer~$m$ such that $\sigma_k^m(p)$ is strongly coprime with~$K$.
\end{lemma}
\begin{proof}
  According to the definition of strong coprimeness, there is one and
  only one of the following three cases true.
	
  \smallskip\noindent {\em Case~1.} $p$ is strongly coprime
  with~$K$. Then the lemma follows by letting~$m = 0$.
	
  \smallskip\noindent {\em Case~2.} There exists an integer~$k\geq 0$
  such that $\sigma_k^k(p)\mid u$. Then for every integer~$\ell$, we
  have $\gcd(\sigma_k^\ell(p),v)=1$, since~$K$ is shift-reduced
  w.r.t.~$k$. Let
  \[m = \max\{i\in \bN\mid \sigma_k^i(p)\mid u\} +1.\] 
  One can see that~$\sigma_k^m(p)$ is strongly coprime with~$K$.
	
  \smallskip\noindent {\em Case~3.} There exists an integer~$k\leq 0$
  such that $\sigma_k^k(p)\mid v$. Then for every integer~$\ell$, we
  have $\gcd(\sigma_k^\ell(p),u)=1$, since~$K$ is shift-reduced
  w.r.t.~$k$. Letting
  \[m = \min\{i\in \bN\mid \sigma_k^i(p)\mid v\}-1\] 
  yields that~$\sigma_k^m(p)$ is strongly coprime with~$K$.
\end{proof}
The next lemma shows that for any integer-linear polynomial in
$\set K[n,k]$, the number of shift-equivalence classes w.r.t.~$k$
produced by shifting the polynomial as a univariate one is finite.
\begin{lemma}\label{LEM:shiftequi}
  Let~$q\in \set K[n,k]$ be integer-linear, and then $q = P(\lambda n +
  \mu k)$ for $P \in \set K[z]$ and $\lambda, \mu \in \bZ$ not both
  zero. Then any shift of $q$ w.r.t.~$n$ or $z = \lambda n + \mu k$ is
  shift-equivalent to $\delta^j(q)$ w.r.t.~$k$ for $\delta =
  \delta^{(\lambda, \mu)}$ and $0\leq j\leq \mu -1$. More precisely,
  let
  \[S = \{\delta^j(q) \mid j = 0, \dots, \mu-1\},\ S_1 =
  \{\sigma_n^i(q)\mid i \in \bN\} \ \text{and} \ S_2 = \{\delta^j(q)
  \mid j\in \bN\}.\] 
  Then for any element $f$ in $S_1\cup S_2$, there
  exists~$g\in S$ such that $f\sim_k g$.
\end{lemma}
\begin{proof}
  Assume that $f\in S_1\cup S_2$. Since $\sigma_n = \delta^{\lambda}$,
  there exists a nonnegative integer~$i$ such that
  \[f = \delta^i(q)= P(\lambda n + \mu k + i).\] 
  By Euclidean division, there exist unique integers~$j,\ell$ with~$0\leq j\leq
  \mu-1$, such that~$i = \ell \mu + j$. It follows that
  \[f = \sigma_k^\ell(P(\lambda n + \mu k + j))
  =\sigma_k^\ell(\delta^j(q)).\] 
  Letting~$g= \delta^j(q)$ completes the proof.
\end{proof}
Now we are ready to compute a common multiple as mentioned before.
\begin{theorem}\label{THM:commonden}\index{integer-linear}
  With Convention~\ref{CON:convention}, assume that
  \begin{equation}\label{EQ:initialred}
    T = \Delta_k(gH) + \left(\frac{a}{b} + \frac{q}{v} \right) H,
  \end{equation}
  where~$g\in \set K(n,k)$, $a,b, q\in \set K(n)[k]$, $\deg_k(a) < \deg_k(b)$,
  $\gcd(a,b)=1$, $b$ is shift-free w.r.t.~$k$ and strongly coprime
  with~$K$, and~$q\in \bW_K$. 
  Further assume that~$b$ is integer-linear and has a univariate representation
  $$(c, (U_1, \ldots, U_s)), \quad \text{where} \ U_j=(P_j, (\lambda_j, \mu_j), \xi_j),\ j=1, \ldots, s.$$
  Then there exists~$B \in \mathbb{K}(n)[k]$ such that $b \mid B$ and for all~$i \in \bN$,
  \begin{equation}\label{EQ:ithdecom}
      \sigma_n^i(T) = \Delta_k(g_iH) + \left(\frac{a_i}{B} + \frac{q_i}{v}\right)H
   \end{equation}
  for some~$g_i \in \mathbb{K}(n,k), a_i \in \mathbb{K}(n)[k]$ with~$\deg_k(a_i) < \deg_k(B)$, and~$q_i \in \mathbb{W}_K$.
  Moreover, 
  \begin{itemize}
  	\item[(i)] $B$ is shift-free w.r.t.~$k$ and strongly coprime with~$K$;
  	\item[(ii)] $\deg_k(B) =  \sum_{j=1}^s \mu_j m_j \deg_k(P_j)$, where~$m_j$ is the maximum of the coefficients of~$\xi_j$.
  \end{itemize}
%
\end{theorem}
\begin{proof}
  Since the shift-homogeneous components of~$b$ are coprime to each other, 
  it suffices to consider the case when $b$ is shift-homogeneous. 
  W.l.o.g., assume that $b$ is shift-homogeneous and has
  a univariate representation $(P, \{\lambda, \mu\}, \xi)$
  such that
   \[b = P(\lambda n + \mu k)^\xi.\] 
   Write $\xi=\sum_{i=0}^d m_i' \delta^i$ where $d \in \bN$, $m_i^\prime\in \bZ$ and 
   $\delta = \delta^{(\lambda, \mu)}$.

  If $\mu=0$ then $b\in \set K(n)$. By the modified \ap reduction we can assume that
  \eqref{EQ:ithmapred} holds for every~$i>0$ and thus~$b_i\in \set K(n)$ by
  Theorem~\ref{THM:brelation}. The assertion follows by
  letting~$B=1$. 
  
  Otherwise we have $\mu>0$.
  By Lemma~\ref{LEM:shiftequi}, for every~$i\in \bN$ there are unique
  integers~$j$ and $\ell_j$ with $0\leq j \leq \mu-1$ such that
  \[P(\lambda n + \mu k)^{\delta^i}= P(\lambda n + \mu k+
  j)^{\sigma_k^{\ell_j}},\] 
  which is equivalent to
  \[P(\lambda n + \mu k+i)= P(\lambda n + \mu k + \mu \ell_j+ j).\]
  Since $P$ is irreducible, we have $i = \mu \ell_j + j$. Let~$m_j''= m'_{\mu
    \ell_j + j}$. Since~$b$ is shift-free w.r.t.~$k$,
  \[b = \prod_{j=0}^{\mu-1}P(\lambda n + \mu
  k+j)^{m_j''\sigma_k^{\ell_j}}.\] 
  For each~$j$, if~$m_j'' \neq 0$ then set $m_j = \ell_j$; 
  otherwise by Lemma~\ref{LEM:stronglyprime},
  let $m_j$ be an integer so that $P(\lambda n+ \mu k+j)^{\sigma_k^{m_j}}$ is
  strongly coprime with~$K$. Let $m = \max_{0\leq j \leq \mu-1}\{m_j''\}$ and
  \begin{equation}\label{EQ:commonden}
    B = \prod_{j=0}^{\mu-1}P(\lambda n + \mu k+j)^{m\,\sigma_k^{m_j}}.
  \end{equation}
  Then $\deg_k(B)= \mu m \deg_k(P)$.
  Since~$m_j = \ell_j$ when~$m_j''\neq 0$, every irreducible factor
  of~$b$ divides~$B$ and thus~$b\mid B$ by the maximum of~$m$. Because
  $0\leq j\leq \mu-1$, so $B$ is shift-free w.r.t.~$k$. Moreover, $B$
  is strongly coprime with~$K$ by the choice of~$m_j$.
	
  It remains to show that~\eqref{EQ:ithdecom} holds for every
  nonnegative integer~$i$. To prove this, we first show
  $\sigma_n(B)\approx_k B$. By~\eqref{EQ:commonden}, we have
  \[B \approx_k \prod_{j=0}^{\mu-1}P(\lambda n + \mu k+j)^m,\] 
  which yields
  \begin{equation*}
    \sigma_n(B) \approx_k\prod_{j=0}^{\mu-1}P(\lambda n + \mu k+j+\lambda)^m.
  \end{equation*}
  By Lemma~\ref{LEM:shiftequi}, there exists a unique integer $\ell$
  with $0\leq \ell \leq \mu-1$ such that
  \[P(\lambda n + \mu k+j+\lambda) \sim_k P(\lambda n + \mu k+\ell).\]
  Conversely, for any~$0\leq \ell\leq \mu-1$, there exists a unique
  integer~$0\leq j\leq \mu-1$ such that the above equivalence
  holds. Thus
  \[\sigma_n(B)\approx_k \prod_{\ell=0}^{\mu-1}P(\lambda n + \mu
  k+\ell)^m\approx_k B.\]

  For~$i=0$, letting $g_0 = g$, $a_0 = aB/b$ and~$q_0 = q$
  gives~\eqref{EQ:ithdecom}.  Since~$\sigma_n(B)\approx_k B$, we have
  $\sigma_n^i(B)\approx_k \sigma_n^{i-1}(B)$ for every positive integer $i$, 
  and then~$\sigma_n^i(B)\approx_k B$.
	
  On the other hand, by the modified \ap reduction
  \eqref{EQ:ithmapred} holds for every~$i\geq 0$, in which~$b_0 =
  b$. According to Theorem~\ref{THM:brelation},
  $b_i\approx_k\sigma_n^i(b_0)$. It follows from~$b\mid B$ that
  $\sigma_n^i(b)\mid \sigma_n^i(B)$.  Consequently, we have
  \[b_i\approx_k \sigma_n^i(b)\mid \sigma_n^i(B)\approx_k B.\] 
  Thus there is $\tilde{b}_{i}\in \set K(n)[k]$ dividing~$B$ so
  that~$\tilde{b}_{i} \approx_k b_{i}$. Moreover, $\tilde{b}_i$ is
  strongly coprime with~$K$ as~$B$ is.  By the shifting property of
  significant denominators (i.e., Lemma~\ref{LEM:rfshift} and
  Remark~\ref{REM:shiftsignificantden}), there exist $\tilde{g_i}\in
  \set K(n,k)$, $\tilde{a}_{i}, \tilde{q}_{i}\in \set K(n)[k]$ with
  $\deg_k(\tilde{a}_{i}) < \deg_k(\tilde{b}_{i})$, and~$\tilde{q}_{i}
  \in \bW_K$ such that~$\sigma_n^i(T) =
  \Delta_k(\tilde{g_i}H)+(\tilde{a}_i/\tilde{b}_i+\tilde{q}_i/v)H$. The
  assertion follows by noticing
  \[\sigma_n^i(T) = \Delta_k(\tilde{g_i}H) + \left(\frac{\tilde{a}_i
      B/\tilde{b}_i}{B} + \frac{\tilde{q}_i}{v}\right)H.\]
\end{proof}
Under the assumptions of Theorem~\ref{THM:commonden}, applying
Algorithm~\ref{ALG:mapred} to~$T$ w.r.t.~$k$ yields
  $T = \Delta_k(g H) + r H,$
where~$g\in \set K(n,k)$ and $r$ is a residual form w.r.t.~$K$.  By
Theorems~\ref{THM:uniqueness} and~\ref{THM:sf}, $b$ and the significant
denominator $r_d$ of~$r$ are shift-related w.r.t.~$k$,
and thus so are the respective shift-homogeneous components.
W.l.o.g., assume that $b$ is shift-homogeneous (then so is $r_d$).
Let $(P_b, \{\lambda_b, \mu_b\}, \xi_b)$ be a univariate representation
of~$b$ and 
$(P_{r_d}, \{\lambda_{r_d}, \mu_{r_d}\},\xi_{r_d})$ be one of~$r_d$.
Definition~\ref{DEF:relatedness} yields that $(\lambda_b, \mu_b)=(\lambda_{r_d}, \mu_{r_d})$
and for each integer $i$, there exists a unique integer~$j$ and another integer $\ell_{ij}$ such that \[P_{b}(z)^{\delta^i}=\sigma_k^{\ell_{ij}}\left(P_{r_d}(z)^{\delta^j}\right)=P_{r_d}(z+\mu_{r_d}\ell_{ij})^{\delta^j} \quad \text{with}\ \delta= \delta^{(\lambda_b, \mu_b)}.\] 
Moreover, the nonzero coefficients of $\xi_b$  are exactly the same as those of $\xi_{r_d}$.
In summary, we have the following remark.
\begin{remark}\label{REM:commonden}
  Although the form of~$B$ in Theorem~\ref{THM:commonden} depends on
  the choice of~$b$, the shift-equivalence classes w.r.t.~$\sim_{n,k}$ as well as the
  degree of $B$ w.r.t.~$k$ depend only on the hypergeometric term~$T$.
\end{remark}

\section{Upper and lower order bounds}\label{SEC:ulbound}%
\index{bound|see {order bound}}\index{order bound|(}
In this section, we show that Theorem~\ref{THM:commonden} implies that 
some residual forms $\{a_i/b_i+q_i/v\}_{i\geq 0}$ satisfying~\eqref{EQ:ithmapred} 
form a finite-dimensional vector space over~$\set K(n)$, 
and then derive an upper bound for the order of minimal telescopers. 
\begin{theorem}\label{THM:upperbound}\index{order bound! upper}
  With the assumptions of Theorem~\ref{THM:commonden}, we have that
  the order of a minimal telescoper for~$T$ w.r.t.~$k$ is no more than
  \begin{align*}\label{EQ:upperbound}
    & \max\{\deg_k(u), \deg_k(v)\} - \llbracket \deg_k(v-u) \leq
    \deg_k(u) -1\rrbracket\nonumber+\sum_{j=1}^s{\mu_j m_j \deg_k(P_j)},
  \end{align*}
  where $\llbracket\varphi\rrbracket$ equals~$1$ if $\varphi$
  is true, otherwise it is~$0$.
\end{theorem}
\begin{proof}
  Let $L=\sum_{i=0}^{\rho}{e_i S_n^i}$ be a minimal telescoper for $T$
  w.r.t.~$k$, where~$\rho \in \bN$ and~$e_0, \ldots ,e_\rho \in
  \set K(n)$ not all zero. By Theorem~\ref{THM:commonden}, there exists
  $B\in \set K(n)[k]$ such that \eqref{EQ:ithdecom} holds for every
  nonnegative integer~$ i$. Then by Theorem~\ref{THM:redct}, the
  residual forms $\{a_i/B+q_i/v\}_{i=0}^\rho$ are linearly dependent
  over $\set K(n)$; equivalently, the following linear system with
  unknowns $e_0,\dots, e_\rho$
  \begin{equation}\label{EQ:system}
    \begin{cases}
      A_\rho = e_0 a_0 \ +\ e_1 a_1\ +\ \cdots\ +\ e_\rho a_\rho=0\\[1ex]
      Q_\rho = e_0q_0\ +\ e_1q_1\ +\ \cdots\ +\ e_\rho q_\rho=0
    \end{cases}
  \end{equation}
  has a nontrivial solution in~$\set K(n)^{\rho+1}$. Since $\deg_k(a_i)<
  \deg_k(B)$,
  \begin{equation}\label{EQ:Adeg}
    \deg_k(A_\rho) < \deg_k(B) = \sum_{j=1}^s{\mu_j m_j \deg_k(P_j)}.
  \end{equation}
  Note that $\bW_K$ is a vector space, so $Q_\rho\in \bW_K$. By
  Proposition~\ref{PROP:termbound}, the number of nonzero terms
  w.r.t.~$k$ in $Q_\rho$ is no more than the dimension
  $\dim_{\set K(n)}(\bW_K)$, which is bounded by
  \begin{equation}\label{EQ:Wdim}
    \max\{\deg_k(u), \deg_k(v)\} - \llbracket \deg_k(v-u) \leq \deg_k(u) -1\rrbracket.
  \end{equation}
  Comparing coefficients of like powers of $k$ of the linear system
  \eqref{EQ:system} yields at most
  \begin{equation}\label{EQ:eqno}
    \deg_k(A_\rho)+\dim_{\set K(n)}(\bW_K)+1
  \end{equation}
  equations. Hence this system has nontrivial solutions whenever the
  order $\rho$ exceeds $\deg_k(A_\rho)+\dim_{\set K(n)}(\bW_K)$. It
  implies that the order of a minimal telescoper for~$T$ w.r.t.~$k$ is
  no more than the number~\eqref{EQ:eqno}. Therefore, the theorem
  follows by \eqref{EQ:Adeg} and \eqref{EQ:Wdim}.
\end{proof}
In addition, we can further obtain a lower order bound for telescopers for~$T$.
\begin{theorem}\label{THM:lowerbound}\index{order bound! lower}
  With the assumptions of Theorem~\ref{THM:commonden}, further assume
  that $T$ is not summable w.r.t.~$k$. Then the order of a telescoper
  for $T$ w.r.t.~$k$ is at least
  \begin{equation*}\label{EQ:lowerbound}
    \max_{\begin{array}{c}
        \scriptscriptstyle p \mid b, \, \deg_k(p) > 0\\[-1ex]
        \scriptscriptstyle \text{ multiplicity } \alpha\\[-1ex]
        \scriptscriptstyle \text{ monic {\em \&} irred. }
      \end{array}}\hspace{-5pt}
    \min \left\{\rho\in \bN\setminus\{0\}: 
      \sigma_k^\ell (p)^\alpha \mid \sigma_n^\rho(b)
      \text{ for some } \ell \in \bZ
    \right\}.
  \end{equation*}
\end{theorem}
\begin{proof}
  Let $L=\sum_{i=0}^{\rho}{e_i S_n^i}$ be a minimal telescoper for $T$
  w.r.t.~$k$, where~$\rho \in \bN$ and~$e_0, \ldots ,e_\rho \in \set
  K(n)$ not all zero. Since~$T$ is not summable w.r.t.~$k$, we have
  $\rho \geq 1$. By the modified \ap reduction, \eqref{EQ:ithmapred}
  holds for~$1\leq i \leq \rho$.  Since $L$ is a minimal telescoper,
  $e_0 \neq 0 $ and by Theorem~\ref{THM:redct},
  \[e_0\frac{a}{b} \ +\ e_1\frac{a_1}{b_1}\ +\ \cdots\ +\ e_\rho
  \frac{a_\rho}{b_\rho}=0.\] 
  By partial fraction decomposition, 
  for any monic irreducible factor $p$ of $b$
  with $\deg_k(p) > 0$ and multiplicity~$\alpha>0$, there exists an
  integer~$i$ with $1\leq i \leq \rho$ so that $p^\alpha$ is also a
  factor of~$b_i$. By Theorem~\ref{THM:brelation}, $b_i\approx_k
  \sigma_n^i(b)$. Thus there is a factor $p'$ of~$\sigma_n^{i}(b)$
  with multiplicity at least~$\alpha$ such that $p' \sim_k p$.  Let
  $i_{p}$ be the minimal one with this property. Then the assertion
  follows by the fact that for each factor~$p$ of~$b$ there exists no
  telescoper for~$T$ of order less than~$i_p$.
\end{proof}
Together with the bounds given above, we can further develop a
variant of Algorithm~\ref{ALG:reductionct} by omitting
step~$4.5$ for each loop until the loop index $i$ reaches and exceeds
the lower bound.

\index{creative telescoping}\index{reduction-based}\index{bound}
\begin{algo}[Bound and Reduction-based creative telescoping]\label{ALG:breductionct}\leavevmode\null
	
	\noindent
	{\bf Input}: A hypergeometric term $T$ over~$\bF(k)$.\\
	{\bf Output}: A minimal telescoper for~$T$ w.r.t.~$k$ and a corresponding certificate
	if telescopers exist; \lq\lq No telescoper exists!\rq\rq, otherwise. 
	
	\bigskip
	\step{1--3}{1} Similar to steps~$1$ -- $3$ of Algorithm~\ref{ALG:reductionct}.
	
	\smallskip
	\step{4}0 Compute the upper bound $b_u\in \bN$ and lower bound $b_l\in \bN$ for the order 
	\step{}0 of minimal telescopers for $T$ w.r.t.~$k$, respectively.
	
	\smallskip
	\step{5}0 Set $N = \sigma_n(H)/H$ and $R = \ell_0 r_0$, where~$\ell_0$ is an indeterminate.
	\step{}0 For $i= 1,2,\dots, b_u$ do
	\substep{5.1}{2.1} Similar to steps~$4.1$ -- $4.3$ of Algorithm~\ref{ALG:reductionct}, 
		compute $g_i, r_i\in \set K(n,k)$
	\substep{}{2.1} such that~\eqref{EQ:ithred} holds, and $R + \ell_i r_i$ 
	is a residual form w.r.t.~$K$, 
	\substep{}{2.1} where $\ell_i$ is an indeterminate.
	\substep{5.2}2 Update $R$ to $R+\ell_i	r_i$. If~$i > b_l$ then find $\ell_j\in \bF$ such that $R =0$ 
	\substep{}{2.1} by solving a linear system in $\ell_0, \dots, \ell_i$ over
	$\bF$.
	\substep{}{2.1} If there is a nontrivial solution, return
	$\left(\sum_{j=0}^i \ell_j S_n^j,\ \sum_{j=0}^i\ell_j g_j H\right).$
\end{algo}

\section{Comparison of bounds}\label{SEC:comparison}%
\index{comparison}
In 2005, upper and lower bounds for the order of telescopers for
hypergeometric terms have been studied in \cite{MoZe2005} and
\cite{AbLe2005}, respectively. In this section, we are going to review
these known bounds and also compare them to our bounds.

\subsection{\az upper bound}\label{SUBSEC:azub}%
\index{order bound! \az}\index{\az upper bound}
Let $T$ be a {\em proper}\index{proper term} hypergeometric term over
$\set K(n,k)$, i.e., it can be written in the form
\begin{equation}\label{EQ:properht}
  T = p w^n z^k\prod_{i= 1}^m\frac{(\alpha_i n + \alpha_i' k + \alpha_i''-1)!
  (\beta_i n -\beta_i' k + \beta_i''-1)!}
  {(\mu_i n + \mu_i' k + \mu_i''-1)! (\nu_i n -\nu_i' k + \nu_i''-1)!},
\end{equation}
where $p\in \set K[n,k]$, $w,z\in \set K$, $m\in \bN$ is fixed, $\alpha_i,
\alpha_i', \beta_i, \beta_i', \mu_i, \mu_i', \nu_i, \nu_i'$ are
nonnegative integers and $ \alpha_i'', \beta_i'', \mu_i'', \nu_i''\in
\set K$ . Further assume that there exist no integers~$i$ and $j$
with~$1\leq i, j \leq m $ such that
\begin{align*}
  \begin{array}{cccccc}
    & \left( \alpha_i = \mu_j\vphantom{\alpha_i''-\mu_j'' \in \bN} \right.
    & \text{and} & \alpha_i' = \mu_j' & \text{and} 
    & \left.\alpha_i''-\mu_j'' \in \bN \right)\\[1ex]
    \text{or}
    & \left( \beta_i = \nu_j\vphantom{\beta_i''-\nu_j'' \in \bN} \right. 
    & \text{and} & \beta_i' = \nu_j' &\text{and}  
    & \left.\beta_i''-\nu_j'' \in \bN \right).
  \end{array}
\end{align*}
We refer to this as the \emph{generic} situation\index{generic situation}.
Then Apagodu and Zeilberger \cite{MoZe2005} stated that the order of a
minimal telescoper for $T$ w.r.t.~$k$ is bounded by
\[B_{AZ} = \max\left\{\sum_{i=1}^m(\alpha_i'+\nu_i'),
  \sum_{i=1}^m(\beta'_i+\mu_i')\right\},\] 
and this bound is generically sharp. 

We now show that $B_{AZ}$ is at least the order bound given in
Theorem~\ref{THM:upperbound}.  Reordering the factorial terms
in~\eqref{EQ:properht} if necessary, let $\cS$ be the maximal set of
integers $i$ with~$1\leq i\leq m$ satisfying
\begin{align*}
  \begin{array}{cccccc}
    & \left( \alpha_i = \mu_i\vphantom{\mu_i''-\alpha_i'' \in \bN}\right. 
    & \text{and} & \alpha_i' = \mu_i' & \text{and} 
    & \left.\mu_i''-\alpha_i'' \in \bN \right)\\[1ex]
    \text{or}
    & \left( \beta_i = \nu_i \vphantom{\nu_i''-\beta_i'' \in \bN}\right. 
    & \text{and} & \beta_i' = \nu_i' & \text{and} 
    & \left.\nu_i''-\beta_i'' \in \bN \right).
  \end{array}
\end{align*}
Rewrite $T$ as
\[r w^n z^k \prod_{ \scriptscriptstyle i = 1,\, \scriptscriptstyle i
  \notin \cS }^m\frac{(\alpha_i n + \alpha_i' k +
  \alpha_i''-1)!(\beta_i n -\beta_i' k + \beta_i''-1)!}{(\mu_i n +
  \mu_i' k + \mu_i''-1)! (\nu_i n -\nu_i' k + \nu_i''-1)!},\] 
where $r \in \set K(n,k)$. For $q \in \set K[n,k]$ and $m \in \bN$, let
\[q^{\overline{m}} = q (q+1) (q+2)\cdots (q+m-1)\]
with the convention $q^{\overline{0}} = 1$. By an easy calculation,  
\begin{align}\label{EQ:kernel}
  K = z \prod_{i
  } \frac{(\alpha_i n + \alpha_i' k +
    \alpha_i'')^{\overline{\alpha'_i}} (\nu_i n -\nu_i' k +
    \nu_i''-\mu_i')^{\overline{\nu'_i}}}{(\mu_i n + \mu_i' k +
    \mu_i'')^{\overline{\mu'_i}}(\beta_i n -\beta_i' k +
    \beta_i''-\beta_i')^{\overline{\beta'_i}}}
\end{align}
where the product runs over all $i$ from $1$ to $m$ such that $ i
\notin \cS$, $\alpha_i',\beta_i'>0$ and~$\mu_i', \nu_i' > 0$, is a
kernel of $T$ and $S=r$ is a corresponding shell. Let~$K = u/v$
with~$u,v\in \set K(n)[k]$ and $\gcd(u,v)=1$. Note that the right-hand
side of \eqref{EQ:kernel} already has the reduced form, then a
straightforward calculation yields
\[\deg_k(u) = \sum_{i=1,i\notin\cS}^m (\alpha_i'+\nu_i') \quad 
\text{and} \quad \deg_k(v) = \sum_{i = 1, i\notin \cS}^m(\beta_i'+
\mu_i').\] 

Applying the modified \ap reduction to~$T$ w.r.t.~$k$
yields~\eqref{EQ:initialred}, in which~$b$ is integer-linear. Since
$b$ only comes from the shift-free part of the denominator of~$r$, it
factors into shift-inequivalent integer-linear polynomials of degree
one which are separately shift-equivalent to either~$(\mu_i n+ \mu_i'
k+\mu_i'')$ or $(\beta_i n -\beta_i' k+\beta_i'')$ w.r.t.\ $n,k$ for
some~$i \in \cS$. Note that each $i$ in $\cS$ corresponds to at most
one integer-linear factor of~$b$, and increases the multiplicity of
the corresponding factor in~$b$ by at most~$1$.  Hence, the bound
given in Theorem~\ref{THM:upperbound} is no more than
\begin{align*}
  &\quad \max\{\deg_k(u), \deg_k(v)\} - \llbracket \deg_k(v-u) \leq
  \deg_k(u) -1\rrbracket+ \sum_{i=1,i\in\cS}^m (\beta_i'+\mu_i'),
\end{align*}
which is exactly equal to
\[B_{AZ} - \llbracket \deg_k(v-u) \leq \deg_k(u) -1\rrbracket,\] 
since $\sum_{i=1,i\in\cS}^m (\alpha_i'+\nu_i')=\sum_{i=1,i\in\cS}^m
(\beta_i'+\mu_i')$.

In general, i.e., in the generic situation, the order bound in
Theorem~\ref{THM:upperbound} is almost the same as~$B_{AZ}$, which is
not suprising since $B_{AZ}$ is already generically sharp. However,
our bound can be much better in some special examples.
\begin{example}\label{EX:sharperub}
  Consider a rational function
  \[T=\frac{\alpha^2 k^2+\alpha^2 k-\alpha \beta k+2 \alpha n
    k+n^2}{(n+\alpha k+\alpha)(n+\alpha k) (n+\beta k)},\]
  where $\alpha, \beta$ are positive integers and $\alpha \neq
  \beta$. Rewriting $T$ into the proper form \eqref{EQ:properht}
  yields $B_{AZ} = \alpha + \beta$. On the other hand, $1$ is the only
  kernel of~$T$ since $T$ is a rational function. By the modified \ap
  reduction, $b=n+\beta k$ in~\eqref{EQ:initialred}. By
  Theorem~\ref{THM:upperbound}, a minimal telescoper for~$T$
  w.r.t.~$k$ has order no more than $\beta$, which is in fact the real
  order of minimal telescopers for $T$ w.r.t.~$k$.
\end{example}
\begin{remark}\label{REM:nonproper}
  Together with~\cite[Theorem 10]{Abra2003}, the upper order bound $B_{AZ}$
  on minimal telescopers derived in~\cite{MoZe2005} can be also applied
  to non-proper hypergeometric terms. On the other hand,
  Theorem~\ref{THM:upperbound} can be directly applied to any
  hypergeometric term provided that its telescopers exist.
\end{remark}

\subsection{\al lower bound}\label{SUBSEC:allb}%
\index{order bound! \al}\index{\al lower bound}
With Convention \ref{CON:convention}, further assume that $T$ has the
initial reduction~\eqref{EQ:initialred}, in which~$b$ is
integer-linear. Let $H'  =  H/v$. A direct calculation leads to
\[\frac{\sigma_k(H')}{H'} = \frac{u}{\sigma_k(v)},\]
which can be easily checked to be shift-reduced w.r.t.~$k$. Let $d'\in
\set K(n)[k]$ be the denominator of $\sigma_n(H')/H'$.  Then the
algorithm \textsf{\em LowerBound}%
\index{LowerBound@\textsf{\em LowerBound}} in \cite{AbLe2005} asserts 
that the order of telescopers for $T$ w.r.t.~$k$ is at least
\begin{align*}
  B_{AL} = \hspace{-8pt}\max_{{\begin{array}{c}
        \scriptscriptstyle p \mid b\\[-1ex]
        \scriptscriptstyle \text{ irred.\ \& monic}\\[-1ex]
        \scriptscriptstyle \deg_k(p) > 0
		\end{array}
              }}\hspace{-8pt} \min \left\{\rho\in \bN\setminus\{0\}:
              \begin{array}{c}
                \sigma_k^\ell(p) \mid \sigma_n^{\rho} (b) \\[1ex]
                \text{ or }\\[1ex]
                \sigma_k^\ell(p) \mid \sigma_n^{\rho-1}(d')
                \text{ for some } \ell \in \bZ
              \end{array}\right\}
\end{align*}
Comparing to $B_{AL}$ from above, one easily sees that the lower bound
given in Theorem~\ref{THM:lowerbound} can be better but never worse
than~$B_{AL}$.
\begin{example}\label{EX:sharperlb}
  Consider a hypergeometric term
  \[T = \frac{1}{(n-\alpha k-\alpha)(n-\alpha k-2)!},\] 
  where $\alpha\in \bN$ and $\alpha>1$. By the algorithm \textsf{\em
    LowerBound}, a telescoper for~$T$ w.r.t.~$k$ has order at least
  $2$. On the other hand, a telescoper for $T$ w.r.t.~$k$ has order at
  least~$\alpha$ by Theorem~\ref{THM:lowerbound}. In fact,~$\alpha$ is
  exactly the order of minimal telescopers for~$T$ w.r.t.~$k$.
\end{example}
\index{order bound|)}
\section{Implementation and timings}\label{SEC:boundtiming}
In {\sc Maple~18}\index{Maple@{\sc Maple}}, we have implemented
Algorithm~\ref{ALG:breductionct} and embedded it into the package
\textbf{ShiftReductionCT}\index{ShiftReductionCT@\textbf{ShiftReductionCT}},
under the name of \textsf{\brct}.
For a detailed explanation, one may refer to Appendix~\ref{APP:guide}.

In this section, we focus on the two procedures -- \textsf{\brct} and
{\sf \rct} in the package \textbf{ShiftReductionCT}, and their runtime
is compared. All timings\index{timing} are measured in seconds on a Linux
computer\index{Linux computer} with 388Gb RAM and twelve 2.80GHz Dual
core processors. No parallelism was used in this experiment. Moreover,
a comparison\index{comparison} of the memory requirements%
\index{memory requirement} is given in Appendix~\ref{APP:memory}.  
For brevity, we denote
\begin{itemize}
\item {\sf RCT$_{tc}$}: the procedure~\textsf{\rct} in
  \textbf{ShiftReductionCT},
  \index{ShiftReductionCT@\textbf{ShiftReductionCT}! ReductionCT@\textsf{ReductionCT}}%
  \index{ReductionCT@\textsf{ReductionCT}~{\em (ShiftReductionCT)}}%
   which computes a minimal telescoper and a corresponding normalized
  certificate;

\smallskip
\item \textsf{RCT$_{t}$}: the procedure~\textsf{\rct} in
  \textbf{ShiftReductionCT}, which computes a minimal telescoper
  without constructing a certificate.

\smallskip
\item {\sf BRCT$_{tc}$}: the procedure~\textsf{\brct} in
  \textbf{ShiftReductionCT}, which computes a minimal telescoper and a corresponding normalized certificate;%
  \index{ShiftReductionCT@\textbf{ShiftReductionCT}! BoundReductionCT@\textsf{BoundReductionCT}}%
 \index{BoundReductionCT@\textsf{BoundReductionCT}~{\em (ShiftReductionCT)}}

\smallskip
\item \textsf{BRCT$_{t}$}: the procedure~\textsf{\brct} in
  \textbf{ShiftReductionCT}, which computes a minimal telescoper
  without constructing a certificate.
\item \textsf{LB}: the lower bound for telescopers given in
  Theorem~\ref{THM:lowerbound}.\index{lower bound}

\smallskip
\item \textsf{order}: the order of the resulting minimal telescoper.\index{order}%
\end{itemize}

\begin{example}\label{EX:bctvsct1}
  Consider the same hypergeometric term as in Example~\ref{EX:sharperlb},
  i.e.,
  \[T = \frac{1}{(n-\alpha k-\alpha)(n-\alpha k-2)!},\] 
  where $\alpha$ is an integer greater than $1$.
  For different choices of~$\alpha$, Table~\ref{TAB:bctvsct1} 
  shows the timings of the above procedures. 
  Note that since the term $T$ in this example is very simple,
  there is little difference in the timings for the two procedures
  with and without construction of a certificate.
  \begin{table}[h]
    \tabcolsep10pt
    \begin{center}
      \begin{tabular}{c|rrrr|cc}
        $\alpha$ & {\sf{RCT$_{t}$}} & {\sf{RCT$_{tc}$}} & {\sf{BRCT$_{t}$}}  
        & {\sf{BRCT$_{tc}$}} & {\sf{LB}} & {\sf{order}}\\
        \hline
        20 & 2.00 & 2.02 & 1.07 & 1.13 & 20 & 20 \\[.5ex]
        30 & 7.01 & 7.19 & 2.86 & 2.96 & 30 & 30 \\[.5ex]
        40 & 20.08 & 20.13 & 7.06 & 7.18 & 40 & 40 \\[.5ex]
        50 & 42.15 & 42.68 & 14.96 & 15.05 & 50 & 50 \\[.5ex]
        60 & 104.07 & 106.31 & 25.54 & 25.93 & 60 & 60 \\[.5ex]
        70 & 225.67 & 229.04 & 45.76 & 45.97& 70 & 70 \\
        \hline
      \end{tabular}
    \end{center}
    \vspace{-\smallskipamount}
    \caption{Timing comparison of two reduction-based\index{reduction-based}%
     creative telescoping with and without construction of a certificate for 
     Example~\ref{EX:bctvsct1} (in seconds)}\label{TAB:bctvsct1}
  \end{table}
\end{example}
\begin{example}[Example~6 in~\cite{AbLe2005}]\label{EX:bctvsct2}
  Consider the hypergeometric term
  \[T =\Delta_k\left( T_1\right)+
  T_2,\] 
  where
  \[T_1=\frac{1}{(n k-1)(n-\alpha k-2)^m(2 n + k+3)!}\ \text{and}\ T_2=\frac{1}{(n-\alpha k-2)(2 n + k+3 )!}\]
  for~$\alpha, m$ positive integers.
  For different choices of $\alpha$ and $m$, we
  compare the timings of the procedures from above. 
  Table~\ref{TAB:bctvsct2} shows the final experimental results.
	
  \begin{table}[h]
    \tabcolsep10pt
    \begin{center}
      \begin{tabular}{l|rrrr|cc}
        $(m, \alpha)$ & {\sf{RCT$_{t}$}} & {\sf{RCT$_{tc}$}} & {\sf{BRCT$_{t}$}} 
        & {\sf{BRCT$_{tc}$}} & {\sf{LB}} & {\sf{order}}\\
        \hline
        (1,1) & 0.20 & 0.24 & 0.20 & 0.23 & 1 & 2 \\[.5ex]
        (1,10) & 5.25 & 9.56 & 4.60 & 8.74 & 10 & 11 \\[.5ex]
        (1,15) & 57.06 & 76.01 & 37.73 & 58.69 & 15 & 16 \\[.5ex]
        (1,20) & 538.59 & 656.99 & 264.04 & 324.09 & 20 & 21 \\[.5ex]
        (2,10) & 5.29 & 9.11 & 4.43 & 8.36 & 10 & 11 \\[.5ex]
        (2,15) & 79.34 & 96.48 & 40.26 & 54.85 & 15 & 16 \\[.5ex]
        (2,20) & 574.00 & 658.20 & 282.54 & 377.84 & 20 & 21 \\
        \hline
      \end{tabular}
    \end{center}
    \vspace{-\smallskipamount}
    \caption{Timing comparison of two reduction-based\index{reduction-based}%
      creative telescoping with and without construction of a certificate for 
      Example~\ref{EX:bctvsct2} (in seconds)}\label{TAB:bctvsct2}
  \end{table}
\end{example}
\begin{remark}\label{REM:breductionct}
  Compared to linear dependence, determining linear independence takes
  much less time because with high probability, independence can be
  recognized by a computation in a homomorphic image. For this reason,
  the procedure~{\sf \brct} makes no big difference from the
  procedure~{\sf \rct} if the lower bound is far away from the real
  order of minimal telescopers. In fact, their perform almost the same
  in this case.
\end{remark}

\part[Limits of P-recursive Sequences]{Limits of \\P-recursive sequences}
\chapter[D-finite Functions and P-recursive Sequences]
{D-finite Functions and \\P-recursive Sequences}
\label{CH:dfprelim}

In this chapter, we recall~\cite{FlSe2009,KaPa2011} basic notions
related to the class of D-finite functions and P-recursive sequences,
and also present some useful properties.

\section{Basic concepts}\label{SEC:dfinite}
Recall~\cite{KaPa2011} that a {\em formal power series}%
\index{formal power series} is an infinite series of the form
\[f(z) = \sum_{n=0}^\infty a_n z^n,\] 
where $z$ is a formal indeterminate.  
It generalizes the notions of polynomials and power series in some sense. 
A formal power series differs from a polynomial
in that it allows an infinite number of terms, and it differs from
power series by assuming a formal variable and ignoring analytic
properties.  One way to view a formal power series $f(z)$ is to take
it as an infinite sequence $(a_n)_{n=0}^\infty$, where the powers
indicate the order of terms.  We will also call a formal power series
$f(x)$ the {\em generating function}\index{generating function} of its
coefficient sequence $(a_n)_{n=0}^\infty$.  Note that these three
notions -- formal power series, sequences\index{sequence}, generating
functions -- all refer to the same object.

For a ring $R$, we denote by $R[[z]]$ the ring of formal power series
endowed with termwise addition~($+$) and {\em Cauchy
  product}\index{Cauchy product} ($\,\cdot\,$):
\begin{align*}
  \left(\sum_{n=0}^\infty a_n z^n \right) + \left( \sum_{n=0}^\infty b_n z^n\right) 
  = \sum_{n=0}^\infty\left(a_n+b_n \right)z^n,\\[1ex]
  \left(\sum_{n=0}^\infty a_n z^n \right) \cdot
  \left(\sum_{n=0}^\infty b_nz^n \right) = \sum_{n=0}^\infty
  \left(\sum_{k=0}^n a_k b_{n-k}\right) z^n,
\end{align*}
and by $R^\bN$ the ring of infinite sequences endowed with termwise
addition ($+$) and {\em Hadamard product}\index{Hadamard product}
($\odot$):
\begin{align*}
(a_n)_{n=0}^\infty  +  (b_n)_{n=0}^\infty = (a_n+b_n )_{n=0}^\infty,\\[1ex]
(a_n)_{n=0}^\infty  \odot  (b_n)_{n=0}^\infty = (a_n b_n )_{n=0}^\infty.
\end{align*}

Also recall~\cite{FlSe2009} that a complex function $f(z)$ is called
{\em analytic}\index{analytic! function} at a point~$\zeta\in \bC$ if
for any $z$ in a neighborhood of $\zeta$, it can be represented by a
convergent power series over $\bC$,
\[f(z) = \sum_{n=0}^\infty a_n(z-\zeta)^n, \quad \text{where}\ a_n\in
\bC \ \text{for all}\ n\in \bN.\] 
A function is {\em analytic} in an open set if it is analytic at every point of the set.

Throughout the chapter, let $R$ be a subring of $\bC$ and $\bF$ be a
subfield of $\bC$.  We consider linear operators that act on sequences
or power series and analytic functions.  
Recall from the previous chapters that we write $\sigma_n$ for the shift
operator\index{operator! shift}\index{shift operator} w.r.t.~$n$ which
maps a sequence~$(a_n)_{n=0}^\infty$ to~$(a_{n+1})_{n=0}^\infty$.  
Also we denote by $\bF[n]\<S_n>$ the ring of linear recurrence operators 
of the form $L:=p_0+p_1S_n+\cdots+p_\rho S_n^\rho$%
\index{recurrence operator}\index{operator! recurrence}, with~$p_0,\dots,p_\rho\in\bF[n]$, 
where $S_n r = \sigma_n(r) S_n$ for all $r\in \bF[n]$. 
This ring forms an Ore algebra\index{Ore algebra}.  
Analogously, we write $D_z$ for the derivation operator%
\index{operator! derivation}\index{derivation operator} w.r.t.~$z$ 
which maps a power series or function~$f(z)$ to
its derivative $f'(z)=\frac{d}{dz}f(z)$.  Also the set of linear
operators of the form
$L:=p_0+p_1D_z+\cdots+p_\rho D_z^\rho$%
\index{differential operator}\index{operator! differential}, with
$p_0,\dots,p_\rho\in\bF[z]$, forms an Ore algebra; we denote it by
$\bF[z]\<D_z>$.  For an introduction to Ore algebras and their
actions, please refer to~\cite{BrPe1996}.  When $p_\rho\neq0$, we call
$\rho$ the \emph{order}\index{order} of the operator and~$\lc(L):=p_\rho$
its \emph{leading coefficient}.\index{leading coefficient}

\let\theenumi=\oldtheenumi
\let\labelenumi=\oldlabelenumi
\begin{definition}\label{DEF:dfinite}\leavevmode\null
  \begin{enumerate}
  \item A sequence $(a_n)_{n=0}^\infty\in R^\bN$ is called {\em
      P-recursive} or \emph{D-finite} over~$\bF$%
  \index{P-recursive sequence}\index{D-finite! sequence|see {P-recursive sequence}}\index{sequence! P-recursive} if there 
  exists a nonzero operator $L= \sum_{j=0}^\rho
    p_j(n) S_n^j \in \bF[n]\langle S_n \rangle$ such that
    \[
    L\cdot a_n = p_\rho(n) a_{n+\rho} + \dots + p_1(n) a_{n+1} + p_0(n)
    a_n = 0
    \]
    for all $n\in \bN$.

    \smallskip
  \item A formal power series $f(z)\in R[[z]]$ is called {\em
      D-finite} over $\bF$%
   \index{D-finite! power series|see {D-finite function}}\index{D-finite! function}\index{function! D-finite} if there exists a nonzero
    operator $L= \sum_{j=0}^\rho p_j(z) D_z^j \in \bF[z]\langle
    D_z\rangle$ such that
    \[
    L \cdot f(z) = p_\rho(z) D_z^\rho f(z) + \dots + p_1(z) D_z f(z) +
    p_0(z) f(z) = 0.
    \]
  \item A formal power series $f(z) \in \bF[[z]]$ is called {\em
      algebraic}\index{algebraic! function}\index{function! algebraic}
    over $\bF$ if there exists a nonzero bivariate polynomial $P(z,y)
    \in \bF[z,y]$ such that $P(z,f(z)) = 0$.
  \end{enumerate}
\end{definition}
In general, D-finite power series are called D-finite functions
instead.  A formal power series is D-finite if and only if its
coefficient sequence is P-recursive.  Many elementary functions like
rational functions, exponentials, logarithms, sine, algebraic
functions, etc., as well as many special functions, like
hypergeometric series, the error function, Bessel functions, etc., are
D-finite. Hence their respective coefficient sequences are
P-recursive.

\section{Useful properties}\label{SEC:property}\index{closure properties|(}
The class of D-finite functions (resp.\ P-recursive sequences) is
closed under certain operations: addition, multiplication, derivative
(resp.\ forward shift) and integration (resp.\ summation). In
particular, the set of D-finite functions (resp.\ P-recursive
sequences) forms a left-$\bF[z]\<D_z>$-module (resp.\ a
left-$\bF[n]\<S_n>$-module). Also, if $f$ is a D-finite function and
$g$ is an algebraic function, then the composition $f\circ g$ is
D-finite. These and further closure properties are easily proved by
linear algebra arguments, whose proofs can be found for instance
in~\cite{Stan1980,SaZi1994,KaPa2011}. We will make free use of these
facts.

We will be considering singularities of D-finite functions. Recall
from the classical theory of linear differential
equations~\cite{Ince1944} that a linear differential equation
$p_0(z)f(z)+\cdots+p_\rho(z)f^{(\rho)}(z)=0$ with polynomial coefficients
$p_0,\dots,p_\rho\in\bF[z]$ and~$p_\rho\neq0$ has a basis of analytic
solutions in a neighborhood of every point $\zeta\in\set C$, except
possibly at roots of~$p_\rho$. The roots of $p_\rho$ are therefore called
the \emph{singularities}\index{singularity} of the equation (or the
corresponding linear operator). If $\zeta\in\set C$ is a singularity
of the equation but the equation nevertheless admits a basis of
analytic solutions at this point, we call it an \emph{apparent
  singularity}.\index{apparent singularity} It is
well-known~\cite{Ince1944,CKS2016} that for any given linear
differential equation with some apparent and some non-apparent
singularities, we can always construct another linear differential
equation (typically of higher order) whose solution space contains the
solution space of the first equation and whose only singularities are
the non-apparent singularities of the first equation. This process is
known as desingularization\index{desingularization}.

For later use, we will give a proof of the composition closure property 
for D-finite functions which pays attention to the singularities.

\begin{theorem}\label{THM:algsubs}\index{algebraic! composition}
  Let $P(z,y)\in\bF[z,y]$ be a square-free polynomial of degree~$d$,
  and let $L\in\bF[z]\<D_z>$ be nonzero.  Let~$\zeta\in\bC$ be such
  that $P$ defines $d$ distinct analytic algebraic functions $g(z)$
  with $P(z,g(z))=0$ in a neighborhood of~$\zeta$, and assume that for
  none of these functions, the value $g(\zeta)\in\set C$ is a
  singularity of~$L$.  Fix a solution $g$ of~$P$ and an analytic
  solution $f$ of~$L$ defined in a neighborhood of~$g(\zeta)$. Then
  there exists a nonzero operator $M\in\bF[z]\langle D_z\rangle$ with
  $M\cdot(f\circ g)=0$ which does not have $\zeta$ among its
  singularities.
\end{theorem}
\begin{proof} (borrowed from \cite{KaPo})
  Consider the operator
  $\tilde L=L(g,(g')^{-1}D_z)\in\overline{\bF(z)}\<D_z>$.  It is easy
  to check that $L\cdot f=0$ if and only if $\tilde L\cdot(f\circ
  g)=0$ for every solution~$g$ of $P$ near~$\zeta$.  Therefore,
  if~$f_1,\dots,f_\rho$ is a basis of the solution space of $L$ near
  $g(\zeta)$, then $f_1\circ g,\dots,f_\rho\circ g$ is a basis of the
  solution space of $\tilde L$ near~$\zeta$.
	
  Let $g_1,\dots,g_d$ be all the solutions of $P$ near~$\zeta$, and
  let $M$ be the least common left multiple of all the operators
  $L(g_j,(g_j')^{-1}D_z)$.  Then the solution space of $M$ near
  $\zeta$ is generated by all the functions $f_i\circ g_j$.  Since the
  coefficients of $M$ are symmetric w.r.t.\ the conjugates
  $g_1,\dots,g_d$, they belong to the ground field~$\bF(z)$, and after
  clearing denominators (from the left) if necessary, we may assume
  that $M$ is an operator in $\bF[z]\<D_z>$ whose solution space is
  generated by functions that are analytic at~$\zeta$. Therefore, by
  the remarks made about desingularization, it is possible to replace
  $M$ by an operator (possibly of higher order) which does not have
  $\zeta$ among its singularities.
\end{proof}

By a similar argument, we see that algebraic extensions of the
coefficient field of the recurrence operators are useless. Moreover,
it is also not useful to make $\bF$ bigger than the quotient
field\index{quotient field} of~$R$.

\goodbreak
\begin{lemma}\label{lemma:changeopseq}\leavevmode\null
  \begin{enumerate}
  \item\label{lemma:changeopseq1} If $\set E$ is an algebraic
    extension field of $\bF$ and $(a_n)_{n=0}^\infty$ is P-recursive
    over $\set E$, then it is also P-recursive over~$\set F$.

    \smallskip
  \item\label{lemma:changeopseq2} If $R\subseteq\set F$ and
    $(a_n)_{n=0}^\infty\in R^{\set N}$ is P-recursive over~$\set F$,
    then it is also P-recursive over~$\quot(R)$, the quotient field
    of~$R$.

    \smallskip
  \item\label{lemma:changeopseq3} If $\set F$ is closed under complex
    conjugation\index{complex conjugation} and $(a_n)_{n=0}^\infty$ is
    P-recursive over~$\set F$, then so are $(\bar a_n)_{n=0}^\infty$,
    $(\real(a_n))_{n=0}^\infty$, and $(\imag(a_n))_{n=0}^\infty$.
  \end{enumerate}
\end{lemma}
\begin{proof}
  \begin{enumerate}
  \item Let $L\in\set E[n]\<S_n>$ be an annihilating operator of
    $(a_n)_{n=0}^\infty$. Then, since $L$ has only finitely many
    coefficients, $L\in\set F(\theta)[n]\<S_n>$ for some
    $\theta\in\set E$. Let~$M$ be the least common left multiple of
    all the conjugates of~$L$.  Then~$M$ is an annihilating operator
    of~$(a_n)_{n=0}^\infty$ which belongs to $\set F[n]\<S_n>$. The
    claim follows.

    \medskip
  \item Let us write $\set K=\quot(R)$.  Let $L\in\set F[n]\<S_n>$ be
    a nonzero annihilating operator of~$(a_n)_{n=0}^\infty$.  Since
    $\bF$ is an extension field of~$\bK$, it is a vector space
    over~$\bK$. Write
    \[
    L = \sum_{m=0}^\rho \sum_{j=0}^{d_m} p_{mj} n^j S_n^m,
    \]
    where $r, d_m \in \bN$ and $p_{mj} \in \bF$ not all zero. Then the
    set of the coefficients~$p_{ij}$ belongs to a finite dimensional
    subspace of~$\bF$. Let $\{\alpha_1, \dots, \alpha_s\}$ be a basis
    of this subspace over~$\set K$. Then for each pair $(m,j)$, there
    exists $c_{mj\ell} \in \bK$ such that $p_{mj} = \sum_{\ell = 1}^s
    c_{mj\ell} \alpha_\ell$, which gives
    \[
    0 = L\cdot a_n = \sum_{\ell=1}^s \alpha_\ell
    \underbrace{\left(\sum_{m=0}^\rho\sum_{j=0}^{d_m} c_{mj\ell} n^j
        a_{n+m}\right)}_{=:b_n\in\set K}.
    \]
    For all $n\in \bN$, it follows from the linear independence of
    $\{\alpha_1, \dots, \alpha_s\}$ over~$\bK$ that~$b_n=0$.
    Therefore
    \[
    \sum_{m=0}^\rho\underbrace{\left(\sum_{j=0}^{d_m} c_{mj\ell}
        n^j\right)}_{\in \bK[n]}S_n^m\cdot a_n = 0 
    \quad \text{for all}\ n\in \bN \ \text{and}\ \ell = 1, \dots, s.
    \]
    Thus $(a_n)_{n=0}^\infty$ has a nonzero annihilating operator with
    coefficients in $\bK[n]$.

    \medskip
  \item Since $(a_n)_{n=0}^\infty$ is P-recursive over~$\bF$, there
    exists a nonzero operator $L$ in~$\bF[n]\langle S_n\rangle$ such
    that $L \cdot a_n = 0$.  Hence $\bar L \cdot \bar a_n =0$
    where $\bar L$ is the operator obtained from~$L$ by taking the
    complex conjugate of each coefficient.  Since $\set F$ is closed
    under complex conjugation by assumption, we see that $\bar
    L$~belongs to $\set F[n]\<S_n>$, and hence~$(\bar
    a_n)_{n=0}^\infty$ is P-recursive over~$\bF$.
		
    Because of $\real(a_n) = \frac12(a_n + \bar a_n)$ and $\imag(a_n)
    = \frac1{2i}(a_n - \bar a_n)$ with $i$ the imaginary unit, the
    other two assertions follow by closure properties.
  \end{enumerate}
\end{proof}
\index{closure properties|)}
Of course, all the statements hold analogously for D-finite functions instead
of P-recursive sequences.

If we consider a D-finite function as an analytic complex function defined in a
neighborhood of zero, then this function can be extended by analytic continuation\index{analytic! continuation}
to any point in the complex plane except for finitely many ones, namely the
singularities of the given function.  In this sense, D-finite functions can be
evaluated at any non-singular point by means of analytic continuation. Numerical
evaluation algorithms\index{numerical evaluation} for D-finite functions have been developed
in~\cite{ChCh1990,vdHo1999,vdHo2001,vdHo2007a,Mezz2010,MeSa2010}, where the last
two references also provide a {\sc Maple} implementation, namely the \textbf{NumGfun}\index{NumGfun@\textbf{NumGfun}}
package, for computing such evaluations.
These algorithms perform arbitrary-precision evaluations with full error control.

\chapter[D-finite Numbers]
{D-finite Numbers\protect\footnotemark{}\protect
\footnotetext{The main results in this chapter are 
joint work with M.\ Kauers~\cite{HuKa}.}}
\label{CH:dfinitenos}


As mentioned in the introduction, the class of algebraic numbers and the class of
algebraic functions are naturally connected to each other. For
instance, evaluating an algebraic function over~$\bQ$ at an algebraic
point gives an algebraic number.  Also the values of compositional
inverses of algebraic functions at algebraic points are algebraic. In
particular, roots of an algebraic function over~$\bQ$ are all
algebraic numbers.  Moreover, we will see below that every algebraic
number can appear as a limit of the coefficient sequence of an
algebraic function.  However, the class of algebraic numbers is quite
small.  Almost all real and complex numbers are not algebraic,
including many important numbers like $\pi$ and Euler's number~$\e$.

Motivated by the above relation, we aim to establish a similar
correspondence between numbers and the class of D-finite functions. To
this end, we introduce the following class of numbers.

\begin{definition}\label{DEF:dfinitenos}
  Let $R$ be a subring of~$\set C$ and let $\set F$ be a subfield
  of~$\set C$.
  \begin{enumerate}
  \item A number $\xi\in\set C$ is called \emph{D-finite} (with
    respect to $R$ and~$\set F$)%
    \index{D-finite! number}\index{number! D-finite} if there exists a convergent sequence
    $(a_n)_{n=0}^\infty$ in $R^{\set N}$ with $\lim_{n\to\infty}
    a_n=\xi$ and some polynomials $p_0,\dots,p_\rho\in\set F[n]$,
    $p_\rho\neq0$, not all zero, such that
    \[
    p_0(n)a_n + p_1(n)a_{n+1} + \cdots + p_\rho(n)a_{n+\rho} = 0
    \]
    for all $n\in\set N$.

    \smallskip
  \item The set of all D-finite numbers with respect to $R$ and $\set
    F$ is denoted by $\cD_{R,\set F}$. If $R=\set F$, we also write
    $\cD_{\set F}:=\cD_{\set F,\set F}$ for short.
  \end{enumerate}
\end{definition}
It turns out that the class of D-finite numbers is closely related to
the class of (regular or singular) holonomic
constants~\cite{FlVa2000}\index{holonomic constant}, i.e., the set of
all finite values of D-finite functions at (regular or singular)
algebraic points.

In this chapter, we show that D-finite numbers are in fact holonomic
constants, and conversely, the regular holonomic constants, 
i.e., the values D-finite functions can assume at non-singular
algebraic number arguments, are essentially D-finite numbers over the
Gaussian rational field.  Together with the work on
arbitrary-precision evaluation of D-finite
functions~\cite{ChCh1990,vdHo1999,vdHo2001,vdHo2007a,Mezz2010,MeSa2010},
it follows that D-finite numbers are computable\index{computable} in
the sense that for every D-finite number~$\xi$ there exists an
algorithm which for any given~$n\in\set N$ computes a numeric
approximation\index{numerical evaluation} of $\xi$ with a guaranteed
precision of $10^{-n}$.  Consequently, all non-computable
numbers\index{number! computable} have no chance to be D-finite.
Besides these artificial examples, we do not know of any explicit real
numbers which are not in~$\cD_{\set Q}$, and we believe that it may be
very difficult to find some.

We see from Definition~\ref{DEF:dfinitenos} that the class
$\cD_{R,\bF}$ depends on two subrings of~$\bC$: the ring $R$ where the
sequence lives, and the field $\bF$ over which the difference equation
is defined.  Obviously, different choices of subrings may or may not
lead to different classes of D-finite numbers.  One goal for this
chapter is to investigate what kind of choices of $R$ and $\bF$ can be
made without changing the resulting class of D-finite numbers.

\section{Examples of D-finite numbers}\label{SEC:example}%
\index{D-finite! number}\index{number! D-finite} 
Throughout the chapter, $R$ is a subring of $\bC$ and $\bF$ is a subfield of~$\bC$,
as in Definition~\ref{DEF:dfinitenos} above.  Thanks to many
mathematicians' work, we can easily recognize for many constants that
they in fact belong to $\cD_\bQ$.
\begin{example}\label{EX:dfiniteno}\leavevmode\null
 \begin{enumerate}
 \item Archimedes' constant~$\pi$. \index{Archimedes' constant}Let
   \[f_n = \sum_{k=0}^n\frac1{16^k}\left(\frac4{8k+1}-\frac2{8
       k+4}-\frac1{8k+5}-\frac1{8k+6}\right).\] 
   It is clear that $(f_n)_{n=0}^\infty$ is a P-recursive sequence in~$\bQ$. According
   to the Bailey-Borwein-Plouffe formula~\cite{BBP1997},
   $\lim_{n\rightarrow \infty} f_n=\pi$.

   \medskip
 \item Euler's number~$\e$.\index{Euler's! number} By the Taylor
   series of the exponential function, we have
   \[\lim_{n\rightarrow \infty} f_n = \e\ \ \text{where} \ f_n =
   \sum_{k=0}^n \frac{1}{k!}.\] 
   It is clear that the terms $f_n$ form a P-recursive sequence over~$\bQ$.
 
   \medskip
 \item Logarithmic value~$\log{2}$.  By the Taylor series of the
   natural logarithm, we find a P-recursive sequence
   $(f_n)_{n=0}^\infty\in \bQ^\bN$ with 
   \[f_n = \sum_{k=1}^n \frac{(-1)^{k+1}}{k},\]
   such that $\lim_{n\rightarrow \infty} f_n = \log(2)$.

   \medskip
 \item Pythagoras' constant~$\sqrt{2}$.\index{Pythagoras' constant}
   One easily finds a P-recursive sequence $(f_n)_{n=0}^\infty$
   over~$\bQ$ with
   \[f_n = \sum_{k=0}^n {\frac12 \choose k},\] 
   and we have $\lim_{n\rightarrow \infty} f_n = \sqrt{2}$ by the binomial
   theorem.
   
   \medskip
 \item Ap{\' e}ry's constant~$\zeta(3)$.\index{Ap{\' e}ry's constant}
   By the definition, we see that
   \[\lim_{n\rightarrow \infty} f_n = \zeta(3)\quad \text{with} \ f_n
   = \sum_{k=1}^n \frac{1}{k^3}.\] 
   It is readily seen that $(f_n)_{n=0}^\infty\in \bQ^\bN$ is D-finite.
 	
   \medskip
 \item The number~$1/\pi$. Thanks to Ramanujan, we know that the terms
   \[f_n = \sum_{k=0}^n{2 k\choose k}^3 \frac{42 k + 5}{2^{12 k +
       4}},\] 
   tend to $1/\pi$ as $n\rightarrow \infty$ and form a
   P-recursive sequence over~$\bQ$.
 	
   \medskip
 \item Euler's constant~$\gamma$.\index{Euler's! constant} A desired
   P-recursive sequence is found by Fischler and Rivoal at their
   work~\cite{FiRi2016a}. They showed that
   \[ \lim_{n\rightarrow \infty} f_n = \gamma \quad \text{with}\ f_n =
   \sum_{k=1}^n (-1)^k{n\choose
     k}\frac1k\left(1-\frac1{k!}\right).\]
 	
   \smallskip
 \item Any value of the Gamma function%
	 \index{Gamma function}\index{function! Gamma} to a rational
   number~$\Gamma(\alpha)$ with~$\alpha < 1$ in~$\bQ$. Again, Fischler
   and Rivoal~\cite{FiRi2016a} proved that
   \[\lim_{n\rightarrow \infty} f_n=\Gamma(\alpha) \quad \text{with}\
   f_n = \sum_{k=0}^n{n+\alpha \choose k+\alpha}\frac{(-1)^k}{k!
     (k+\alpha)}.\]
 \end{enumerate}
\end{example}

\section{Algebraic numbers}\label{SEC:algnos}%
\index{algebraic! number|(}\index{number! algebraic|(} 
Before turning to general D-finite numbers, 
let us consider the subclass of algebraic functions.  
We will show that in this case, the possible limits are
precisely the algebraic numbers. For the purpose of this chapter, let
us say that a sequence $(a_n)_{n=0}^\infty\in\set F^{\set N}$ is
\emph{algebraic}\index{algebraic! sequence}\index{sequence! algebraic}
over $\set F$ if the corresponding power series $\sum_{n=0}^\infty
a_nz^n\in\set F[[z]]$ is algebraic in the sense of
Definition~\ref{DEF:dfinite}. Since algebraic functions are D-finite
(Abel's theorem), it is clear that algebraic
sequences are P-recursive.  We will write $\cA_{\set F}$ for the set
of all numbers $\xi\in\set C$ which are limits of convergent algebraic
sequences over~$\set F$.

Recall~\cite{FlSe2009} that two sequences $(a_n)_{n=0}^\infty$,
$(b_n)_{n=0}^\infty$ are called \emph{asymptotically
  equivalent},\index{asymptotically equivalent} written $a_n\sim b_n$
($n\to\infty$), if the quotient $a_n/b_n$ converges to $1$ as
$n\to\infty$. Similarly, two complex functions $f(z)$ and $g(z)$ are
called {\em asymptotically equivalent} at a point~$\zeta\in\set C$,
written $f(z)\sim g(z)$ ($z\to\zeta$), if the quotient $f(z)/g(z)$
converges to~$1$ as $z$ approaches~$\zeta$. These notions are
connected by the following classical theorem.

\begin{theorem}\label{THM:asympt}\leavevmode\null
  \begin{enumerate}
  \item\label{THM:transfer} (Transfer
    theorem~\cite{FlOd1990,FlSe2009})\index{Transfer theorem} For
    every $\alpha\in\set C\setminus\set Z_{\leq 0}$ we have
    \[
    [z^n]\frac1{(1-z)^\alpha}\sim \frac{n^{\alpha-1}}{\Gamma(\alpha)}
    \quad (n\to\infty),
    \]
    where $\Gamma(z)$ stands for the Gamma function and the notation
    $[z^n]f(z)$ refers to the coefficient of $z^n$ in the power series
    $f(z) \in \bF[[z]]$.
		
    \smallskip
  \item\label{THM:abelian} (Basic Abelian
    theorem~\cite{FGS2004})\index{Basic Abelian theorem} Let
    $(a_n)_{n=0}^\infty\in \bF^\bN$ be a sequence that satisfies the
    asymptotic estimate
    \[
    a_n\sim n^{\alpha}\quad (n\rightarrow\infty),
    \]
    where $\alpha \geq 0$.  Then the generating function $f(z) =
    \sum_{n=0}^\infty a_n z^n$ satisfies the asymptotic estimate
    \[
    f(z) \sim \frac{\Gamma(\alpha + 1)}{(1-z)^{\alpha+1}} \quad
    (z\rightarrow 1^{-}).
    \]
    This estimate remains valid when $z$ tends to $1$ in any sector
    with vertex at~$1$ symmetric about the horizontal axis, and with
    opening angle less than~$\pi$.
  \end{enumerate}
\end{theorem}

To show that $\cA_\bF$ is in fact a field, we need the following
lemma.  It indicates that depending on whether $\set F$ is a real
field or not, every real algebraic number or every algebraic number
can appear as a limit.
\begin{lemma}\label{LEM:minpoly}
  Let $p(z)\in \bF[z]$ be an irreducible polynomial of degree $d$.
  Then there is a square-free\index{square-free} polynomial $P(z,y)\in
  \bF[z,y]$ of degree~$d$ in~$y$ and admitting~$d$ distinct analytic
  algebraic functions~$f(z)\in \bF[[z]]$ with $P(z,f(z))=0$ in a
  neighborhood of $0$ such that $1$ is the only dominant singularity
  of each $f$ and
  \begin{enumerate}
  \item\label{LEM:minpoly1} if $\bF\subseteq \bR$, then for each root
    $\xi\in \bar \bF\cap \bR$ of $p(z)$ there exists a solution $f(z)$
    of~$P(z,y)$ with~$\lim_{n\rightarrow\infty}[z^n]f(z)=\xi$;
		
    \smallskip
  \item\label{LEM:minpoly2} if $\bF\setminus\bR\neq \emptyset$, then
    for each root $\xi\in \bar \bF$ of~$p(z)$ there exists a solution
    $f(z)$ of~$P(z,y)$ with~$\lim_{n\rightarrow\infty}[z^n]f(z)=\xi$.
  \end{enumerate}
\end{lemma}
\begin{proof}
  The two assertions can be proved simultaneously as follows.
	
  Let $\varepsilon>0$ be such that any two (real or complex) roots of
  $p$ have a distance of more than~$\varepsilon$ to each other.  Such
  an $\varepsilon$ exists because $p$ is a polynomial, and polynomials
  have only finitely many roots.  The roots of a polynomial depend
  continuously on its coefficients.  Therefore there exists a real
  number~$\delta>0$ so that perturbing the coefficients by up to
  $\delta$ won't perturb the roots by more than~$\varepsilon/2$.  Any
  positive smaller number than $\delta$ will have the same property.
  By the choice of~$\varepsilon$, any such perturbation of the
  polynomial will have exactly one (real or complex) root in each of
  the balls of radius $\varepsilon/2$ entered at the roots of~$p$.
	
  Let $\xi$ be a root of $p$. If $\xi=0$, then $p(z) = z$. Letting
  $P(z,y) = y$ yields the assertions.  Now assume that $\xi \neq 0$.
  Let $m\in \bF$ be the maximal modulus of coefficients of~$p$.  Then
  $m\neq 0$ since $p$ is irreducible.  Therefore, we always can find a
  number $a_0\in \bF$ such that~$|a_0-\xi|< \delta/m$, with
  the~$\delta$ from above.  Indeed, we have the following case
  distinction.
	
  \noindent For part~\ref{LEM:minpoly1} where $\bF\subseteq
  \bR$, we only consider $\xi\in \bar \bF\cap\bR$. In this case, $\bF$
  is dense in~$\bR$ since $\bF\supseteq\bQ$. Hence such $a_0 \in
  \bF\subseteq \bR$ exists.
	
  \noindent For part~\ref{LEM:minpoly2} where
  $\bF\setminus\bR \neq \emptyset$, there exists a non-real complex
  number~$\alpha$ in $\bF$. Therefore, $\bQ(\alpha)$ is dense in
  $\bC$. Since $\bQ(\alpha)\subseteq \bF$, such $a_0 \in \bF$ is
  guaranteed by the density of $\bF$ in $\bC$.
	
  After finding $a_0\in\bF$ with $|a_0-\xi|< \delta/m$, for both
  cases, we have
  \[ |p(a_0)|= |p(a_0)-p(\xi)| \leq m |a_0-\xi|<\delta.\] 
  Replace this $\delta$ by $|p(a_0)|$ for such a choice of~$a_0$.  
  The remaining argument works for both cases.
	
  Consider the perturbation $\tilde p(y)=p(y)-p(a_0)(1-z)$.  For any
  $z\in[0,1]$,
  \[|{-}p(a_0)(1-z)|<|p(a_0)|=\delta.\] 
  Therefore, as $z$ moves from $0$ to~$1$, each root of $p(y)-p(a_0)$ 
  moves to the corresponding root of~$p(y)$, which belongs to the same ball.  
  In particular, the root $a_0$ of $\tilde p|_{z=0}$ will move to 
  the root~$\xi$ of $\tilde p|_{z=1}$.  Define
  \[P(z,y)= p((1-z) y) -p(a_0) (1-z)\in \bF[z,y].\]
	
  We claim that $P(z,y)$ determines an analytic algebraic
  function~$f(z)$ in~$\bF[[z]]$ with the dominant singularity $1$ and
  whose coefficient sequence converges to $\xi$.  To prove this, we
  make an ansatz
  \[f(z) = \sum_{n=0}^\infty a_nz^n,\] 
  where the $a_0$ is from above and $(a_n)_{n=1}^\infty$ are to
  determined.  Observe that for any $c(z) \in \bF[z]$, $c(z)/(1-z)$ is
  a root of $P(z,y)$ if and only if $c(z)$ is a root of~$\tilde p(y)$,
  so $f(z)$ admits the following Laurent expansion at $z = 1$,
  \[f(z) = \frac{\xi}{1-z} + \sum_{n=0}^\infty b_n (1-z)^n \quad
  \text{for} \ b_n \in \bC.\] 
  Hence $z = 1$ is a singularity of $f(z)$ as $\xi\neq 0$.
  \goodbreak
  The above argument also implies that $z=1$ is the only dominant
  singularity of~$f(z)$.  Indeed, note that $z=1$ is the only root of
  the leading coefficient of~$P(z,y)$ w.r.t.\ $y$, so the other
  singularities of $f(z)$ could only be branch points, i.e., roots of
  discriminant of $P(z,y)$ w.r.t.~$y$.  However, the choices
  of~$\varepsilon$ and $\delta$ make it impossible for~$f(z)$ to have
  branch points in the disk~$|z|\leq 1$, because in order to have a
  branch point, two roots of the polynomial~$P(z,y)$ w.r.t.~$y$ would
  need to touch each other, and we have ensured that they are always
  separated by more than~$\varepsilon$.  Consequently, $z=1$ is the
  dominant singularity of $f(z)$, which gives $a_n \sim \xi$
  as~$n\rightarrow \infty$ by part~\ref{THM:transfer} of
  Theorem~\ref{THM:asympt}.  Therefore $\lim_{n\rightarrow \infty} a_n
  = \xi$ since~$\xi \neq 0$.
	
  To complete the proof, it remains to show that the coefficients of
  $f(z)$ are indeed in $\bF$.  This is observed by plugging the ansatz
  of $f(z)$ into~$P(z,y)$ and comparing the coefficients of like
  powers of $z$ to zero.  Since $p(z)$ is irreducible and $\xi$ is
  arbitrary, one sees that $P(z,y)$ admits $d$ distinct analytic
  solutions in~$\bF[[z]]$ in a neighborhood of $0$.
\end{proof}
The following theorem clarifies the converse direction for algebraic
sequences. \index{algebraic! sequence}\index{sequence! algebraic} It
turns out that every element in $\cA_\bF$ is algebraic over $\bF$.

\begin{theorem}\label{THM:algnos}%
  \index{algebraic! function}\index{function! algebraic} 
  Let $\bF$ be a subfield of $\bC$.
  \begin{enumerate}
  \item\label{THM:algnos1} If $\bF\subseteq \bR$, then $\cA_\bF=\bar
    \bF\cap\bR$.
		
    \smallskip
  \item\label{THM:algnos2} If $\bF\setminus \bR \neq \emptyset$, then
    $\cA_\bF = \bar \bF$.
  \end{enumerate}
\end{theorem}
\begin{proof}
  \begin{enumerate}
  \item Let $\xi\in \bar\bF\cap \bR$.  Then there is an irreducible
    polynomial $p(z) \in\bF[z]$ such that $p(\xi) =0$.  By
    part~\ref{LEM:minpoly1} of Lemma~\ref{LEM:minpoly}, $\xi$ is in
    fact a limit of an algebraic sequence in~$\bF^\bN$, which implies
    $\xi \in \cA_\bF$.
    
    \smallskip
    To show the converse inclusion, we let $\xi \in \cA_\bF$.
    When $\xi = 0$, there is nothing to show. Assume that $\xi \neq
    0$.  Then there is an algebraic sequence $(a_n)_{n=0}^\infty \in
    \bF^\bN$ such that $\lim_{n\rightarrow \infty}a_n = \xi$.  Since
    $\xi \neq 0$, $a_n\sim \xi$ $(n\rightarrow \infty)$.
    
    \smallskip
    Let $f(z) = \sum_{n=0}^\infty a_n z^n$. Clearly $f(z)$ is an
    algebraic function over~$\bF$.  By part~\ref{THM:abelian} of
    Theorem~\ref{THM:asympt}, $f(z) \sim \xi/(1-z)$ $(z\rightarrow
    1^-)$, implying that $z=1$ is a simple pole of $f(z)$ and
    \[f(z) = \frac{\xi}{1-z} + \sum_{n=0}^\infty b_n(1-z)^n \quad
    \text{for} \ (b_n)_{n=0}^\infty \in \bC^\bN.\] 
    Setting $g(z) = f(z)(1-z)$ establishes that $g(z) = \xi +
    \sum_{n=0}^\infty b_n (1-z)^{n+1},$ and then $g(z)$ is analytic at
    $1$.  Sending $z$ to $1$ gives $g(1)=\xi$.  By closure properties,
    $g(z)$ is again an algebraic function over $\bF$.  Thus $\xi =
    g(1)\in \bar \bF\cap \bR$ as $\bF\subseteq \bR$.
		
    \medskip
  \item By part~\ref{LEM:minpoly2} of Lemma~\ref{LEM:minpoly} and a
    similar argument as above, we have $\cA_\bF = \bar\bF$.
  \end{enumerate}
\end{proof}

If we were to consider the class $\cC_{\set F}$ of limits of
convergent sequences in $\set F$ satisfying linear difference
equations with constant coefficients over~$\set F$, sometimes called
C-finite sequences\index{C-finite sequence}\index{sequence! C-finite},
then an argument analogous to the above proof would imply that
$\cC_{\set F}\subseteq\set F$, because the power series corresponding
to such sequences are rational functions, and the values of rational
functions over $\set F$ at points in~$\set F$ evidently gives values
in~$\set F$.  The converse direction $\set F\subseteq\cC_{\set F}$ is
trivial, so $\cC_{\set F}=\set F$.

\begin{corollary}\label{COR:qcase}
  If $\bF\subseteq \bR$, then $\bar \bF = \cA_{\bF(i)} = \cA_\bF[i]
  =\cA_\bF + i \cA_\bF$, where $i$ is the imaginary
  unit\index{imaginary unit}.
\end{corollary}
\begin{proof}
  Since $\cA_\bF$ is a ring and $i^2 = -1\in\bF\subseteq \cA_\bF$, we
  have $\cA_\bF[i] = \cA_\bF + i \cA_\bF$.  Since~$i\in \bar \bF$
  and~$\bF\subseteq \bR$, $\bar \bF$ is closed under complex
  conjugation and then
  \[\bar \bF = (\bar \bF\cap\bR) + i(\bar \bF\cap \bR)= \cA_{\bF} +
  i\cA_{\bF},\] 
  by part~\ref{THM:algnos1} of Theorem~\ref{THM:algnos}.
  It follows from part~\ref{THM:algnos2} of Theorem~\ref{THM:algnos}
  that $\cA_{\bF(i)}=\overbar{\bF(i)}$.  Since $\cA_{\bF}\subseteq
  \cA_{\bF(i)}$ and $i\in \cA_{\bF(i)}$, we have
  \[\bar \bF= \cA_\bF + i \cA_\bF\subseteq
  \cA_{\bF(i)}=\overbar{\bF(i)}=\bar \bF.\] 
  The assertion holds.
\end{proof}

The following lemma says that every element in $\bar \bF$ can be
represented as the value at~$1$ of an analytic algebraic function
vanishing at zero, provided that $\bF$ is dense in~$\bC$.  This will
be used in the next section to extend the evaluation domain.

\begin{lemma}\label{LEM:algseq}%
 \index{algebraic! function}\index{function! algebraic} 
  Let $\bF$ be a subfield of $\bC$ with $\bF\setminus \bR \neq \emptyset$. 
  Let $p(z)\in \bF[z]$ be an irreducible polynomial of degree $d$.  
  Assume that $\xi_1, \dots, \xi_d$ are all the (distinct) roots of $p$ in
  $\bar\bF$.  Then there is a square-free\index{square-free}
  polynomial $P(z,y)\in \bF[z,y]$ of degree~$d$ in~$y$ and admitting
  $d$ distinct analytic algebraic functions~$g_1(z), \dots, g_d(z)$
  with~$P(z,g_j(z))=0$ in a neighborhood of~$0$ such that all $g_j$'s
  are analytic in the disk $|z|\leq 1$ with $g_j(0)=0$ and, after
  reordering (if necessary), $g_j(1)=\xi_j$.
\end{lemma}
\begin{proof}
  By part~\ref{LEM:minpoly2} of Lemma~\ref{LEM:minpoly}, there exists
  a bivariate square-free polynomial $\tilde P (z,y)\in \bF[z,y]$ of
  degree~$d$ in~$y$ and admitting $d$ distinct analytic algebraic
  functions $f_1(z),\dots, f_d(z)$ with $P(z,f_j(z)) =0$ in a
  neighborhood of $0$ such that $1$ is the only dominant singularity
  of each $f_j(z)$ and, after reordering (if necessary),
  \[\lim_{n\rightarrow \infty}[z^n]f_j(z) = \xi_j, \quad j = 1, \dots
  d.\]
	
  If $\xi_j =0$ for some $j$ then $p(z) = z$.  Letting $P(z,y) =y$
  yields the assertion.  Otherwise all roots $\xi_1, \dots, \xi_d$ are
  nonzero, and thus $[z^n]f_j(z) \sim \xi_j\ (n\rightarrow \infty)$
  for each~$j $.  By part~\ref{THM:abelian} of
  Theorem~\ref{THM:asympt}, 
  \[f_j(z) \sim \frac{\xi_j}{1-z}\ (z\rightarrow 1^-),\]
  which implies that $z=1$ is a simple pole of each~$f_j$.  Let
  $g_j(z) = f_j(z)z(1-z)$.  Then $g_1(z), \dots, g_d(z)$ are distinct
  and each $g_j(z)\in \bF[[z]]$ is analytic for any $z$ in the disk $|z|\leq 1$.
  Moreover, $g_j(0)=0$ and $g_j(1)=\xi_j$.  By closure properties, $g_j(z)$
  is again algebraic over~$\bF$.  Define a square-free polynomial
  \[P(z,y) = \prod_{j=1}^d(y-g_j(z)) = \prod_{j=1}^d \left(y-f_j(z)
    z(1-z)\right)\in \overbar{\bF(z)}[y].\] 
  Then $P\in\bF[z,y]$ since $P$ is symmetric in $f_1, \dots, f_d$. The lemma follows.
\end{proof}
\index{algebraic! number|)}\index{number! algebraic|)}

\section{Rings of D-finite numbers}\label{SEC:dfinitenos}%
\index{D-finite! number|(} 
Let us now return to the study of D-finite numbers.  Let $R$ be a
subring of~$\bC$ and $\bF$ be a subfield of $\bC$.  Recall that by
Definition~\ref{DEF:dfinitenos}, the elements of $\cD_{R,\bF}$ are
exactly limits of convergent sequences in $R^\bN$ which are
P-recursive over~$\bF$.  Some facts about P-recursive sequences
translate directly into facts about $\cD_{R,\bF}$.

\begin{prop}\label{PROP:property}\leavevmode\null
  \begin{enumerate}
  \item\label{PROP:0} $R\subseteq\cD_{R,\bF}$ and $\cA_{\set
      F}\subseteq\cD_{\set F}$.
		
    \smallskip
  \item\label{PROP:1} If $R_1\subseteq R_2$ then $\cD_{R_1,\set
      F}\subseteq\cD_{R_2,\set F}$, and if $\set F\subseteq\set E$
    then $\cD_{R,\set F}\subseteq\cD_{R,\set E}$.
		
    \smallskip
  \item\label{PROP:2} $\cD_{R,\set F}$ is a subring of~$\set C$.
    Moreover, if $R$ is an $\set F$-algebra then so is $\cD_{R,\set
      F}$.
		
    \smallskip
  \item\label{PROP:3} If $\set E$ is an algebraic extension
    field\index{algebraic! extension} of~$\set F$, then $\cD_{R,\set
      F}=\cD_{R,\set E}$.
		
    \smallskip
  \item\label{PROP:4} If $R\subseteq\set F$, then $\cD_{R,\set
      F}=\cD_{R,\quot(R)}$.
		
    \smallskip
  \item\label{PROP:5} If $R$ and $\set F$ are closed under complex
    conjugation\index{complex conjugation}, then so is $\cD_{R,\bF}$.
		
    In this case, we have $ \cD_{R,\bF}\cap \bR=\cD_{R\cap
      \bR, \bF}$.
		
    Moreover, if the imaginary unit\index{imaginary unit}
    $i\in \cD_{R,\bF}$ then $\cD_{R,\bF} =\cD_{R\cap \bR, \bF} + i
    \cD_{R\cap \bR, \bF}$.
  \end{enumerate}
\end{prop}
\begin{proof}
  \begin{enumerate}
  \item The first inclusion is clear because every element of $R$ is
    the limit of a constant sequence, and every constant sequence is
    P-recursive.  The second inclusion follows from the fact that
    algebraic functions are D-finite, and the coefficient sequences of
    D-finite functions are P-recursive.
		
    \medskip
  \item Clear.
		
    \medskip
  \item Follows directly from the corresponding closure properties for
    P-recursive sequences.
		
    \medskip
  \item Follows directly from part~\ref{lemma:changeopseq1} of
    Lemma~\ref{lemma:changeopseq}.
		
    \medskip
  \item Follows directly from part~\ref{lemma:changeopseq2} of
    Lemma~\ref{lemma:changeopseq}.
		
    \medskip
  \item For any convergent sequence $(a_n)_{n=0}^\infty\in R^\bN$, we
    have
    \[
    \real\left(\lim_{n\rightarrow \infty}a_n\right)=\lim_{n\rightarrow
      \infty}\real(a_n), \quad \imag\left(\lim_{n\rightarrow
        \infty}a_n\right)=\lim_{n\rightarrow \infty}\imag(a_n),
    \]
    and thus $\overbar{\lim_{n\rightarrow \infty}a_n}=
    \lim_{n\rightarrow \infty}\bar{a}_n.$ Hence the first assertion
    follows by $(\bar a_n)_{n=0}^\infty\in R^\bN$ and
    part~\ref{lemma:changeopseq3} of Lemma~\ref{lemma:changeopseq}.
		
    \smallskip
    Since $R$ is closed under complex conjugation,
    $(\real(a_n) )_{n=0}^\infty\in (R\cap \bR)^\bN$.  Then the
    inclusion $\cD_{R,\bF} \cap \bR\subseteq \cD_{R \cap \bR, \bF}$
    can be shown similarly as the first assertion. The converse
    direction holds by part~\ref{PROP:1}.  Thus $ \cD_{R,\bF}\cap
    \bR=\cD_{R\cap \bR, \bF}$.
    
    \smallskip
    If $i \in \cD_{R,\bF}$, then $\cD_{R\cap \bR,\bF}
    + i \cD_{R\cap \bR,\bF}\subseteq \cD_{R,\bF}$ since $\cD_{R\cap
      \bR,\bF} \subseteq \cD_{R,\bF}$.  To show the converse
    inclusion, let $\xi \in \cD_{R,\bF}$.  Then $\overbar{\xi}\in
    \cD_{R,\bF}$ by the first assertion. Since $i \in \cD_{R,\bF}$ and
    $R$ is closed under complex conjugation, $\real(\xi),\imag(\xi)$
    both belong to $\cD_{R,\bF}\cap\bR = \cD_{R\cap \bR,\bF}$ by the
    second assertion.  Therefore we have $\xi=\real(\xi) +
    i\imag(\xi)\in \cD_{R\cap \bR,\bF} + i \cD_{R\cap \bR,\bF}$.
  \end{enumerate}
\end{proof}

\begin{example}\leavevmode\null
  \begin{enumerate}
  \item We have $\cD_{\set Q(\sqrt2),\set Q(\pi,\sqrt2)}= \cD_{\set
      Q(\sqrt2),\set Q(\sqrt2)}=\cD_{\set Q(\sqrt2),\set Q}$.  The
    first identity holds by part~\ref{PROP:4}, the second by
    part~\ref{PROP:3} of the proposition.
		
    \smallskip
  \item We have $\cD_{\bar{\set Q},\set Q}=\cD_{\bar{\set Q},\set R}$.
    The inclusion ``$\subseteq$'' is clear by part~\ref{PROP:1}.  For
    the inclusion~``$\supseteq$'', let~$\xi\in\cD_{\bar{\set Q},\set
      R}$.  Then $\xi=a+ib$ for some $a,b\in\set R$, and there exists
    a sequence $(a_n+ib_n)_{n=0}^\infty$ in~$\bar{\set Q}^\bN$ and a
    nonzero operator $L\in\set R[n]\<S_n>$ such that $L\cdot
    (a_n+ib_n)=0$ and $\lim_{n\to\infty}(a_n+ib_n)=a+ib$.  Since the
    coefficients of~$L$ are real, we then have $L\cdot a_n=0$ and
    $L\cdot b_n=0$.  Furthermore, we see that
    $\lim_{n\to\infty}a_n=a$ and $\lim_{n\to\infty}b_n=b$.  Therefore,
    \[
    a,b\in\cD_{\bar{\set Q}\cap\set R,\set R}
    \stackrel{\text{part~\ref{PROP:4}}}=\cD_{\bar{\set Q}\cap\set
      R,\bar{\set Q}\cap\set R}
    \stackrel{\text{part~\ref{PROP:3}}}=\cD_{\bar{\set Q}\cap\set
      R,\set Q},
    \]
    which implies $a+ib\in\cD_{\bar{\set Q}\cap\set R,\set Q}+
    i\cD_{\bar{\set Q}\cap\set R,\set
      Q}\stackrel{\text{part~\ref{PROP:5}}}=\cD_{\bar{\set Q},\set
      Q}$, as claimed.
  \end{enumerate}
\end{example}

Lemma~\ref{LEM:algseq} motivates the following theorem, which says
that every D-finite number is essentially the value at~$1$ of an
analytic D-finite function, and thus a holonomic constant.\index{holonomic constant}

\begin{theorem}\label{THM:eval}%
\index{D-finite! function}\index{function! D-finite}
  Let $R$ be a subring of $\bC$ and $\bF$ be a subfield of $\bC$.
  Then for any~$\xi \in \cD_{R,\bF}$, there exists $g(z) \in R[[z]]$
  D-finite over $\bF$ and analytic at $1$ such that $\xi = g(1)$.
\end{theorem}
\begin{proof}
  The statement is clear when $\xi = 0$. Assume that $\xi$ is nonzero.
  Then there exists a sequence $(a_n)_{n=0}^\infty\in R^\bN$,
  P-recursive over $\bF$, such that $\lim_{n\rightarrow \infty} a_n =
  \xi$.  Since~$\xi$ is nonzero, we have $a_n \sim \xi $ $(n
  \rightarrow \infty)$.  Let $f(z)=\sum_{n=0}^\infty a_nz^n$.  Then by
  Theorem~\ref{THM:asympt}, we see that
  \[f(z)\sim \frac{\xi}{1-z}\ (z\rightarrow 1^{-}),\]
  which implies that $z=1$ is a simple pole
  of~$f(z)$. Let~$g(z)=f(z)(1-z)$.  Then $g(z)$ belongs to $R[[z]]$
  and is analytic at $z=1$. Write
  \[
  f(z) = \frac{\xi}{1-z} + \sum_{n=0}^\infty b_n (1-z)^n \quad
  \text{with} \ b_n \in \bC.
  \]
  It follows that
  $
  g(z)=f(z)(1-z) = \xi + \sum_{n=0}^\infty b_n(1-z)^{n+1},
  $
  which gives $\xi = g(1)$.  The assertion follows by noticing that
  $g(z)$ is D-finite over~$\bF$ due to closure properties.
\end{proof}

\begin{example}\label{EX:evaluation}
  We have $\zeta(3)=\sum_{n=1}^\infty\frac1{n^3}=\Li_3(1)$, where
  $\Li_3(z)=\sum_{n=1}^\infty\frac{1}{n^3}z^n$ is the
  polylogarithm function, D-finite over~$\set Q$ and analytic
  at~$1$.
\end{example}

Note that the above theorem implies that D-finite numbers are
computable when the ring $R$ and the field $\set F$ consist of
computable numbers\index{computable}\index{number! computable}. This
allows the construction of (artificial) numbers that are not D-finite.

Some kind of converse of Theorem~\ref{THM:eval} can be proved for the
case when $\bF$ is not a subfield of $\bR$, namely $\bF\setminus\bR
\neq \emptyset$.  Note that this condition is equivalent to saying
that $\bF$ is dense in~$\bC$.  To this end, we first need to develop
several lemmas.

The following lemma says that the value of a D-finite function at any
non-singular point in~$\bar{\set F}$ can be represented by the value
at~$1$ of another D-finite function.%
\index{D-finite! function}\index{function! D-finite}
\begin{lemma}\label{LEM:evalFbar}
  Let $\bF$ be a subfield of~$\bC$ with $\bF\setminus \bR \neq
  \emptyset$ and $R$ be a subring of~$\bC$ containing~$\bF$.  Assume
  that $f(z)\in \cD_{R,\bF}[[z]]$ is analytic and annihilated by a
  nonzero operator $L \in \bF[z]\langle D_z\rangle$ with zero an
  ordinary point.  Then for any non-singular point $\zeta \in
  {\bar\bF}$ of $L$, there exists an analytic function $h(z)\in
  \cD_{R,\bF}[[z]]$ and a nonzero operator $M\in \bF[z]\langle
  D_z\rangle$ with $0$ and $1$ ordinary points such that $M\cdot h(z)
  = 0$ and~$f(\zeta)=h(1)$.
\end{lemma}
\begin{proof}
  Let $\zeta \in \bar\bF$ be a non-singular point of~$L$.  Then there
  exists an irreducible polynomial $p(z)\in \bF[z]$ such that
  $p(\zeta)=0$.  Let~$\zeta_1=\zeta, \dots, \zeta_d$ be all the roots
  of~$p$ in~$\bar \bF$.  By Lemma~\ref{LEM:algseq}, there exists a
  square-free polynomial $P(z,y)\in \bF[z,y]$ of degree $d$ in~$y$ and
  admitting $d$ distinct analytic algebraic functions $g_1(z), \dots,
  g_d(z)$ with $P(z,g_j(z))=0$ in a neighborhood of~$0$.  Moreover,
  $g_1(z), \dots, g_d(z)$ are all analytic in the disk $|z|\leq 1$
  with $g_j(1)=\zeta_j$ and $g_j(0)=0$.
	
  Since $g_1(1) =\zeta$ is not a singularity of $L$ by assumption,
  none of $g_j(1)=\zeta_j$ is a singularity of~$L$.  Suppose otherwise
  that for some $2\leq \ell \leq d$, the point $g_\ell(1) = \zeta_\ell$ is a
  root of~$\lc(L)$.  Since $\lc(L)\in \bF[z]$ and $p$ is the
  minimal polynomial of~$\zeta_\ell$ over~$\bF$, 
  we know that $p$ divides $\lc(L)$ over~$\bF$.  
  Thus $\zeta$ is also a root of $\lc(L)$, a contradiction.
	
  Note that $g_1(z),\dots, g_d(z)$ are analytic in the disk $|z|\leq
  1$ and $g_j(0)=0$.  By Theorem~\ref{THM:algsubs}, there exists a
  nonzero operator $M\in \bF[z]\langle D_z\rangle$ with $M\cdot(f\circ
  g_1) =0$ which does not have $0$ or $1$ among its singularities.
  By part~\ref{PROP:0} of Proposition~\ref{PROP:property},
  $\bF\subseteq R\subseteq \cD_{R,\bF}$.  Since $f(z) \in
  \cD_{R,\bF}[[z]]$ and~$g_1(z)\in \bF[[z]]$ with $g_1(0)=0$, we
  have~$f(g_1(z))\in \cD_{R,\bF}[[z]]$.  Setting $h(z) = f(g_1(z))$
  completes the proof.
\end{proof}

With the above lemma, it suffices to consider the case when the
evaluation point is in~$R\cap \bF$.  This is exactly what the next two
lemmas are concerned about.
\begin{lemma}\label{LEM:coeffRinsideconverg}
  Assume that $f(z) = \sum_{n=0}^\infty a_n z^n \in R[[z]]$ is
  D-finite over $\bF$ and convergent in some neighborhood of~$0$.  Let
  $\zeta \in R\cap\bF$ be in the disk of convergence.  Then
  $f^{(k)}(\zeta) \in\cD_{R,\bF}$ for all~$k \in \bN$.
\end{lemma}
\begin{proof}
  For $k\in \bN$, it is well-known that $f^{(k)}(z)\in R[[z]]$ is also
  D-finite and has the same radius of convergence at zero as $f(z)$.
  Note that since $f(z)$ is D-finite over~$\bF$, so is $f^{(k)}(z)$.
  Thus to prove the lemma, it suffices to show the case when~$k=0$,
  i.e., $f(\zeta)\in \cD_{R,\bF}$.
	
  Since $f(z)$ is D-finite over $\bF$, the coefficient sequence
  $(a_n)_{n=0}^\infty$ is P-recursive over~$\bF$.  Note that $\zeta
  \in R\cap\bF$ is in the disk of convergence of $f(z)$ at zero, so
  \[
  f(\zeta) = \sum_{n=0}^\infty a_n \zeta^n = \lim_{n\rightarrow
    \infty} \sum_{\ell = 0}^n a_\ell \zeta^\ell.
  \]
  Since $(\zeta^n)_{n=0}^\infty$ is P-recursive over~$\bF$, the
  assertion follows by noticing that the sequence $(\sum_{\ell = 0}^n
  a_\ell \zeta^\ell)_{n=0}^\infty\in R^\bN$ is P-recursive over~$\bF$
  due to closure properties.
\end{proof}
\begin{example}
  Since $\exp(z)=\sum_{n=0}^\infty\frac1{n!}z^n\in\set Q[[z]]$ is
  D-finite over~$\set Q$, and converges everywhere, we get from the
  lemma that the numbers $\e, 1/\e, \sqrt{\e}$ belong to~$\cD_{\set
    Q,\set Q}$.  More precisely, since we are currently only
  considering non-real fields~$\set F$, we could say that the function
  $\exp(z)$ is D-finite over $\bar{\set Q}$, therefore
  $\e,1/\e,\sqrt{\e}$ all belong to $\cD_{\set Q,\bar{\set Q}}$, but
  by Proposition~\ref{PROP:property}, $\cD_{\set Q,\bar{\set Q}}=\cD_{\set Q,\set Q}$.
\end{example}

\begin{lemma}\label{LEM:coeffDinsideconverg}
  Let $R$ be a subring of $\bC$ containing $\bF$.  Let $f(z) =
  \sum_{n=0}^\infty a_nz^n$ in~$\cD_{R,\bF}[[z]]$ be an analytic
  function.  Assume that there exists a nonzero operator $L \in
  \bF[z]\langle D_z\rangle$ with zero an ordinary point such that
  $L\cdot f(z) =0$.  Let~$r>0$ be the smallest modulus of roots of
  $\lc(L)$ and let $\zeta \in\bF$ with~$|\zeta|<r$.  Then
  $f^{(k)}(\zeta) \in \cD_{R,\bF}$ for all $k \in \bN$.
\end{lemma}
\begin{proof}
  Let $\rho$ be the order of $L$. Since zero is an ordinary point of~$L$, 
  there exist P-recursive sequences
  $(c_n^{(0)})_{n=0}^\infty, \dots, (c_n^{(\rho-1)})_{n=0}^\infty$ in
  $\bF^\bN\subseteq R^\bN$ with~$c_j^{(m)}$ equal to the Kronecker
  delta $\delta_{mj}$ for $m, j= 0,\dots, \rho-1$, so that the set
  $\{\sum_{n=0}^\infty c_n^{(m)} z^n \}_{m=0}^{\rho -1}$ forms a basis
  of the solution space of~$L$ near zero.  Note that the
  singularities of solutions of~$L$ can only be roots of $\lc(L)$.
  Hence the power series $f(z) = \sum_{n=0}^\infty a_n z^n$ as well as
  $\sum_{n=0}^\infty c_n^{(m)} z^n$ for $m = 0, \dots, \rho-1$ are
  convergent in the disk $|z|<r$.  It follows from $|\zeta|<r$ 
  and Lemma~\ref{LEM:coeffRinsideconverg} that the set
  $\{\sum_{n=0}^\infty c_n^{(m)} \zeta^n\}_{m=0}^{\rho -1}$ belongs to~$\cD_{R,\bF}$. 
  Since $a_0,\dots,a_{\rho-1}\in \cD_{R,\bF}$,
  \[f(\zeta)=\sum_{n=0}^\infty a_n \zeta^n = a_0 \sum_{n=0}^\infty
  c_n^{(0)} \zeta^n + \dots + a_{\rho-1}\sum_{n=0}^\infty
  c_n^{(\rho-1)}\zeta^n\] 
  is D-finite by closure properties.  
  In the same vein, we find that for $k>0$, the
  derivative $f^{(k)}(\zeta)$ also belongs to $\cD_{R,\bF}$.
\end{proof}

\goodbreak
\begin{example}\leavevmode\null\label{EX:inside}
  \begin{enumerate}
  \item
    We know from Proposition~\ref{PROP:property} that $\sqrt{2}\in\cD_{\set Q}$.
    The series
    \[
    (z+1)^{\sqrt2}=1 + \sqrt2z+(1-\frac1{\sqrt2})z^2 + \cdots\in\set Q(\sqrt2)[[z]]\subseteq\cD_{\set Q}[[z]]
    \]
    is D-finite over~$\set Q$, an annihilating operator is $(z+1)^2D_z^2+(z+1)D_z-2$.
    Here we have the radius $r=1$. Taking $\zeta=\sqrt2-1$, the lemma implies 
    that~$\sqrt2^{\sqrt2}$ belongs to $\cD_{\set Q}$. 

    \smallskip
  \item\label{EX:bessel} 
    Observe that the lemma refers to the singularities of the operator rather than to the
    singularities of the particular solution at hand. For example, it does not imply
    that $J_1(1)\in\cD_{\set Q,\set Q}$, where $J_1(z)$ is the first Bessel function,
    because its annihilating operator is $z^2D_z^2+zD_z+(z^2-1)$, which has a singularity at~$0$.
    It is not sufficient that the particular solution $J_1(z)\in\set Q[[z]]$ is analytic at~$0$.
    Of course, in this particular example we see from the series representation
    $J_1(1)=\frac12\sum_{n=0}^\infty \frac{(-1/4)^n}{(n+1)n!^2}$ that the value belongs to~$\cD_{\set Q,\set Q}$. 

    \smallskip
  \item
    The hypergeometric function $f(z):={_2F_1}(\frac13,\frac12,1,z+\frac12)$ can be viewed as an element of
    $\cD_{\set Q,\set Q}[[z]]$:
    \begin{alignat*}1
    f(z) &=
      \underbrace{\sqrt[3]2\sum_{n=0}^\infty\frac{(1/3)_n(1/2)_n}{n!^2}(-1)^n}_{\in\cD_{\set Q}}
      + \underbrace{\frac{\sqrt[3]2}3\sum_{n=0}^\infty\frac{(1/2)_n(4/3)_n}{(2)_nn!}(-1)^n}_{\in\cD_{\set Q}} z\\[1ex]
      &\qquad{}
      + \underbrace{\frac{2\sqrt[3]2}3\sum_{n=0}^\infty\frac{(1/2)_n(7/3)_n}{(3)_nn!}(-1)^n}_{\in\cD_{\set Q}} z^2 + \cdots.
    \end{alignat*}
    The function $f$ is annihilated by the operator
    \[
      L = 3(2z-1)(2z+1)D_z^2 +(22z-1)D_z + 2.
    \]
    This operator has a singularity at $z=1/2$, and there is no
    annihilating operator of $f$ which does not have a singularity
    there. Although
    \[f(1/2)=\frac{\Gamma(1/6)}{\Gamma(1/2)\Gamma(2/3)}\]
    is a finite and specific value,
    the lemma does not imply that this value is a D-finite number.
  \end{enumerate}
\end{example}

\begin{theorem}\label{THM:coeffDoutsideconverg}
  Let $\bF$ be a subfield of~$\bC$ with $\bF\setminus \bR\neq
  \emptyset$ and let $R$ be a subring of~$\bC$ containing~$\bF$.
  Assume that $f(z) \in \cD_{R,\bF}[[z]]$ is analytic and there exists
  a nonzero operator $L \in \bF[z]\langle D_z\rangle$ with zero an
  ordinary point such that $L\cdot f(z) =0$.  Further assume
  that~$\zeta \in\bar \bF$ is not a singularity of~$L$.  Then
  $f^{(k)}(\zeta) $ belongs to $\cD_{R,\bF}$ for all~$k \in \bN$.
\end{theorem}
\begin{proof}
  By Lemma~\ref{LEM:evalFbar}, it suffices to show the assertion holds
  for $\zeta = 1$ (or more generally $\zeta \in \bF$).  Now assume
  that $\zeta \in \bF$.  We apply the method of analytic continuation.
  \index{analytic! continuation}
	
  Let $\mathcal{P}$ be a simple path with a finite cover
  $\bigcup_{j=0}^s \cB_{r_j}(\beta_j)$, where $s\in \bN$, $\beta_0=0$,
  $\beta_s = \zeta$, $\beta_j \in \bF$, $r_j>0$ is the distance
  between~$\beta_j$ and the zero set of~$\lc(L)$,
  and~$\cB_{r_j}(\beta_j)$ is the open circle centered at~$\beta_j$
  and with radius $r_j$.  Moreover, $\beta_{j+1}$ is inside
  $\cB_{r_j}(\beta_j)$ for each~$j$ (as illustrated by
  Figure~\ref{FIG:ac}).  Such a path exists because $\bF$ is dense in
  $\bC$ and the zero set of~$\lc(L)$ is finite.  Since the path $\mathcal{P}$
  avoids all roots of~$\lc(L)$, the function $f(z)$ is analytic along
  $\mathcal{P}$.  We next use induction on the index $j$ to show
  that~$f^{(k)}(\beta_j) \in \cD_{R,\bF}$ for all $k \in \bN$.
  \begin{figure}[t!]
    \centering
    \begin{tikzpicture}[scale=1.3]
	\draw[dashed,thick](0,0) circle (1); 
	\fill (0,0) circle (2pt) node[below right] {$\beta_0=0$}; 
	\fill (0.5,.75) circle (2pt) node[below right] {$\beta_1$}; 
	\draw[dashed,](0.5,0.75) circle (.7); 
	\fill (1,1) circle (2pt) node[below right] {$\beta_2$};
	\draw[dashed](1,1) circle (.65); 
	\fill (1.5,.915) circle (2pt) node[below right] {$\beta_3$}; 
	\fill (2.8,1.76) circle (2pt) node[below right] {$\beta_{s-2}$}; 
	\draw[dashed](2.8,1.76) circle (.6); 
	\fill (3,2.04) circle (2pt) node[below right] {$\beta_{s-1}$}; 
	\draw[dashed](3,2.04) circle (.6); 
	\fill (3.2,2.36) circle (2pt) node[below right] {$\beta_s =\zeta$};
	\node at (2.3,.8) {$\mathcal{P}$}; 
	\draw[->] (0,0)--(-.8,-.6);  
	\node at (-.6,-.3) {$r_0$}; 
	\draw[->] (.5,.75)--(-.2,.75); 
	\node at (.2,.55) {$r_1$}; 
	\draw[->] (1,1)--(1.3,1.57663); 
	\node at (1.4,1.3) {$r_2$}; 
	\draw[->] (2.8,1.76)--(2.2,1.76); 
	\node at (1.9,1.8) {$r_{s-1}$}; 
	\draw[->] (3,2.04)--(2.8,2.60569); 
	\node at (2.9,2.8) {$r_s$}; 
	\draw[thick, domain = 0:1.2] plot (\x,{1-(\x-1)^2}); 
	\draw[thick,domain=1.2:1.9] plot (\x,{1/2*(\x-1.5)^2+.915}); 
	\draw[thick,dashed,domain=1.9:2.5] plot (\x, {1/2*(\x-1.5)^2+.915}); 
	\draw[thick,domain=2.5:3.2] plot (\x, {1/2*(\x-1.5)^2+.915}); 
	\draw[thick] (-1.5,0)--(3,0);
	\draw[thick] (0,-1.5)--(0,3); 
	\draw plot[mark = x,mark options = {color = red,scale=1.3}] (1,0); 
	\draw plot[mark = x,mark options = {color = red,scale=1.3}] (-.8,.6); 
	\draw plot[mark = x,mark options = {color = red,scale=1.3}] (.530329,1.44934); 
	\draw plot[mark = x,mark options = {color = red,scale=1.3}] (-.7,-1);
	\draw plot[mark = x,mark options = {color = red,scale=1.3}] (1,-.6); 
	\draw plot[mark = x,mark options = {color = red,scale=1.3}] (1.5,.4); 
	\draw plot[mark = x,mark options = {color = red,scale=1.3}] (2.43226,2.23409); 
	\draw plot[mark = x,mark options = {color = red,scale=1.3}] (2.8,.9);
      \end{tikzpicture}
      \caption{a simple path $\mathcal{P} $ with a finite cover
        $\bigcup_{j=0}^s \cB_{r_j}(\beta_j)$
        ({\protect\tikz\protect\draw plot[mark = x,mark options =
          {color = red,scale=1.3}] (0,0);} stands for the roots of
        $\lc(L)$)}\label{FIG:ac}
   \end{figure}

   It is trivial when $j=0$ as $f^{(k)} (\beta_0) =f^{(k)} (0)\in
   \cD_{R,\bF}$ for $k\in \bN$ by assumption.  Assume now that $0<
   j\leq s$ and $f^{(k)}(\beta_{j-1})\in \cD_{R,\bF}$ for all $k \in
   \bN$.  We consider~$f(\beta_{j})$ and its derivatives.
	
   Recall that $r_{j-1}>0$ is the distance between $\beta_{j-1}$ and
   the zero set of $\lc(L)$.  Since~$f(z)$ is analytic
   at~$\beta_{j-1}$, it is representable by a convergent power series
   expansion
   \[f(z) = \sum_{n=0}^\infty
   \frac{f^{(n)}(\beta_{j-1})}{n!}(z-\beta_{j-1} )^n \quad \text{for
     all} \ |z-\beta_{j-1}|<r_{j-1}.\] 
   By the induction hypothesis, $f^{(n)}(\beta_{j-1})/n! \in
   \cD_{R,\bF}$ for all $n\in \bN$ and thus $f(z)$ belongs to
   $\cD_{R,\bF}[[z-\beta_{j-1}]]$.
	
   Let $Z = z-\beta_{j-1}$, i.e., $z = Z+\beta_{j-1}$.  Define $g(Z) =
   f(Z+\beta_{j-1})$ and~$\tilde{L}$ to be the operator obtained by
   replacing $z$ in $L$ by $Z+\beta_j$.  Since $\beta_{j-1}\in
   \bF\subseteq \cD_{R,\bF}$ and~$D_z = D_Z$, we have $g(Z)\in
   \cD_{R,\bF}[[Z]]$ and $\tilde{L} \in \bF[Z]\langle D_Z\rangle$.
   Note that~$L\cdot f(z) =0$ and $\beta_{j-1}$ is an ordinary point
   of~$L$ as $r_{j-1}>0$.  It follows that $\tilde{L} \cdot g(Z) = 0$
   and zero is an ordinary point of~$\tilde{L}$.  Moreover, we see
   that $r_{j-1}$ is now the smallest modulus of roots of
   $\lc(\tilde{L})$.  Since $|\beta_{j} - \beta_{j-1}| < r_{j-1}$,
   applying Lemma~\ref{LEM:coeffDinsideconverg} to $g(Z)$ yields
   \[f^{(k)}(\beta_{j}) =g^{(k)}(\beta_{j}-\beta_{j-1}) \in
   \cD_{R,\bF}\quad \text{for $k\in \bN$}.\] 
   Thus the assertion holds for $j = s$. The theorem follows.
\end{proof}

\begin{example}
  By the above theorem, $\exp(\sqrt2)$ and $\log(1+\sqrt3)$ both belong to~$\cD_{\set Q}$.
  We also have $\e^\pi \in \cD_{\set Q}$. This is because $\e^\pi = (-1)^{-i}$ 
  with $i$ the imaginary unit\index{imaginary unit},
  is equal to the value of the D-finite function $(z+1)^{-i}\in \set Q(i)[[z]]$ at $z=-2$
  (outside the radius of convergence; analytically continued in consistency with the usual
  branch cut conventions) and then $\e^\pi\in \cD_{\set Q(i)} \cap \set R = \cD_{\set Q}$.
  Furthermore, as remarked in the introduction, the numbers obtained by evaluating 
  a G-function at algebraic numbers which avoid the singularities of its 
  annihilating operator are in~$\cD_{\set Q(i)}$, because G-functions are D-finite.
\end{example}

\section{Open questions}\label{SEC:conclusion}\index{D-finite! number|)}%

We have introduced the notion of D-finite numbers and made some first
steps towards understanding their nature. We believe that, similarly
as for D-finite functions, the class is interesting because it has
good mathematical and computational properties and because it contains
many special numbers that are of independent interest. We conclude
this chapter with some possible directions of future research.

\smallskip\goodbreak\noindent
{\bf Evaluation at singularities.} \index{singularity}
While every singularity of a D-finite function must also be a
singularity of its annihilating operator, the converse is in general
not true.  We have seen above that evaluating a D-finite function at a
point which is not a singularity of its annihilating operator yields a
D-finite number.  It would be natural to wonder about the values of a
D-finite function at singularities of its annihilating operator,
including those at which the given function is not analytic but its
evaluation is finite.  Also, we always consider zero as an ordinary
point of the annihilating operator.  If this is not the case, the
method used in this chapter fails, as pointed out by part~\ref{EX:bessel}
of Example~\ref{EX:inside}.

\smallskip\goodbreak\noindent
{\bf Quotients of D-finite numbers.}%
\index{D-finite! number}\index{number! D-finite} The set of algebraic
numbers\index{algebraic! number}\index{number! algebraic} forms a
field, but we do not have a similar result for D-finite numbers. It is
known that the set of D-finite functions does not form a
field. Instead, Harris and Sibuya~\cite{HaSi1985} showed that a
D-finite function~$f$ admits a D-finite multiplicative inverse if and
only if~$f'/f$ is algebraic.  This explains for example why both $\e$
and $1/\e$ are D-finite, but it does not explain why both $\pi$ and
$1/\pi$ are D-finite. It would be interesting to know more precisely
under which circumstances the multiplicative inverse of a D-finite
number is D-finite. Is $1/\log(2)$ a D-finite number? Are there
choices of~$R$ and $\set F$ for which~$\cD_{R,\set F}$ is a field?

\smallskip\goodbreak\noindent
{\bf Roots of D-finite functions.}%
\index{D-finite! function}\index{function! D-finite} A similar pending analogy
concerns compositional inverses. We know that if $f$ is an algebraic
function, then so is its compositional inverse~$f^{-1}$. The analogous
statement for D-finite functions is not true. Nevertheless, it could
still be true that the values of compositional inverses of D-finite
functions are D-finite numbers, although this seems somewhat
unlikely. A particularly interesting special case is the question
whether (or under which circumstances) the roots of a D-finite
function are D-finite numbers.

\smallskip\goodbreak\noindent
{\bf Evaluation at D-finite number arguments.}%
\index{D-finite! number}\index{number! D-finite} We see
that the class $\cC_\bF$ of limits of convergent C-finite
sequences\index{C-finite sequence}\index{sequence! C-finite} is the
same as the values of rational functions at points in~$\bF$, namely
the field~$\bF$.  Similarly, the class $\cA_\bF$ of limits of
convergent algebraic sequences%
\index{algebraic! sequence}\index{sequence! algebraic} essentially consists of the
values of algebraic functions%
\index{algebraic! function}\index{function! algebraic} at points in~$\bar \bF$.
Continuing this pattern, is the value of a D-finite function at a
D-finite number again a D-finite number? If so, this would imply that
also numbers like $\e^{\e^{\e^{\e}}}$ are D-finite. Since $1/(1-z)$ is
a D-finite function, it would also imply that D-finite numbers form a
field.

\cleardoublepage
\phantomsection
\appendix
\renewcommand\chaptername{Appendix}
\numberwithin{equation}{section} 
\addcontentsline{toc}{part}{Appendices}
\begin{appendices}
	\pagestyle{fancy}
	\fancyhf{}
	\fancyhead[re]{\slshape\nouppercase{\leftmark}}
	\fancyhead[lo]{\slshape\nouppercase{\leftmark}}
	\fancyhead[le,ro]{\thepage}

\chapter{The \textsf{ShiftReductionCT} Package}
\label{APP:guide}\index{ShiftReductionCT@\textbf{ShiftReductionCT}|(}
	
In order to be able to experiment with the algorithms proposed in the
first part of this thesis, we have implemented all of them and
encapsulated the procedures as a Maple package, namely the
\textbf{ShiftReductionCT} package. This package was developed for {\sc
  Maple~18}\index{Maple@{\sc Maple}} and it is available upon request
from the author.  Here is a description of the package.

\bigskip
\begin{maplegroup}
  \begin{mapleinput}
    \mapleinline{active}{1d}{eval(ShiftReductionCT); }{%
    }
  \end{mapleinput}
	
  \mapleresult
  \begin{maplelatex}
    \mapleinline{inert}{2d}{module(~)}{%
      \hspace{0pt} {\bf module}(~) } \mapleinline{inert}{2d}{option\
      package;}{%
      \hspace{13pt} {\bf option}\ package; }
    \mapleinline{inert}{2d}{export \ ReductionCT, BoundReductionCT,}{%
      \hspace{13pt} {\bf export}\ $\mathit{ReductionCT},\,
      \mathit{BoundReductionCT},$ }
    \mapleinline{inert}{2d}{ModifiedAbramovPetkovsekReduction,
      ShiftMAPReduction, IsSummable,}{%
      \hspace{13pt}
      $\mathit{ModifiedAbramovPetkovsekReduction},\mathit{ShiftMAPReduction},\,
      \mathit{IsSummable},$ } \mapleinline{inert}{2d}{ShellReduction,
      KernelReduction, PolynomialReduction,}{%
      \hspace{13pt} $\mathit{ShellReduction},\,
      \mathit{KernelReductionCT}, \, \mathit{PolynomialReduction}, $ }
    \mapleinline{inert}{2d}{TranslateDRF, VerifyMAPReduction,
      VerifyRCT;}{%
      \hspace{13pt} $\mathit{TranslateDRF},\,
      \mathit{VerifyMAPReduction}, \, \mathit{VerifyRCT};$ }
    \mapleinline{inert}{2d}{description}{%
      \hspace{13pt} {\bf description} }
    \mapleinline{inert}{2d}{"Creative Telescoping for Bivariate
      Hypergeometric Terms via the Modified Abramov-Petkovsek
      Reduction";}{%
      \hspace{13pt} "Creative Telescoping for Bivariate Hypergeometric
      Terms via } \mapleinline{inert}{2d}{"Creative Telescoping for
      Bivariate Hypergeometric Terms via the Modified
      Abramov-Petkovsek Reduction";}{%
      \hspace{13pt} the Modified Abramov-Petkovsek Reduction"; }
    \mapleinline{inert}{2d}{end module}{%
      \hspace{0pt} {\bf end module} }
  \end{maplelatex}
\end{maplegroup}

\bigskip\noindent 
This appendix is intended to give a detailed instruction for the
package.  All export commands will be discussed in the order of their
first appearance in the thesis, but only some of them will be
emphasized particularly.  By applying them to some concrete examples,
we show the usage of the package as well as its applications. These
examples are chosen to take virtually no computation time.

The appendix contains a whole Maple session. The inputs are given
exactly in the way how the commands need to be used in Maple and
displayed in the type of Maple notation, while the outputs are
displayed in $2$-D math notation.  To start with, we load the package
in Maple.

\medskip
\begin{maplegroup}
  \begin{mapleinput}
    \mapleinline{active}{1d}{read(ShiftReductionCT):
      with(ShiftReductionCT): }{%
    }
  \end{mapleinput}
	
\end{maplegroup}

\bigskip\bigskip\goodbreak\noindent
{\bf Commands related to Chapter~\ref{CH:apreduction}}

\bigskip
We first consider univariate hypergeometric terms. Let $T$ be the
hypergeometric term in Example~\ref{EX:nonsummable} (or
Example~\ref{EX:mapnonsummable}).

\medskip
\begin{maplegroup}
  \begin{mapleinput}
    \mapleinline{active}{1d}{T:=k^2*k!/(k+1); }{%
    }
  \end{mapleinput}

  \mapleresult
  \begin{maplelatex}
    \mapleinline{inert}{2d}{T:=k^2*k!/(k+1)}{%
      \[
      T:= {\displaystyle \frac {k^2 k!}{ k+1}}
      \]
    }
  \end{maplelatex}
\end{maplegroup}

\medskip\noindent
By commands from the built-in Maple package
\textbf{SumTools[Hypergeometric]}, we find a kernel $K=k+1$ and its
corresponding shell $S=k^2/(k+1)$ of $T$.

The command \textsf{\shellred} performs Algorithm~\ref{ALG:shellred} and returns a decomposition of
the form \eqref{EQ:sap} for the shell~$S$ with respect to its
kernel~$K$.%
\index{ShiftReductionCT@\textbf{ShiftReductionCT}! ShellReduction@\textsf{ShellReduction}}

\medskip
\begin{maplegroup}
  \begin{mapleinput}
    \mapleinline{inert}{1d}{res:=ShellReduction(numer(K),denom(K),numer(S),denom(S),k);
    }{%
    }
  \end{mapleinput}
	
  \mapleresult
  \begin{maplelatex}
    \mapleinline{inert}{2d}{res:=[[-1/(k+1)],-1,k+2,k]}{%
      \[
      res:= {\displaystyle \left[\left[-\frac{1}{k+1}\right],\,-1,\,
          k+2,\, k\right]}
      \]
    }
  \end{maplelatex}
\end{maplegroup}

\medskip\noindent
Using the notations in \eqref{EQ:sap}, we check the correctness by

\medskip
\begin{maplegroup}
  \begin{mapleinput}
    \mapleinline{inert}{1d}{ S1:=add(res[1][i],i=1..nops(res[1])):
      a:=res[2]: b:=res[3]: p:=res[4]:
      normal(K*subs(y=y+1,S1)-S1+(a/b+p/denom(K))-S); }{%
    }
  \end{mapleinput}

  \mapleresult
  \begin{maplelatex}
    \mapleinline{inert}{2d}{0}{%
      \[
      0
      \]
    }
  \end{maplelatex}
\end{maplegroup}

\medskip
The command \textsf{PolynomialReduction},%
\index{ShiftReductionCT@\textbf{ShiftReductionCT}! PolynomialReduction@\textsf{PolynomialReduction}}
namely Algorithm~\ref{ALG:polyred}, projects a polynomial onto the
image space of the map for polynomial reduction with respect to a
shift-reduced rational function, and the standard complement of the
image space.

\medskip
\begin{maplegroup}
  \begin{mapleinput}
    \mapleinline{inert}{1d}{res:=PolynomialReduction(p,numer(K),denom(K),k);
    }{%
    }
  \end{mapleinput}
	
  \mapleresult
  \begin{maplelatex}
    \mapleinline{inert}{2d}{res:=[1], 0}{%
      \[
      res:={\displaystyle [1],\, 0}
      \]
    }
  \end{maplelatex}
\end{maplegroup}

\medskip\noindent
Using the notations in Algorithm~\ref{ALG:polyred}, we check the
correctness by

\medskip
\begin{maplegroup}
  \begin{mapleinput}
    \mapleinline{inert}{1d}{f:=add(res[1][i],i=1..nops(res[1])):
      q:=res[2]: normal(numer(K)*subs(k=k+1,f)-denom(K)*f+q-p); }{%
    }
  \end{mapleinput}

  \mapleresult
  \begin{maplelatex}
    \mapleinline{inert}{2d}{0}{%
      \[
      0
      \]
    }
  \end{maplelatex}
\end{maplegroup}

\medskip
The built-in Maple command \textsf{SumDecomposition}, which is in the
package \textbf{SumTools[Hypergeometric]}, is implemented based on the
\ap reduction. It computes a minimal additive decomposition described
in Section~\ref{SEC:apred} for a given hypergeometric term.%
\index{SumTools[Hypergeometric]@\textbf{SumTools[Hypergeometric]}! SumDecomposition@\textsf{SumDecomposition}}%
\index{SumDecomposition@\textsf{SumDecomposition}~{\em (SumTools[Hyper-\\geometric])}}

\medskip
\begin{maplegroup}
  \begin{mapleinput}
    \mapleinline{inert}{1d}{SumTools[Hypergeometric][SumDecomposition](T,k);}{%
    }
  \end{mapleinput}
	
  \mapleresult
  \begin{maplelatex}
    \mapleinline{inert}{2d}{[k*(Product(_k+1, _k=1 .. k-1))/(k+1),-(Product(_k+1, _k=1 .. k-1))/(k+2)]
    }{%
      \[
      \left[{\displaystyle \frac{k{\displaystyle
              \prod_{\_k=1}^{k-1}(\_k+1)}}{k+1},\,-\, \frac{
            {\displaystyle \prod_{\_k=1 }^{k-1}(\_k+1)}}{k+2}}\right]
      \]
    }
  \end{maplelatex}
\end{maplegroup}

\medskip To avoid solving any auxiliary recurrence equations
explicitly, we present a modified version of the \ap reduction, namely
Algorithm~\ref{ALG:mapred}, and implement it as the command
\textsf{\map}.  This command can be used in the following (default)
way.%
\index{ShiftReductionCT@\textbf{ShiftReductionCT}! ModifiedAbramovPetkovsekReduction@\textsf{ModifiedAbramovPetkovsekReduction}} %
\index{ModifiedAbramovPetkovsekReduction@\textsf{ModifiedAbramovPetkovsekReduction}\\{\em (ShiftReductionCT)}}%

\medskip
\begin{maplegroup}
  \begin{mapleinput}
    \mapleinline{inert}{1d}{res:=ModifiedAbramovPetkovsekReduction(T,k);
    }{%
    }
  \end{mapleinput}

  \mapleresult
  \begin{maplelatex}
    \mapleinline{inert}{2d}{res:=[[k+1,k^2/(k+1)], [k/(k+1),-1/(k+2)], k!]}{%
      \[
      res:= {\displaystyle
        \left[\left[\frac{k}{k+1},-\frac{1}{k+2}\right],\, k!\right]}
      \]
    }
  \end{maplelatex}
\end{maplegroup}
 
\medskip\noindent
Using the notations in Algorithm~\ref{ALG:mapred}, we have

\medskip
\begin{maplegroup}
  \begin{mapleinput}
    \mapleinline{inert}{1d}{f:=res[1][1]: r:=res[1][2]: H:=res[2]: }{%
    }
  \end{mapleinput}
\end{maplegroup}

\medskip\noindent
The package also provides the command \textsf{VerifyMAPReduction}
\index{ShiftReductionCT@\textbf{ShiftReductionCT}! VerifyMAPReduction@\textsf{VerifyMAPReduction}}%
to verify the reduction.  This command is used according to the
presented form of the result.  In the default case, we say

\medskip
\begin{maplegroup}
  \begin{mapleinput}
    \mapleinline{inert}{1d}{VerifyMAPReduction(res,T,k); }{%
    }
  \end{mapleinput}

  \mapleresult
  \begin{maplelatex}
    \mapleinline{inert}{2d}{true}{%
      \[
      \mathit{true}
      \]
    }
  \end{maplelatex}
\end{maplegroup}

\medskip\noindent
Moreover, we can change the outputs of \textsf{\map}\ by specifying
the third argument. For example, we would like to display the result
in terms of hypergeometric terms,

\medskip
\begin{maplegroup}
  \begin{mapleinput}
    \mapleinline{inert}{1d}{res:=ModifiedAbramovPetkovsekReduction(T,k,output=
      hypergeometric);
      VerifyMAPReduction(res,T,k,output=hypergeometric); }{%
    }
  \end{mapleinput}

  \mapleresult
  \begin{maplelatex}
    \mapleinline{inert}{2d}{res:=[k*k!/(k+1),-k!/(k+2)]}{%
      \[
      res:=\left[{\displaystyle \frac{k
            k!}{k+1},\,-\frac{k!}{k+2}}\right]
      \]
    }
  \end{maplelatex}
  \begin{maplelatex}
    \mapleinline{inert}{2d}{true}{%
      \[
      \mathit{true}
      \]
    }
  \end{maplelatex}
\end{maplegroup}

\medskip\noindent
or we can also perform it as a list of functions, which specifies the
standard form of the residual forms.

\medskip
\begin{maplegroup}
  \begin{mapleinput}
    \mapleinline{inert}{1d}{res:=ModifiedAbramovPetkovsekReduction(T,k,output=list);
      VerifyMAPReduction(res,T,k,output=list); }{%
    }
  \end{mapleinput}

  \mapleresult
  \begin{maplelatex}
    \mapleinline{inert}{2d}{res:=[[[-1/(k+1), 0, 1], [-1, k+2, 0]], k!]}{%
      \[
      res :=\left[\left[\left[{\displaystyle-\frac1{k+1}}, 0,
            1\right], \left[-1, k+2, 0\right]\right], k!\right]
      \]
    }
  \end{maplelatex}
\begin{maplelatex}
	\mapleinline{inert}{2d}{true}{%
		\[
		\mathit{true}
		\]
	}
\end{maplelatex}
\end{maplegroup}

\medskip
As mentioned in Section~\ref{SEC:timing}, we also implement a
procedure based on the modified \ap reduction, which is only used to
determine hypergeometric summability and performs in a similar way as
Gosper's algorithm, namely the command~\textsf{IsSummable}.%
\index{ShiftReductionCT@\textbf{ShiftReductionCT}! IsSummable@\textsf{IsSummable}} %
\index{IsSummable@\textsf{IsSummable}~{\em (ShiftReductionCT)}}%
	
\medskip
\begin{maplegroup}
  \begin{mapleinput}
    \mapleinline{inert}{1d}{IsSummable(T,k); }{%
    }
  \end{mapleinput}

  \mapleresult
  \begin{maplelatex}
    \mapleinline{inert}{2d}{false}{%
      \[
      \mathit{false}
      \]
    }
  \end{maplelatex}
\end{maplegroup}

\medskip
The built-in Maple command for Gosper's algorithm is \textsf{Gosper}
in the package \textbf{SumTools[Hypergeometric]}.%
 \index{SumTools[Hypergeometric]@\textbf{SumTools[Hypergeometric]}! Gosper@\textsf{Gosper}}
\index{Gosper@\textsf{Gosper}~{\em (SumTools[Hypergeometric])}}
\index{Gosper's algorithm}

\medskip
\begin{maplegroup}
  \begin{mapleinput}
    \mapleinline{active}{1d}{SumTools[Hypergeometric][Gosper](T,k);
    }{%
    }
  \end{mapleinput}
	
  \mapleresult
  \begin{maplelatex}
    \mapleinline{inert}{2d}{Error, (in
      SumTools:-Hypergeometric:-Gosper) no solution found }{%
      \hspace{-38pt}\underline{Error, (in
        SumTools:-Hypergeometric:-Gosper) no solution found}
    }
  \end{maplelatex}
\end{maplegroup}

\bigskip\bigskip\noindent
{\bf Commands related to Chapter~\ref{CH:rfproperties}}

\bigskip
In Chapter~\ref{CH:rfproperties}, we showed that the sum of two
residual forms is congruent to a residual form (see
Theorem~\ref{THM:rfsum}), which plays an important role in developing
the reduction-based creative telescoping algorithm for hypergeometric
terms (i.e., Algorithm~\ref{ALG:reductionct}).

To prove Theorem~\ref{THM:rfsum}, we introduced two congruences in
Lemma~\ref{LEM:kernelreduction}. These two congruences stand for two
types of kernel reduction in the shift case, that is, denominator type
and numerator type, respectively. We implemented them by the command
\textsf{KernelReduction}.  To call it in Maple, using the notations
from Lemma~\ref{LEM:kernelreduction}, one just says%
\index{ShiftReductionCT@\textbf{ShiftReductionCT}! KernelReduction@\textsf{KernelReduction}}%

\medskip
\begin{maplegroup}
  \begin{mapleinput}
    \mapleinline{inert}{1d}{KernelReduction(p1,numer(K),denom(K),m,k,type=denominator);
    }{%
    }
  \end{mapleinput}
\end{maplegroup}

\medskip\noindent
or

\medskip
\begin{maplegroup}
  \begin{mapleinput}
    \mapleinline{inert}{1d}{KernelReduction(p2,numer(K),denom(K),m,k,type=numerator);
    }{%
    }
  \end{mapleinput}
\end{maplegroup}

\medskip
The key idea of Algorithm~\ref{ALG:transferdrf} is to move the
significant denominator of a residual form to a required form 
according to a given residual form, so
that the resulting sum is again a residual form. This process was
implemented as the command \textsf{TranslateDRF}. We also provide a
command named \textsf{SignificantDenom} to extract the significant
denominator of a residual form.%
\index{ShiftReductionCT@\textbf{ShiftReductionCT}! TranslateDRF@\textsf{TranslateDRF}}%
\index{ShiftReductionCT@\textbf{ShiftReductionCT}! SignificantDenom@\textsf{SignificantDenom}}%

Now let's consider Example \ref{EX:rfsum}. For $K=1/k$ shift-reduced,
we have two residuals form w.r.t.~$K$: $ r= 1/(2k+1)$ and
$s=1/(2k+3)$.

\medskip
\begin{maplegroup}
  \begin{mapleinput}
    \mapleinline{inert}{1d}{K:=1/k: r:=1/(2*k+1): s:=1/(2*k+3): }{%
    }
  \end{mapleinput}
	
\end{maplegroup}

\medskip\noindent
One can compute a residual form of $r+s$ in terms of the significant
denominator of~$r$ by
 
\medskip
\begin{maplegroup}
  \begin{mapleinput}
    \mapleinline{inert}{1d}{res:=TranslateDRF(s, SignificantDenom(r,K,k), K, k);
      S1:=res[1]: a:=res[2][1]: b:=res[2][2]: q:=res[2][3]:
      new:=r+normal(a/b)+q/denom(K); # evaluate the sum
      normal(K*subs(k=k+1, S1)-S1+new-r-s); # check the result
    }{%
    }
  \end{mapleinput}
  
  \mapleresult
  \begin{maplelatex}
    \mapleinline{inert}{2d}{b:=k+1/2}{%
      \[
      b:={\displaystyle k+\frac12}
      \]
    }
  \end{maplelatex}
  \begin{maplelatex}
    \mapleinline{inert}{2d}{res:=[-3/(2*(2*k+1)),[-3/2, 2*k+1, 1/2]]}{%
      \[
      res:={\displaystyle \left[-\frac{3}{2(2k+1)},\,
          \left[-\frac34,\, k+\frac12,\, \frac12\right]\right]}
      \]
    }
  \end{maplelatex}
  \begin{maplelatex}
    \mapleinline{inert}{2d}{new:=-1/(2*(2*k+1))+1/(2*k)}{%
      \[
      new:={\displaystyle-\frac1{2(2k+1)}+\frac1{2k}}
      \]
    }
  \end{maplelatex}
  \begin{maplelatex}
    \mapleinline{inert}{2d}{0}{%
      \[
      0
      \]
    }
  \end{maplelatex}
\end{maplegroup}

\medskip\noindent
This confirms the result given in Example~\ref{EX:rfsum}.  Of course,
one can also compute a residual form of $r+s$ in terms of the
significant denominator of $s$,

\medskip
\begin{maplegroup}
  \begin{mapleinput}
    \mapleinline{inert}{1d}{b:=SignificantDenom(s,K,k);
      res:=TranslateDRF(r, b, K, k);
      new:=s+normal(res[2][1]/b)+res[2][2]/denom(K);
      normal(K*subs(k=k+1, res[1])-res[1]+new-r-s);
    }{%
    }
  \end{mapleinput}
  
  \mapleresult
  \begin{maplelatex}
    \mapleinline{inert}{2d}{res:=[-1/(2*k+1), [-2/3, 2*k+3, 1/3]]}{%
      \[
      res:={\displaystyle \left[-\frac{1}{2k+1},\, \left[-\frac13,\,
            k+\frac32,\, \frac13\right]\right]}
      \]
    }
  \end{maplelatex}
  \begin{maplelatex}
    \mapleinline{inert}{2d}{new:=1/(3*(2*k+3))+1/(3*k)}{%
      \[
      new:={\displaystyle \frac1{3(2k+3)}+\frac1{3k}}
      \]
    }
  \end{maplelatex}
  \begin{maplelatex}
    \mapleinline{inert}{2d}{0}{%
      \[
      0
      \]
    }
  \end{maplelatex}
\end{maplegroup}

\bigskip\bigskip\noindent
{\bf Commands related to Chapter~\ref{CH:telescoping}}

\bigskip
Now let's turn our attention to bivariate hypergeometric
terms. Consider the following hypergeometric term from
Example~\ref{EX:reductionct}.

\medskip
\begin{maplegroup}
  \begin{mapleinput}
    \mapleinline{active}{1d}{T:=binomial(n,k)^3; 
    }{%
    }
  \end{mapleinput}
  
  \mapleresult
  \begin{maplelatex}
    \mapleinline{inert}{2d}{T:=binomial(n,k)^3;}{%
      \[
      T:={\displaystyle \binomial(n,k)^3}
      \]
    }
  \end{maplelatex}
\end{maplegroup}

\medskip
Based on the modified \ap reduction, Algorithm~\ref{ALG:reductionct}
is implemented in the command \textsf{ReductionCT}, which (by default)
returns the (monic) minimal telescoper for a given hypergeometric
term.%
\index{ShiftReductionCT@\textbf{ShiftReductionCT}! ReductionCT@\textsf{ReductionCT}}%
\index{ReductionCT@\textsf{ReductionCT}~{\em (ShiftReductionCT)}}%

\medskip
\begin{maplegroup}
  \begin{mapleinput}
    \mapleinline{inert}{1d}{ReductionCT(T,n,k,Sn);
    }{%
    }
  \end{mapleinput}
  
  \mapleresult
  \begin{maplelatex}
    \mapleinline{inert}{2d}{-(8*(n^2+2*n+1))/(n^2+4*n+4)-(7*n^2+21*n+16)*Sn/(n^2+4*n+4)+Sn^2}{%
      \[
      {\displaystyle-\frac{8(n^2+2n+1)}{n^2+4n+4}-\frac{(7n^2+21n+16)Sn}{n^2+4n+4}+Sn^2}
      \]
    }
  \end{maplelatex}
\end{maplegroup}

\medskip\noindent
As illustrated by the following commands, if a fifth argument is
specified then the command also returns a corresponding certificate,
whose form depends on the specification.  More precisely, we get a
certificate as a list of a normalized rational function and a
hypergeometric term by saying

\medskip
\begin{maplegroup}
  \begin{mapleinput}
    \mapleinline{inert}{1d}{res:=ReductionCT(T,n,k,Sn,output=normalized); 
    }{%
    }
  \end{mapleinput}
  
  \mapleresult
  \begin{maplelatex}
    \mapleinline{inert}{2d}{[-(8*(n^2+2*n+1))/(n^2+4*n+4)-(7*n^2+21*n+16)*Sn/(n^2+4*n+4)+Sn^2,%
      [-k^3*(14*n^5-27*n^4*k+18*n^3*k^2-4*n^2*k^3+102*n^4-147*n^3*k+66*n^2*k^2-8*n*k^3+290*n^3%
      -291*n^2*k+78*n*k^2-4*k^3+402*n^2-249*n*k+30*k^2+272*n-78*k+72)/((n+1-k)^3*(n^2+4*n+4)*(n+2-k)^3),%
      binomial(n, k)^3]]}{%
      \[
      \hspace{28pt}
      \maplemultiline{res:=\left[{\displaystyle-\frac{8(n^2+2n+1)}{n^2+4n+4}-\frac{(7n^2+21n+16)Sn}{n^2+4n+4}+Sn^2},
          \, \vphantom{{\displaystyle-\frac{8(n^2+2n+1)}{n^2+4n+4}}}\right.\\
        \\
        \qquad
        {\displaystyle \left[\frac{1}{(-n-1+k)^3(n^2+4n+4)(-n-2+k)^3}\left(k^3(4k^3n^2-18k^2n^3+27kn^4\right.\right.}\\
        \\
        \qquad
        {\displaystyle\vphantom{{\displaystyle-\frac{8(n^2+2n+1)}{n^2+4n+4}}}
          -14n^5+8k^3n-66k^2n^2+147kn^3-102n^4+4k^3-78k^2n+291kn^2}\\
        \\
        \qquad
        {\displaystyle\left.\left.\left.\vphantom{{\displaystyle-\frac{8(n^2+2n+1)}{n^2+4n+4}}}
                -290n^3-30k^2+249kn-402n^2+78k-272n-72)\right),\,
              \binomial(n,k)^3\vphantom{{\displaystyle-\frac{8(n^2+2n+1)}{n^2+4n+4}}}
            \right]\right]} }
      \]
    }
  \end{maplelatex}
\end{maplegroup}

\medskip\noindent
or get one as a list of a linear combination of several simple
rational functions and a hypergeometric term by

\medskip
\begin{maplegroup}
  \begin{mapleinput}
    \mapleinline{inert}{1d}{res:=ReductionCT(T,n,k,Sn,output=unnormalized);
    }{%
    }
  \end{mapleinput}
  
  \mapleresult
  \begin{maplelatex}
    \mapleinline{inert}{2d}{res:=[-(8*(n^2+2*n+1))/(n^2+4*n+4)-(7*n^2+21*n+16)*Sn/(n^2+4*n+4)+Sn^2, %
      [(4*(n^2+2*n+1))/(n^2+4*n+4)+(7*n^2+21*n+16)*(n+1)^3/((n^2+4*n+4)*(n+1-k)^3)-%
      (n+1)^3*(10*n^2-15*n*k+6*k^2+25*n-18*k+16)/((n^2+4*n+4)*(n+1-k)^3)+%
      (11*n^5-12*n^4*k+12*n^3*k^2+62*n^4-48*n^3*k+36*n^2*k^2+140*n^3-72*n^2*k%
      +36*n*k^2+158*n^2-48*n*k+12*k^2+89*n-12*k+20)/((n^2+4*n+4)*(n+1-k)^3)%
      -((12*(n^3+3*n^2+3*n+1))*k^2/(n^2+4*n+4)-(12*(n^4+4*n^3+6*n^2+4*n+1))*k%
      /(n^2+4*n+4)+(11*n^5+62*n^4+140*n^3+158*n^2+89*n+20)/(n^2+4*n+4))%
      /(n+1-k)^3-(n+2)^3*(n+1)^3/((n+2-k)^3*(n+1-k)^3), binomial(n, k)^3]]}{%
      \[
      \hspace{5pt}
      \maplemultiline{res:=\left[{\displaystyle-\frac{8(n^2+2n+1)}{n^2+4n+4}
            -\frac{(7n^2+21n+16)Sn}{n^2+4n+4}+Sn^2},\, \left[{\displaystyle\frac{4(n^2+2n+1)}{n^2+4n+4}}
          \right.\right.\\
        \\
        \qquad
        {\displaystyle-\frac{(7n^2+21n+16)(n^3+3n^2+3n+1)}{(n^2+4n+4)(-n-1+k)^3}}\\
        \\
        \qquad
        {\displaystyle-\frac{(n+1)^3(6k^2+3 k n+n^2+6k+4n+4)}{(n^2+4n+4)(-n-1+k)^3}}\\
        \\
        \qquad
        +{\displaystyle\frac{1}{(n^2+4n+4)(-n-1+k)^3}(12k^2n^3-12kn^4+11n^5+36k^2n^2}\\
        \\
        \qquad
        {\displaystyle\vphantom{{\displaystyle-\frac{8(n^2+2n+1)}{n^2+4n+4}}}
          -48kn^3+62n^4+36k^2n-72kn^2+140n^3+12k^2-48kn}\\
        \\
        \qquad
        {\displaystyle\vphantom{{\displaystyle-\frac{8(n^2+2n+1)}{n^2+4n+4}}}
          +158n^2-12k+89n+20)}\\
        \\
        \qquad
        {\displaystyle-\frac{((n+1)^3+3(n+1)^2+3n+4)(n+1)^3}{(-n-2-k)^3(-n-1+k)^3},}\\
        \\
        \qquad {\displaystyle\left.\left.\binomial(n,
              k)^3\vphantom{{\displaystyle-\frac{8(n^2+2n+1)}{n^2+4n+4}}}
            \right]\right]} }
      \]
    }
  \end{maplelatex}
\end{maplegroup}

\medskip
%
%

The result returned by the command \textsf{ReductionCT} can be
verified by the command \textsf{VerifyRCT}.%
\index{ShiftReductionCT@\textbf{ShiftReductionCT}! VerifyRCT@\textsf{VerifyRCT}}%

\medskip
\begin{maplegroup}
  \begin{mapleinput}
    \mapleinline{active}{1d}{VerifyRCT(res,T,n,k,Sn);
    }{%
    }
  \end{mapleinput}
  
  \mapleresult
  \begin{maplelatex}
    \mapleinline{inert}{2d}{true}{%
      \[
      \mathit{true}
      \]
    }
  \end{maplelatex}
\end{maplegroup}

%
%

\medskip
Maple's implementation for Zeilberger's algorithm is the command
\textsf{Zeilberger}, which is also in the package
\textbf{SumTools[Hypergeometric]}.%
\index{SumTools[Hypergeometric]@\textbf{SumTools[Hypergeometric]}! Zeilberger@\textsf{Zeilberger}}\index{Zeilberger's algorithm}%
\index{Zeilberger@\textsf{Zeilberger}~{\em (SumTools[Hypergeometric])}}%

\medskip
\begin{maplegroup}
  \begin{mapleinput}
    \mapleinline{inert}{1d}{SumTools[Hypergeometric][Zeilberger](T,n,k,Sn);
    }{%
    }
  \end{mapleinput}
  
  \mapleresult
  \begin{maplelatex}
    \mapleinline{inert}{2d}{[Sn^2*(n^2+4*n+4)+(-7*n^2-21*n-16)*Sn-8*n^2-16*n-8, %
      (k^3+(-(9/2)*n-15/2)*k^2+((27/4)*n^2+(93/4)*n+39/2)*k-(7/2)*n^3-(37/2)*n^2-32*n-18)%
      *k^3*binomial(n, k)^3*(4*n^2+8*n+4)/((n+2-k)^3*(n+1-k)^3)]
    }{%
      \[
      \hspace{-55pt} \maplemultiline{ {\displaystyle
          \left[\vphantom{{\displaystyle-\frac{8(n^2+2n+1)}{n^2+4n+4}}}
            (n^2+4n+4)\, Sn^2+(-7n^2-21n-16)\, Sn-8n^2-16n-8,\, \right.}\\
        \\
        \\
        \qquad
        {\displaystyle\frac{1}{(-n-2+k)^3(-n-1+k)^3}\left(\left( k^3+\left(-\frac92n-\frac{15}{2}\right)k^2\right.\right.}\\
        \\
        \\
        \qquad
        {\displaystyle\left.+\left(\frac{27}{4}n^2+\frac{93}{4}n+\frac{39}{2}\right)k-\frac{7}{2}n^3-\frac{37}{2}n^2-32
            n-18\right)}
        \\
        \\
        \qquad {\displaystyle\left.\left.k^3\binomial(n,
              k)^3(4n^2+8n+4)\right)\vphantom{{\displaystyle-\frac{8(n^2+2n+1)}{n^2+4n+4}}}
          \right]} }
      \]
    }
  \end{maplelatex}
\end{maplegroup}

\medskip
In view of Remark~\ref{REM:simple}, we introduce the command
\textsf{ShiftMAPReduction}, which performs the same function as
applying \textsf{\map} with respect to~$k$ to the $m$-th shift
$\sigma_n^m(T)$ for a bivariate hypergeometric term $T(n,k)$ but in a
faster way as pointed out by the remark. Moreover, this command always
uses the same kernel independent of the value of $m$. Note that when
$m=0$ the command is the same as the command \textsf{\map}.%
\index{ShiftReductionCT@\textbf{ShiftReductionCT}! ShiftMAPReduction@\textsf{ShiftMAPReduction}}%

To illustrate this command, we consider the same hypergeometric term
$T$ as before.

\medskip
\begin{maplegroup}
  \begin{mapleinput}
    \mapleinline{active}{1d}{T:=binomial(n,k)^3:
    }{%
    }
  \end{mapleinput}
  
\end{maplegroup}

\medskip\noindent
Then it has a minimal additive decomposition

\medskip
\begin{maplegroup}
  \begin{mapleinput}
    \mapleinline{active}{1d}{
      ModifiedAbramovPetkovsekReduction(T,k);
    }{%
    }
  \end{mapleinput}
  
  \mapleresult
  \begin{maplelatex}
    \mapleinline{inert}{2d}{[[-1/2, (1/2)*(3*k^2*n-3*k*n^2+n^3+3*k^2+3*k+1)/(k+1)^3], 
      binomial(n, k)^3]}{%
      \[
      {\displaystyle\left[\left[-\frac12, \frac12\, \frac{3k^2n-3kn^2+n^3+3k^2+3k+1}{(k+1)^3}\right], 
          \binomial(n, k)^3\right]}
      \]
    }
  \end{maplelatex}  
\end{maplegroup}

\medskip\noindent
For the first shift of $T$ w.r.t.~$n$, we have

\medskip
\begin{maplegroup}
  \begin{mapleinput}
    \mapleinline{active}{1d}{ModifiedAbramovPetkovsekReduction(subs(n=n+1,T),k);
    }{%
    }
  \end{mapleinput}
  
  \mapleresult
  \begin{maplelatex}
    \mapleinline{inert}{2d}{[[-1/2, (1/2)*(3*k^2*n-3*k*n^2+n^3+6*k^2-6*k*n+3*n^2+3*n+2)/(k+1)^3], 
      binomial(n+1, k)^3]
    }{%
      \[\hspace{22pt}
      {\displaystyle\left[\left[-\frac12, \frac12\, \frac{3k^2n-3kn^2+n^3+6k^2-6kn+3n^2+3n+2}{(k+1)^3}\right], 
          \binomial(n+1, k)^3\right]}
      \]
    }
  \end{maplelatex}
\end{maplegroup}
%
%

\medskip\noindent
On the other hand, applying the command \textsf{ShiftMAPReduction} gives

\medskip
\begin{maplegroup}
	\begin{mapleinput}
		\mapleinline{active}{1d}{ShiftMAPReduction(T,n,k,1);
		}{%
	}
\end{mapleinput}

\mapleresult
\begin{maplelatex}
	\mapleinline{inert}{2d}{
		[[(n^3+3*n^2+3*n+1)/(-n-1+k)^3, (n^3+3*n^2+3*n+1)/(k+1)^3], binomial(n, k)^3]
	}{%
	\[\hspace{0pt}
	{\displaystyle\left[\left[\frac{n^3+3n^2+3n+1}{(-n-1+k)^3},\frac{n^3+3n^2+3n+1}{(k+1)^3}\right], \binomial(n, k)^3\right]}
	\]
}
\end{maplelatex}

\end{maplegroup}

\bigskip\bigskip\noindent
{\bf Commands related to Chapter~\ref{CH:bounds}}

\bigskip
Combining the bounds given in Chapter~\ref{CH:bounds}, we implemented
Algorithm~\ref{ALG:breductionct} as the command
\textsf{BoundReductionCT}.  The function of this command is
illustrated as follows.

Consider Example~\ref{EX:sharperub} with $\alpha=5$.

\medskip
\begin{maplegroup}
  \begin{mapleinput}
    \mapleinline{active}{1d}{alpha:=5: T:=1/((n-alpha*k-alpha)*(n-alpha*k-2)!);
    }{%
    }
  \end{mapleinput}
  
  \mapleresult
  \begin{maplelatex}
    \mapleinline{inert}{2d}{T:=1/((n-5*k-5)*(n-5*k-2)!);
    }{%
      \[
      T:={\displaystyle \frac{1}{(-5 k+n-5)(-5k+n-2)!}}
      \]
    }
  \end{maplelatex}  
\end{maplegroup}

\medskip
In Maple, the built-in command for the algorithm {\em
  LowerBound}~\cite{AbLe2005} is also named \textsf{LowerBound} in the
package \textbf{SumTools[Hypergeometric]}. With only three arguments,
it returns a lower order bound of the telescopers for a given
hypergeometric term,
\index{SumTools[Hypergeometric]@\textbf{SumTools[Hypergeometric]}! LowerBound@\textsf{LowerBound}}%

\medskip
\begin{maplegroup}
  \begin{mapleinput}
    \mapleinline{inert}{1d}{SumTools[Hypergeometric][LowerBound](T,n,k);
    }{%
    }
  \end{mapleinput}
  
  \mapleresult
  \begin{maplelatex}
    \mapleinline{inert}{2d}{2
    }{%
      \[
      {\displaystyle 2}
	\]
      }
    \end{maplelatex}
  \end{maplegroup}
  
\medskip\noindent
Moreover, by specifying a fourth and a fifth argument, the command
also gives information about telescopers as well as certificates.
  
\medskip
\begin{maplegroup}
  \begin{mapleinput}
    \mapleinline{inert}{1d}{SumTools[Hypergeometric][LowerBound](T,n,k,Sn,'Zpair'); 
      Zpair;
    }{%
    }
  \end{mapleinput}
  
  \mapleresult
  \begin{maplelatex}
    \mapleinline{inert}{2d}{2
    }{%
      \[
      {\displaystyle 2}
      \]
    }
  \end{maplelatex}
  \begin{maplelatex}
    \mapleinline{inert}{2d}{[Sn^5-1, (Product(-(5*_k-n-3)*(5*_k-n-2)*(5*_k-n-1)*(5*_k-n+1)*(5*_k-n),
      _k=0 .. k-1))/(GAMMA(n+4)*(5*k-n))]
    }{%
      \[\hspace{-20pt}
      \maplemultiline{
        \left[{\displaystyle Sn^5-1,\, \left[\frac{1}{\Gamma(n+4)\, (5k-n)}
              \left(\prod_{ \, \_k=0}^{k-1}(-(5 \, \_k-n-3)(5 \, \_k-n-2)\right.\right.}\right.\\
        \\
        \\
        \qquad
        \left. {\displaystyle\left. \vphantom{\prod_{ \, \_k=0}^{k-1}(-(5 \, \_k-n-3)(5 \, \_k-n-2)}
              (5\, \_k-n-1)(5 \, \_k-n+1)(5 \, \_k-n)\right)}\right]
      }
      \]
    }
  \end{maplelatex}
\end{maplegroup}

\medskip
In the same fashion, our implementation for Algorithm~\ref{ALG:breductionct}, namely the command \textsf{BoundReductionCT}, with three arguments specified returns an upper bound as well as a lower bound for the order of minimal telescopers for a given hypergeometric term.%
\index{ShiftReductionCT@\textbf{ShiftReductionCT}! BoundReductionCT@\textsf{BoundReductionCT}}%
\index{BoundReductionCT@\textsf{BoundReductionCT}~{\em (ShiftReductionCT)}}%

\medskip
\begin{maplegroup}
	\begin{mapleinput}
		\mapleinline{inert}{1d}{BoundReductionCT(T,n,k);
		}{%
	}
\end{mapleinput}

\mapleresult
\begin{maplelatex}
	\mapleinline{inert}{2d}{}{%
		\[
		{\displaystyle [5,\, 10]}
		\]
	}
\end{maplelatex}
\end{maplegroup}

\medskip
In addition, depending on the numbers of specified arguments and the
specifications, the command performs in the same manner as the command
\textsf{ReductionCT} introduced above. To be precise, we have the
following commands.

\medskip
\begin{maplegroup}
  \begin{mapleinput}
    \mapleinline{inert}{1d}{BoundReductionCT(T,n,k,Sn);
    }{%
    }
  \end{mapleinput}

  \mapleresult
  \begin{maplelatex}
    \mapleinline{inert}{2d}{Sn^5-1
    }{%
      \[
      \maplemultiline{
        {\displaystyle Sn^{5}	-1}\\
      }
      \]
    }
  \end{maplelatex}
\end{maplegroup}

\medskip
\begin{maplegroup}
  \begin{mapleinput}
    \mapleinline{inert}{1d}{res:=BoundReductionCT(T,n,k,Sn,output=normalized): 
    }{%
    }
  \end{mapleinput}
  
  \mapleresult
  \begin{maplelatex}
    \mapleinline{inert}{2d}{res:=[Sn^5-1, [5/((n-5*k)^2*(n-5*k+3)*(n+2-5*k)
      *(n+1-5*k)*(n-1-5*k)),-1/(5*factorial(n-5*k-2))]]
    }{%
      \[
      \hspace{35pt}
      \maplemultiline{res:=\left[{\displaystyle Sn^{5}-1},\, 
          \left[{\displaystyle \frac{5}{(5k-n)^2(5k-3-n)(5k-n-2)(5k-1-n)(5k-n+1)}},\,
          \right.\right.\\
        \\
        \\
        \qquad
        {\displaystyle\left.\left.-\, \frac{1}{5(-5k+n-2)!} \right]\right]}
      }
      \]
    }
  \end{maplelatex}
\end{maplegroup}

\medskip
\begin{maplegroup}
  \begin{mapleinput}
    \mapleinline{inert}{1d}{res:=BoundReductionCT(T,n,k,Sn,output=unnormalized); 
    }{%
    }
  \end{mapleinput}
  
  \mapleresult
  \begin{maplelatex}
    \mapleinline{inert}{2d}{res:=[Sn^5-1, [-5/(n-5*k-5)+5/(n-1-5*k)+20/((n-5*k-5)*(n-1-5*k))
      +5/((n-5*k)^2*(n-5*k+3)*(n+2-5*k)*(n+1-5*k)*(n-1-5*k)),-1/(5*factorial(n-5*k-2))]]
    }{%
      \[
      \hspace{5pt}
      \maplemultiline{res:=\left[{\displaystyle Sn^{5}-1},\, 
          \left[{\displaystyle\frac{5}{5k-n+5}-\frac{5}{5k-n+1}+\frac{20}{(5k-n+5)(5k-n+1)}}
          \right.\right.\\
        \\
        \\
        \qquad
        {\displaystyle+\, \frac{5}{(5k-n)^2(5k-3-n)(5k-n-2)(5k-1-n)(5k-n+1)}},\,\\
        \\
        \\
        \qquad
        {\displaystyle\left.\left.-\, \frac{1}{5(-5k+n-2)!} \right]\right]}
      }
      \]
    }
  \end{maplelatex}
\end{maplegroup}
\index{ShiftReductionCT@\textbf{ShiftReductionCT}|)}

\chapter[Comparison of Memory Requirements]{Comparison of \\Memory Requirements}
\label{APP:memory}
	
In this section, we collect all comparisons\index{comparison} of
memory requirements between our new procedures from the
\textbf{ShiftReductionCT} package (see Appendix~\ref{APP:guide}) and
Maple's implementations of known
algorithms. \index{ShiftReductionCT@\textbf{ShiftReductionCT}} All
memory requirements\index{memory requirement} are obtained by the
Maple command

\medskip
\begin{maplegroup}
  \begin{mapleinput}
    \mapleinline{inert}{1d}{kernelopts("bytesused");
    }{%
    }
  \end{mapleinput}
\end{maplegroup}

\medskip\noindent
and measured in bytes on a Linux computer\index{Linux computer} with
388Gb RAM and twelve 2.80GHz Dual core processors.  Recall that
\begin{itemize}
\item \textsf{G}: the procedure \textsf{Gosper} in
  \textbf{SumTools[Hypergeometric]}, which is based on Gosper's
  algorithm;
  \index{SumTools[Hypergeometric]@\textbf{SumTools[Hypergeometric]}! Gosper@\textsf{Gosper}}

\smallskip
\item \textsf{AP}: the procedure \textsf{SumDecomposition} in
  \textbf{SumTools[Hypergeometric]},
  \index{SumTools[Hypergeometric]@\textbf{SumTools[Hypergeometric]}! SumDecomposition@\textsf{SumDecomposition}} which is based on the
  \ap reduction;

\smallskip
\item \textsf{Z}: the procedure
  \textbf{SumTools[Hypergeometric]}[\textsf{Zeilberger}], which is
  based on Zeilberger's algorithm;
  \index{SumTools[Hypergeometric]@\textbf{SumTools[Hypergeometric]}! Zeilberger@\textsf{Zeilberger}}

\smallskip
\item \textsf{S}: the procedure~\textsf{IsSummable}
  \index{ShiftReductionCT@\textbf{ShiftReductionCT}! IsSummable@\textsf{IsSummable}} in \textbf{ShiftReductionCT},
  which determines hypergeometric summability in a similar way as
  Gosper's algorithm;

\smallskip
\item \textsf{MAP}: the
  procedure~\textsf{ModifiedAbramovPetkovsekReduction}
  \index{ShiftReductionCT@\textbf{ShiftReductionCT}! ModifiedAbramovPetkovsekReduction@\textsf{ModifiedAbramovPetkovsekReduction}}
  in \textbf{ShiftReductionCT}, which is based on the modified
  reduction.

\smallskip
\item \textsf{RCT$_{tc}$}: the procedure~\textsf{ReductionCT} in
  \textbf{ShiftReductionCT}, which computes a minimal telescoper and
  a corresponding normalized
  certificate;\index{ShiftReductionCT@\textbf{ShiftReductionCT}! ReductionCT@\textsf{ReductionCT}}

\smallskip
\item \textsf{RCT$_{t}$}: the procedure~\textsf{ReductionCT} in
  \textbf{ShiftReductionCT}, which computes a minimal telescoper
  without constructing a certificate.

\smallskip
\item {\sf BRCT$_{tc}$}: the procedure~\textsf{\brct} in
  \textbf{ShiftReductionCT},
  \index{ShiftReductionCT@\textbf{ShiftReductionCT}! BoundReductionCT@\textsf{BoundReductionCT}} which computes a
  minimal telescoper and a corresponding normalized certificate;

\smallskip
\item \textsf{BRCT$_{t}$}: the procedure~\textsf{\brct} in
  \textbf{ShiftReductionCT}, which computes a minimal telescoper
  without constructing a certificate.

\smallskip
\item \textsf{LB}: the lower bound for telescopers given in
  Theorem~\ref{THM:lowerbound}.

\smallskip
\item \textsf{order}: the order of the resulting minimal telescoper.
  \index{order}
\end{itemize}

\bigskip\noindent
{\bf Tables for Example~\ref{EX:generaltest} and
  Example~\ref{EX:summabletest}.}

\medskip
\begin{table}[!ht]
  \tabcolsep10pt
  \begin{center}
    \begin{tabular}{c|rr|rr}
      $(\lambda, \mu)$ & {\sf{G}}\hspace{20pt} & {\sf{AP}}\hspace{15pt}  
      & {\sf{S}}\hspace{20pt} & {\sf{MAP}} \hspace{8pt} \\ \hline
      $(0, 0)$ & 1.80015e7 & 2.79579e7 & 2.19643e7 & 2.20057e7 \\[1ex]
      $(5, 5)$ & 6.92148e7 & 5.45788e8 & 8.31337e7 & 1.00876e8 \\[1ex]
      $(10, 10)$ & 1.06237e8 & 1.74321e9 & 1.63963e8  & 2.23078e8 \\[1ex]
      $(10, 20)$ & 3.67295e8  & 4.22563e9 & 3.41155e8 & 7.14421e8 \\[1ex]
      $(10, 30)$ & 9.08446e8 & 2.06166e10 & 5.73637e8 & 2.07008e9 \\[1ex]
      $(10, 40)$ & 1.79107e9  & 3.74146e10  & 8.60492e8 & 5.01724e9 \\[1ex]
      $(10, 50)$ & 3.19600e9 & 4.98811e10 & 1.16624e9 & 9.80644e9 \\
      \hline
    \end{tabular}
  \end{center}
  \caption{Memory comparison of Gosper's algorithm, the Abramov-Petkov\v sek reduction 
    and the modified version for random hypergeometric terms (in bytes)}\label{TAB:mgeneraltest}
\end{table}

\begin{table}[!ht]
  \tabcolsep10pt
  \begin{center}
    \begin{tabular}{c|rr|rr}
      $(\lambda, \mu)$ & {\sf{G}}\hspace{20pt} & {\sf{AP}}\hspace{15pt}  
      & {\sf{S}}\hspace{20pt} & {\sf{MAP}} \hspace{8pt} \\ \hline
      $(0, 0)$ & 1.49566e8  & 3.83358e8 & 1.96563e8 & 1.97086e8 \\[1ex]
      $(5, 5)$ & 2.76453e8 & 9.42523e8 & 2.40684e8 & 2.40927e8  \\[1ex]
      $(10, 10)$ &  3.15859e8 & 1.86511e9 & 2.50334e8 & 2.50661e8 \\[1ex]
      $(10, 20)$ & 6.81883e8 & 4.15802e9 & 3.19633e8 & 3.20250e8 \\[1ex]
      $(10, 30)$ & 1.48580e9 & 7.60674e9 & 3.61856e8 & 3.60798e8  \\[1ex]
      $(10, 40)$ & 2.66329e9 & 1.24394e10 & 3.81800e8 & 3.82879e8 \\[1ex]
      $(10, 50)$ & 4.96349e9  & 2.22568e10 & 4.15063e8 & 4.14124e8 \\
      \hline
    \end{tabular}
  \end{center}
  \caption{Memory comparison of Gosper's algorithm, the Abramov-Petkov\v sek reduction 
    and the modified version for summable hypergeometric terms (in bytes)}\label{TAB:msummabletest}
\end{table}

\newpage\bigskip\noindent
{\bf Tables for Example~\ref{EX:test}.}

\bigskip
\begin{table}[!ht]
  \tabcolsep10pt
  \begin{center}
    \begin{tabular}{l|r|rr|c}
      $(d_1, d_2, \alpha, \lambda, \mu)$ & {\sf{Z}}\hspace{20pt} & {\sf{RCT$_{tc}$}}\hspace{8pt} 
      & {\sf{RCT$_{t}$}}\hspace{9pt} & {\sf{order}}\\
      \hline
      $(1, 0, 1, 5, 5)$            &     2.05992e9 &     5.36111e8 &    1.58646e8 & 4     \\[1ex]
      $(1, 0, 2, 5, 5)$            &      6.13485e9 &   3.33929e9 &     9.01651e8 & 6     \\[1ex]
      $(1, 0, 3, 5, 5)$            &   2.05569e10 &    1.12736e10 &    2.59005e9 &  7     \\[1ex]
      $(1, 8, 3, 5, 5)$            &    2.84955e10 &   1.46063e10 &    3.24756e9 &  7     \\[1ex]
      $(2, 0, 1, 5, 10)$           &   3.58374e10 &    6.87524e9 &    6.90891e8 & 4     \\[1ex]
      $(2, 0, 2, 5, 10)$           &   3.03599e10 &   4.30070e10 &   7.44379e9 & 6     \\[1ex]
      $(2, 0, 3, 5, 10)$           &   6.95166e10 &  1.29853e11  &   2.56292e10 &  7     \\[1ex]
      $(2, 3, 3, 5, 10)$           &   7.63196e10 &  1.34622e11 &   2.78371e10 &  7     \\[1ex]
      $(2, 0, 1, 10, 15)$          &  1.72175e11 &   2.44536e10 &    1.52217e9 & 4     \\[1ex]
      $(2, 0, 2, 10, 15)$          &   8.27362e10 &  1.38827e11 &   2.09677e10 & 6     \\[1ex]
      $(2, 0, 3, 10, 15)$          &   1.79564e11 &  4.57813e11 &  1.04973e11 & 7     \\[1ex]
      $(2, 5, 3, 10, 15)$          &  2.01763e11 &   4.49569e11 &   1.06872e11 & 7     \\[1ex]
      $(3, 0, 1, 5, 10)$           &  7.48174e11 &   4.17901e10 &    5.18114e9 & 6     \\[1ex]
      $(3, 0, 2, 5, 10)$           &  3.63162e11 &  2.25463e11 &   5.19205e10  & 8     \\[1ex]
      $(3, 0, 3, 5, 10)$           & 7.60572e11 &  6.16676e11 &  1.78310e11 & 9     \\ \hline
    \end{tabular}
  \end{center}
  \vspace{-\smallskipamount}
  \caption{Memory comparison of Zeilberger's algorithm to reduction-based creative telescoping 
    with and without construction of a certificate (in bytes)}\label{TAB:mcttest}
\end{table}

\newpage\bigskip\noindent
{\bf Tables for Example~\ref{EX:bctvsct1} and
  Example~\ref{EX:bctvsct2}.}

\bigskip
\begin{table}[h]
  \tabcolsep8pt
  \begin{center}
    \begin{tabular}{c|rrrr|cc}
      $\alpha$ & {\sf{RCT$_{t}$}}\hspace{9pt} & {\sf{RCT$_{tc}$}}\hspace{8pt} 
      & {\sf{BRCT$_{t}$}}\hspace{6pt} & {\sf{BRCT$_{tc}$}}\hspace{5pt} & {\sf{LB}} & {\sf{order}}\\
      \hline
      20 & 2.53275e8 & 2.57797e8 & 1.42371e8 & 1.46826e8 & 20 & 20 \\[1ex]
      30 & 1.04691e9 & 1.05593e9 & 4.73815e8 & 4.83413e8 & 30 & 30 \\[1ex]
      40 & 3.16905e9 & 3.18565e9 & 1.31468e9 & 1.33395e9  & 40 & 40 \\[1ex]
      50 & 7.69274e9 & 7.71999e9 & 3.12029e9 & 3.15161e9 & 50 & 50 \\[1ex]
      60 & 1.62442e10 & 1.62819e10 & 6.24941e9 & 6.28674e9 & 60 & 60 \\[1ex]
      70 & 3.15561e10 & 3.16084e10 & 1.19886e10  & 1.20418e10 & 70 & 70 \\
      \hline
    \end{tabular}
  \end{center}
  \vspace{-\smallskipamount}
  \caption{Memory comparison of two reduction-based creative telescoping 
    with and without construction of a certificate for Example~\ref{EX:bctvsct1} 
    (in bytes)}\label{TAB:mbctvsct1}
\end{table}

\bigskip
\begin{table}[h]
  \tabcolsep8pt
  \begin{center}
    \begin{tabular}{l|rrrr|cc}
      $(m, \alpha)$ & {\sf{RCT$_{t}$}}\hspace{9pt} & {\sf{RCT$_{tc}$}}\hspace{8pt} 
      & {\sf{BRCT$_{t}$}}\hspace{6pt} & {\sf{BRCT$_{tc}$}}\hspace{5pt} & {\sf{LB}} & {\sf{order}}\\
      \hline
      (1,1) & 2.64768e7 & 3.12387e7 & 2.64914e7 & 3.12548e7 & 1 & 2 \\[1ex]
      (1,10) & 9.91388e8  & 1.62603e9 & 8.94051e8  & 1.50416e9 & 10 & 11 \\[1ex]
      (1,15) & 2.01112e10 & 2.32990e10 & 1.33834e10 & 1.75427e10 & 15 & 16 \\[1ex]
      (1,20) &  2.23859e11 & 2.43209e11 & 1.13767e11 & 1.29430e11 & 20 & 21 \\[1ex]
      (2,10) & 1.03547e9 & 1.65297e9 & 9.12084e8 & 1.52683e9 & 10 & 11 \\[1ex]
      (2,15) &  2.70850e10 & 3.02579e10 & 1.38753e10 & 1.64594e10 & 15 & 16 \\[1ex]
      (2,20) & 2.37348e11 & 2.48004e11 & 1.29174e11 & 1.41685e11  & 20 & 21 \\
      \hline
    \end{tabular}
  \end{center}
  \vspace{-\smallskipamount}
  \caption{Memory comparison of two reduction-based creative telescoping 
    with and without construction of a certificate for Example~\ref{EX:bctvsct2} 
    (in bytes)}\label{TAB:mbctvsct2}
\end{table}
\end{appendices}

\backmatter 
\phantomsection
\addcontentsline{toc}{chapter}{Bibliography}
\bibliographystyle{plain}
\def\cprime{$'$} \def\cprime{$'$} \def\cprime{$'$} \def\cprime{$'$}
\def\cprime{$'$} \def\cprime{$'$} \def\cprime{$'$} \def\cprime{$'$}
\def\polhk#1{\setbox0=\hbox{#1}{\ooalign{\hidewidth
			\lower1.5ex\hbox{`}\hidewidth\crcr\unhbox0}}} \def\cprime{$'$}

\cleardoublepage
\phantomsection
\addcontentsline{toc}{chapter}{Notation}



\renewcommand{\nomname}{Notation}

\renewcommand{\nompreamble}{The following list describes the most important 
	mathematical notations that have been used in this thesis.
	For each group, the order follows roughly the order of first appearance in the text.}
\newcommand\Nomenclature[3][X]{\nomenclature[#1#3]{#2}{#3}}

\markboth{\nomname}{\nomname}  
\setlength{\nomlabelwidth}{3.5cm}



	
	\Nomenclature[N1]{$\set N$, $\set Z$, $\set Q$, $\set R$, $\set C$}
	{Sets of natural, integer, rational, real, complex numbers}
	\Nomenclature[N2]{$\set Q(i)$}{The Gaussian rational field}
	\Nomenclature[N3]{$\cal D_{R,\set F}$}
	{The set of D-finite numbers with respect to $R$ and $\set F$}
	\Nomenclature[N4]{$\cal D_{\set F}$}
	{The set $\cal D_{\set F,\set F}$}
	\Nomenclature[N5]{$\cal A_{\set F}$}
	{The set of limits of convergent algebraic sequences over~$\set F$}
	\Nomenclature[N6]{$\emptyset$}{The empty set}
	\Nomenclature[N7]{$\cal C_{\set F}$}
	{The set of limits of convergent C-finite sequences over~$\set F$}
	
	\Nomenclature[R1]{$\quot(R)$}{The quotient field of the ring $R$}
	\Nomenclature[R2]{$\set K$}{A field of characteristic zero}
	\Nomenclature[R3]{$\set F$}{A field of characteristic zero, or the field ${\set K}(n)$ (Chapter~\ref{CH:telescoping}), or a subfield of $\set C$ (Chapter~\ref{CH:dfinitenos})}
	\Nomenclature[R4]{$\set F(k)$}{The field of univariate rational functions in $k$ over~$\set F$}
	\Nomenclature[R5]{$\set F[k]$}{The ring of univariate polynomials in $k$ over~$\set F$}
	\Nomenclature[R6]{$\set D$}{A difference ring extension of $\set \bF(k)$}
	\Nomenclature[R7]{$\set K(n,k)$}{The field of bivariate rational functions in $n,k$ over $\set K$}
	\Nomenclature[R8]{$\set F[n]\langle S_n\rangle$}{The Ore algebra of linear recurrence operators with polynomial coefficients w.r.t.~$n$}
	\Nomenclature[R9]{$\set K[n,k]$}{The ring of bivariate polynomials in $n,k$ over $\set K$}
	\Nomenclature[Ra]{$\set K(n)[k]$}{The ring of polynomials in $k$ over the field $\set K(n)$}
	\Nomenclature[Rb]{$R$}{A subring of $\set C$}
	\Nomenclature[Rc]{$R[[z]]$}{The ring of formal power series over~$R$}
	\Nomenclature[Rd]{$R^{\set N}$}{The ring of all sequences from $\set N$ to~$R$}
	\Nomenclature[Re]{$\set F[z]\langle D_z\rangle$}{The Ore algebra of linear differential operators with polynomial coefficients wr.t.~$z$}
	\Nomenclature[Rf]{$\bar{\set F}$}{The algebraic closure of the field $\set F$}
	
	\Nomenclature[O1]{$\sigma_k$}{The shift operator w.r.t.~$k$ which maps $r(k)$ to $r(k+1)$ for every rational function $r\in \set F(k)$}
	\Nomenclature[O2]{$\Delta_k$}{The difference of $\sigma_k$ and the identity map}
	\Nomenclature[O3]{$S_n$}{The operator in the ring of linear recurrence operators over~$\set F$ which satisfies $S_n r= \sigma_n(r) S_n$ for all $r \in \set F$.}
	\Nomenclature[O4]{$\sum_{j=0}^{\rho} p_j S_n^j$}{A recurrence operator with polynomial coefficients $p_j$}
	\Nomenclature[O5]{$D_z$}{The derivation operator w.r.t.~$z$ which maps a power series or function~$f(z)$ to its derivative $f'(z)=\frac{d}{dz}f(z)$}
	\Nomenclature[O6]{$\sum_{j=0}^{\rho} p_j D_z^j$}{A differential operator with polynomial coefficients $p_j$}

	\Nomenclature[A1]{$\gcd$}{The greatest common divisor}
	\Nomenclature[A2]{$\min$}{The minimum}
	\Nomenclature[A3]{$\max$}{The maximum}
	\Nomenclature[A4]{$p\mid q$}{A polynomial $p$ divides a polynomial $q$ over the domain where the polynomials live}
	\Nomenclature[A5]{$p\nmid q$}{A polynomial $p$ does not divide a polynomial $q$ over the domain where the polynomials live}
	\Nomenclature[A6]{$\log$}{The natural logarithm}
	\Nomenclature[A7]{$\exp$}{The exponential function}
	
	
	\Nomenclature[Sa]{$\sum_{j=a}^b f(k)$}{The sum $f(a) + f(a+1) + \dots + f(b)$}
	\Nomenclature[Sb]{$\deg_k(p)$}{Degree of a polynomial $p$ w.r.t.~$k$}
	\Nomenclature[Sc]{$\lc_k(p)$}{Leading coefficient of a polynomial $p$ w.r.t.~$k$}
	\Nomenclature[Sd]{$k"!$, ${\displaystyle {n\choose k}}$}{Factorial $k"! = 1\cdot 2 \cdot 3 \dots (k-1)\cdot k$ and binomial coefficient ${n\choose k} = n(n-1)\dots(n-k+1)/k"!$}
	\Nomenclature[Sda]{$A\setminus B$}{The relative complement of a set $B$ with respect to a set $A$}
	\Nomenclature[Se]{$\set U_T$}{The union of $\{0\}$ and the set of summable hypergeometric terms that are similar to a hypergeometric term~$T$}
	\Nomenclature[Sf]{$\set V_K$}{The set $\{K\sigma_k(r)-r\mid r\in \set F(k)\}$ where $K$ is a shift-reduced rational function in $\set F(k)$}
	\Nomenclature[Sg]{$A \equiv_k B \mod C_k$}{The expression $A-B$ belongs to a set $C_k$}
	\Nomenclature[Sh]{$\phi_K$}{The map for polynomial reduction with respect to a shift-reduced rational function $K$}
	\Nomenclature[Si]{$\im(\phi_K)$}{The image space of the map $\phi_K$}
	\Nomenclature[Sj]{$\set W_K$}{The standard complement of $\im(\phi_K)$}
	\Nomenclature[Sk]{$A\oplus B$}{The direct sum of two vector spaces $A$ and $B$}
	\Nomenclature[Sl]{$A\cap B$}{The intersection of two sets $A$ and $B$}
	\Nomenclature[Sm]{$A\cup B$}{The union of two sets $A$ and $B$}
	\Nomenclature[Sn]{$\lvert \cP\rvert$}{The number of elements of the set~$\cP$}
	\Nomenclature[So]{$\llbracket\varphi\rrbracket$}{The Iversion bracket, namely $\llbracket\varphi\rrbracket$ equals~$1$ if the expression~$\varphi$ is true, otherwise it is~$0$.}
	\Nomenclature[Sp]{$\prod_{j=a}^b f(k)$}{The product $f(a)f(a+1)\dots f(b)$}
	\Nomenclature[Sq]{$p\sim_k q$}{A polynomial $p$ is shift-equivalent to a polynomial $q$ w.r.t.~$k$}
	\Nomenclature[Sr]{$p\approx_k q$}{A shift-free polynomial $p$ is shift-related to a shift-free polynomial $q$ w.r.t.~$k$}
	\Nomenclature[Ss]{$L(T)$}{The application of a recurrence operator $L$ to a hypergeometric term~$T$}
	\Nomenclature[St]{$p\sim_{n,k} q$}{A polynomial $p$ is shift-equivalent to a polynomial $q$ w.r.t.~$n$ and $k$}
	\Nomenclature[Su]{$\lvert\xi\rvert$}{The modulus of a complex number $\xi$}
	\Nomenclature[Sv]{$\delta^{(\lambda, \mu)}$}{The operator $\sigma_n^\alpha \sigma_k^\beta$ where $\lambda,\mu$ are coprime integers and $\alpha\lambda + \beta \mu =1$ with $\lvert\alpha\rvert<\lvert\mu\rvert$ and $\lvert\beta\rvert<\lvert\lambda\rvert$}
	\Nomenclature[Sva]{$\dim_{\bK(n)}(\bW_K)$}{The dimension of the vector space $\bW_K$ over the field $\bK(n)$}
	\Nomenclature[Sw]{$\sum_{n=0}^\infty a_n z^n$}{A power series with the coefficient sequence $(a_n)_{n=0}^\infty$}
	\Nomenclature[Sx1]{$(a_n)_{n=0}^\infty$}{An infinite sequence $a_0, a_1, a_2, \dots$}
	\Nomenclature[Sx]{$f'(z)$}{The first derivative of a power series or function~$f(z)$ w.r.t.~$z$}
	\Nomenclature[Sy]{$\lc(L)$}{The leading coefficient of an operator $L$}
	\Nomenclature[Sz0]{$L\cdot a_n$}{The application of a recurrence operator $L$ to an infinite sequence~$(a_n)_{n=0}^\infty$}
	\Nomenclature[Sza]{$L\cdot f(z)$}{The application of a differential operator $L$ to a power series~$f$}
	\Nomenclature[Szb]{$f\circ g$}{The composition $f(g)$ of functions $f$ and $g$}
	\Nomenclature[Szc]{$A\subseteq B$}{A set $A$ is contained by a set $B$}
	\Nomenclature[Szd]{$\bar \xi$}{The complex conjugation of a complex number $\xi$}
	\Nomenclature[Sze]{$\real(\xi)$}{The real part of a complex number $\xi$}
	\Nomenclature[Szf]{$\imag(\xi)$}{The imaginary part of a complex number $\xi$}
	\Nomenclature[Szg]{$a_n\sim b_n$ ($n\to\infty$)}{The quotient $a_n/b_n$ converges to $1$ as $n\to\infty$}
	\Nomenclature[Szh]{$f(z)\sim g(z)$ ($z\to\zeta$)}{The quotient $f(z)/g(z)$ converges to~$1$ as $z$ approaches~$\zeta$}
	\Nomenclature[Szi]{$[z^n]f(z)$}{The coefficient of $z^n$ in a power series $f(z)\in\bF{[[z]]}$}
	\Nomenclature[Szj]{$f^{(k)}(z)$}{The $k$th derivative of a power series or function~$f(z)$ w.r.t.~$z$}
	
	\printnomenclature

\cleardoublepage
\phantomsection
\addcontentsline{toc}{chapter}{Index}
\setlength{\columnsep}{.5cm}
\printindex

\chapter*{Curriculum Vitae}
\label{CH:cv}
\markboth{Curriculum Vitae}{}


\section*{Personal data}

\begin{tabular}{ll}
	Name & Hui HUANG\\
	Gender & Female\\
    Date of birth & August 28, 1989\\
    Place of birth & Fujian province, China\\
    Nationality & People's Republic of China
\end{tabular}

\section*{Contact}

\begin{tabular}{ll}
	Address & Institute for Algebra\\
	& Johannes Kepler University Linz\\
	& Altenberger Stra{\ss}e 69\\
	& 4040 Linz, Austria\\
	Email & huanghui@amss.ac.cn
\end{tabular}

\section*{Education}
	\begin{tabularx}{\textwidth}{lX}
		2013--2017  & Doctorate studies in symbolic computation at the FWF Doctoral Program \lq\lq Computational Mathematics\rq\rq\ (Johannes Kepler University, Linz, Austria) and  Key Laboratory of Mathematics  Mechanization (University of Chinese Academy of Sciences, Beijing, China)\\
		& Co-supervisors:  Manuel Kauers and Ziming Li\\[1ex]
		2011--2013  &  Studies in symbolic computation at Key Laboratory of
		Mathematics  Mechanization,
		University of Chinese Academy of Sciences, Beijing, China\\
		& Supervisor: Ziming Li\\[1ex]
		2007--2011 & Studies in applied mathematics at Xiamen University, Fujian, China
	\end{tabularx}

\section*{Awards}

\begin{tabular}{ll}
	2016 & Distinguished Female Student Award at ISSAC 2016\\[1ex]
    2014 & Distinguished Poster Award at ISSAC 2014 \\
    &(together with S.\ Chen and Z.\ Li)
\end{tabular}

\section*{Scientific Work}
\subsection*{Refereed Publications}
\begin{enumerate}[leftmargin=0.6cm]
	\item 
	Hui Huang.
	\newblock New bounds for hypergeometric creative telescoping.
	\newblock In {\em I{SSAC} 2016---{P}roceedings of the 41st {I}nternational
	{S}ymposium on {S}ymbolic and {A}lgebraic {C}omputation}, pages 279--286. ACM, New York, 2016.
	
	\medskip
	\item
	Shaoshi Chen, Hui Huang, Manuel Kauers, and Ziming Li.
	\newblock A modified {A}bramov-{P}etkov{\v s}ek reduction and creative telescoping for hypergeometric terms.
	\newblock In {\em  I{SSAC} 2015---{P}roceedings of the 40th {I}nternational
              {S}ymposium on {S}ymbolic and {A}lgebraic {C}omputation}, pages 117--124. ACM, New York, 2015.
\end{enumerate}

\subsection*{Posters}
\begin{itemize}[leftmargin=0.6cm]
	\item
	Shaoshi Chen, Hui Huang, and Ziming Li.
	\newblock Improved {A}bramov-{P}etkov{\v s}ek's Reduction and Creative Telescoping for Hypergeometric Terms (Poster at ISSAC 2014).
	\newblock In {\em ACM Commun. Comput. Algebra}, 48(3/4):106-108. ACM, New York, 2014.
\end{itemize}

\subsection*{Talks}

\begin{tabularx}{\textwidth}{lX}
	July 2016
	&\emph{New Bounds for Hypergeometric Creative Telescoping}.
	ISSAC 2016, Waterloo, Canada.\\[1ex]

    June 2016
	&\emph{Reduction and Creative Telescoping for Hypergeometric Terms}.
	Center for Combinatorics Seminar, Tianjin, China.\\[1ex]


    November 2015
	&\emph{Two Applications of the Modified Abramov-Petkov\v{s}ek Reduction}.
	CM 2015, Hefei, China.\\[1ex]

    July 2015
	&\emph{A Modified {A}bramov-{P}etkov{\v s}ek Reduction and Creative Telescoping for Hypergeometric Terms}.
	ISSAC 2015, Bath, United Kingdom.\\[1ex]

    June 2015
	&\emph{Abramov-Petkov\v{s}ek Reduction and Creative Telescoping for Rational Functions}.
	CanaDAM 2015, Saskatoon, Canada.\\[1ex]




    August 2013
	&\emph{An Improved Abramov-Petkov{\v s}ek Reduction for Hypergeometric Terms}.
	CM 2013, Changchun, China.
\end{tabularx}

\end{document}